\documentclass[twocolumn,english,aps,prb,twocolum,superscriptaddress,bibnotes,amsmath,amssymb,floatfix]{revtex4-2}
\pdfoutput=1
\usepackage{amsfonts}
\usepackage{amsmath}
\usepackage{amssymb}
\usepackage[colorlinks=true,citecolor=blue,linkcolor=blue,breaklinks=true]{hyperref}
\usepackage{graphicx}

\usepackage{soul}
\usepackage{url}
\usepackage{dcolumn}
\usepackage{bm}
\usepackage{stmaryrd}
\usepackage[final]{changes}
\usepackage{multirow}
\usepackage{rotating}
\usepackage{titlesec}
\usepackage{afterpage}

\usepackage{titlesec}
\renewcommand{\thesection}{\arabic{section}}
\titleformat{\section}
{\normalfont\large\bfseries}{Supplementary Note~\thesection.}{1em}{}

\begin{document}

\title{Topological lattices realized in superconducting circuit optomechanics}

\author{Amir Youssefi}\thanks{These authors contributed equally.}
\author{Shingo Kono} \thanks{These authors contributed equally.}
\author{Andrea Bancora}\thanks{These authors contributed equally.}
\author{Mahdi Chegnizadeh}
\author{Jiahe Pan}
\author{Tatiana Vovk}
\author{Tobias~J.~Kippenberg}
\email[]{tobias.kippenberg@epfl.ch}
\affiliation{Laboratory of Photonics and Quantum Measurement (LPQM), Swiss Federal Institute of Technology Lausanne (EPFL), Lausanne, Switzerland}
\affiliation{Center for Quantum Science and Engineering, EPFL, Lausanne, Switzerland}

\maketitle

\textbf{Cavity optomechanics enables controlling mechanical motion via radiation pressure interaction~\cite{RMP_optomechanics}, and has contributed to the quantum control of engineered mechanical systems ranging from kg scale LIGO mirrors to nano-mechanical systems, enabling ground-state preparation~\cite{teufel2011sideband,chan2011laser},
	entanglement~\cite{kotler2021direct,ockeloen2018stabilized}, squeezing of mechanical objects~\cite{wollman2015quantum}, position measurements at the standard quantum limit~\cite{teufel2009nanomechanical}, and quantum transduction~\cite{andrews2014bidirectional}.
	Yet, nearly all prior schemes have employed single- or few-mode optomechanical systems. In contrast, novel dynamics and applications are expected when utilizing optomechanical lattices~\cite{peano2015topological}, which enable to synthesize non-trivial band structures, and have been actively studied in the field of circuit QED~\cite{carusotto2020photonic}.
	Superconducting microwave optomechanical circuits~\cite{teufel2011sideband} are a promising platform to implement such lattices, but have been compounded by strict scaling limitations.
	Here, we overcome this challenge and demonstrate topological microwave modes in 1D circuit optomechanical chains realizing the Su-Schrieffer-Heeger (SSH) model~\cite{asboth2016short,ozawa2019topological}. Furthermore, we realize the strained graphene model~\cite{pereira2009tight,naumis2017electronic} in a 2D optomechanical honeycomb lattice.
	Exploiting the embedded optomechanical interaction, we show that it is possible to directly measure the mode functions of the hybridized modes without using any local probe~\cite{underwood2016imaging,wang2019mode}. This enables us to reconstruct the full underlying lattice Hamiltonian and directly measure the existing residual disorder.
	Such optomechanical lattices, accompanied by the measurement techniques introduced, offers an avenue to explore collective~\cite{heinrich2011collective,xuereb2012strong}, quantum many-body~\cite{ludwig2013quantum}, and quench~\cite{raeisi2020quench} dynamics, topological properties~\cite{peano2015topological,zangeneh2020topological} and more broadly, emergent nonlinear dynamics in complex optomechanical systems with a large number of degrees of freedoms~\cite{akram2012photon,sanavio2020nonreciprocal,tomadin2012reservoir}.
}

Mechanical oscillators can exhibit modes with ultra-low mechanical dissipation and compact form factors due to the slow velocity of acoustic waves, and are already used in applications ranging from timing to wireless filters. Over the past decade, novel ways in which mechanical systems can be quantum controlled have been developed, based on either coupling to electromagnetic cavities in quantum optomechanics~\cite{RMP_optomechanics} or superconducting qubits in quantum acoustics~\cite{o2010quantum}. The former route has utilized the coupling to electromagnetic cavities both in the optical and microwave domains, and enabled to reach a regime where the quantum nature of the optomechanical interaction becomes relevant~\cite{teufel2009nanomechanical}. This has allowed a host of manipulations of mechanical systems, including cooling mechanical systems to the ground state~\cite{teufel2011sideband,chan2011laser}, state transfer~\cite{palomaki2013coherent}, remote entanglement of mechanical oscillators~\cite{riedinger2018remote}, or generation of squeezed mechanical states ~\cite{wollman2015quantum}. In addition, such systems have been employed for quantum microwave to optical transduction ~\cite{andrews2014bidirectional}.

The majority of optomechanical systems, which have probed classical as well as quantum properties and dynamics, have utilized few-mode optomechanics, i.e.,\ systems that employ a small number of optomechanical degrees of freedom.
Pioneering theoretical works have predicted that significantly richer and novel dynamics can be accessed in optomechanical lattices including topological phases of light and sound~\cite{peano2015topological},  collective~\cite{heinrich2011collective,xuereb2012strong,roque2017anderson} and quench~\cite{raeisi2020quench} dynamics, quantum many-body dynamics~\cite{ludwig2013quantum} and entanglement~\cite{akram2012photon}, non-reciprocity~\cite{sanavio2020nonreciprocal}, reservoir engineering~\cite{tomadin2012reservoir}, and topological phonon transport~\cite{ren2020topological}.
To date, such optomechanical lattices have only been realized via mechanically mediated coupling~\cite{safavi2014two} - similar to studies that employ mechanical meta-materials~\cite{yang2015topological}. Indeed, while the coupling of mechanical oscillators for acoustic metamaterials has been successfully achieved~\cite{huber2016topological,surjadi2019mechanical}, implementing optomechanical lattices has been a long-lasting challenge.  Topological phonon transport has recently been reported in optomechanical crystals~\cite{ren2020topological}, consisting of coupled mechanical oscillators. However, site-by-site engineerable optomechanical lattices have not yet been realized due to the stringent requirements on identical individual optomechanical sites. To realize optomechanical lattices that include photon transport, it is imperative that the disorder in the optical (or microwave) cavity is sufficiently small to enable building lattice models.

Superconducting circuit optomechanical systems based on vacuum-gap capacitors~\cite{cicak2010low} are a very promising platform to realize such lattices, and have been employed in a wide range of experiments, including ground-state cooling~\cite{teufel2011sideband}, mechanical squeezing~\cite{wollman2015quantum}, entanglement~\cite{ockeloen2018stabilized,kotler2021direct, de2021quantum} of mechanical motion, dissipative quantum reservoir engineering~\cite{Toth2017}, the realization of hybrid qubit-mechanical systems~\cite{palomaki2013coherent, pirkkalainen2013hybrid, reed2017faithful}, as well as non-reciprocal microwave devices~\cite{Bernier2017}. While microwave planar resonators have been coupled and been used to create topological waveguides~\cite{mirhosseini2018superconducting,kim2021quantum}, it has to date not been possible to realize optomechanical lattices in a similar fashion due to the technical challenge of reliably fabricating multiple vacuum-gap capacitors, with identical mechanical and microwave properties.
Here we overcome this challenge and demonstrate circuit optomechanical lattices. We use them to implement a 1D chain with a topological band structure exhibiting topologically protected edge states~\cite{asboth2016short,ozawa2019topological}, as well as a 2D honeycomb lattice realizing the strained graphene model with edge states~\cite{pereira2009tight,naumis2017electronic,ni2008uniaxial,rechtsman2013topological,delplace2011zak}.
Using the on-site optomechanical interactions, we are able to perform a direct measurement of the collective microwave modeshapes and reconstruct the full Hamiltonian of such a multimode system, addressing an experimental challenge in large-scale multimode superconducting circuits where only indirect approaches were performed by near field scanning probes~\cite{underwood2016imaging}, laser scanning microscopy~\cite{wang2019mode,morvan2021bulk}, or dispersive coupling to qubits~\cite{kim2021quantum}.

\subsection*{Multimode optomechanics in lattices}
\begin{figure*}[ht]
	\includegraphics[width=\textwidth]{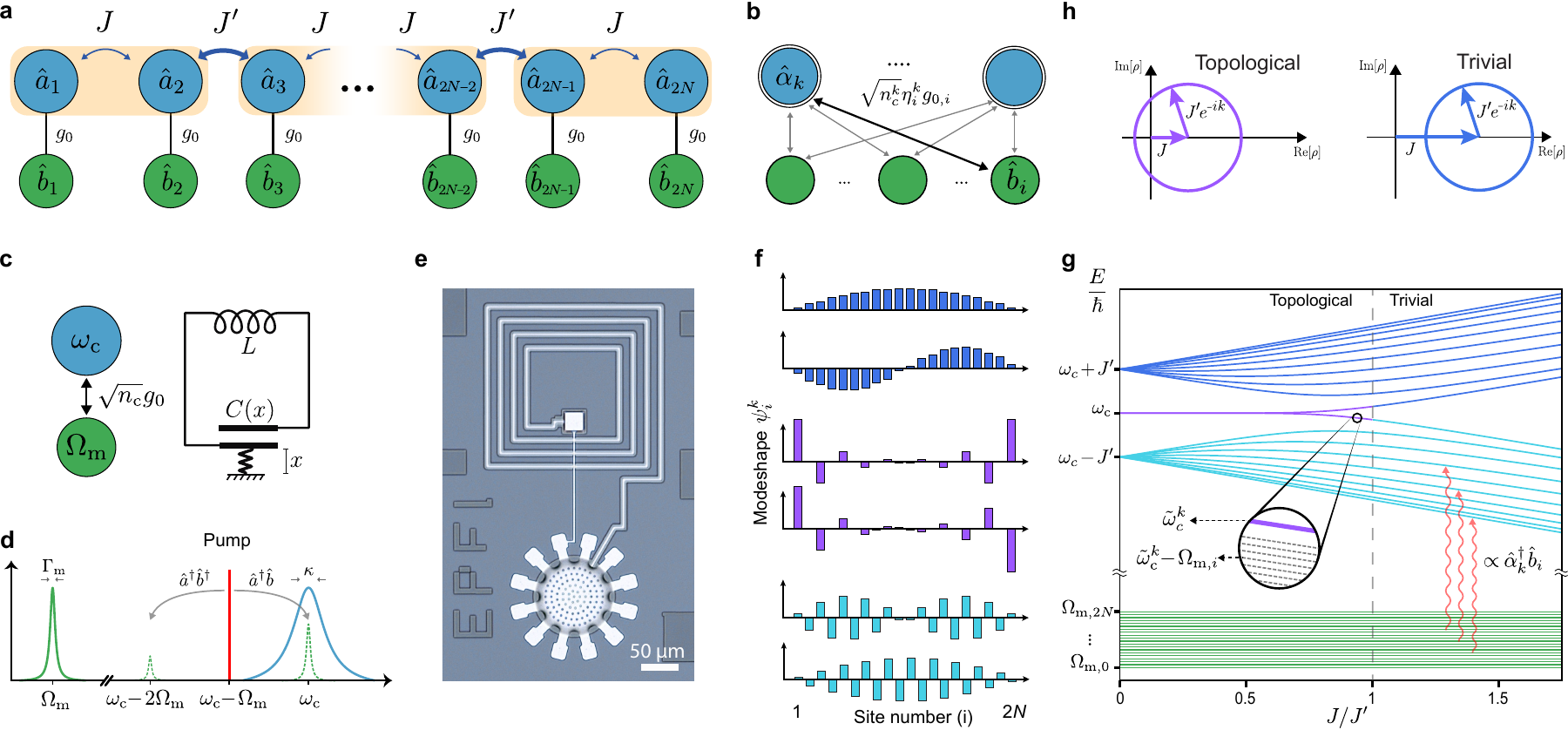}
	\caption{\footnotesize \linespread{1} \textbf{Optomechanical lattices composed of superconducting circuit optomechanical systems.} \textbf{a}, Mode diagram of an optomechanical array with staggered mutual couplings demonstrating the Su-Schrieffer-Heeger model. Electromagnetic and mechanical modes are shown by blue and green circles correspondingly. \textbf{b}, Equivalent mode diagram in the collective electromagnetic basis. The collective electromagnetic modes are coupled to all the mechanical resonators by the effective optomechanical coupling rates which are proportional to their energy participation ratio, $\eta_i^k$. \textbf{c}, The equivalent circuit representation of an optomechanical site. \textbf{d}, Red-detuned pumping on the lower sideband, generating thermomechanical sidebands \textbf{e}, Microscope image of an optomechanical circuit with a mechanically compliant capacitor. \textbf{f,g},~Modeshapes and energy levels of a 1D SSH chain versus mutual coupling rates ratio. The optomechanical damping of mechanical oscillator $i$ can be induced by sideband driving collective electromagnetic mode $k$. Collective electromagnetic modeshape examples of the two highest and lowest frequency modes as well as topological edge states shown in \textbf{f} are calculated for $J/J' = 0.5$. \textbf{h}, Off-diagonal element $\rho(k)$ of the bulk Hamiltonian for the SSH model in the cases of the topological and trivial phases, respectively.\label{fig:1}}
\end{figure*}
As shown in Fig.~\ref{fig:1}a, we first theoretically consider an arbitrary lattice composed of optomechanical systems, whereby electromagnetic modes are mutually coupled with coupling rates $J_{ij}$. Each optomechanical system consists of a mechanical oscillator with a frequency $\Omega_{\mathrm{m},i}$ and an electromagnetic mode with a frequency $\omega_{\mathrm{c},i}$, coupled via radiation-pressure force with a single-photon optomechanical coupling rate $g_{0,i}$, where the photon-phonon interaction can be induced by pumping the electromagnetic mode, leading to the effective optomechanical coupling rate $g = \sqrt{n_\mathrm{c}} g_0$~\cite{RMP_optomechanics}, enhanced by the mean intracavity photon number $n_\mathrm{c}$~(see Figs.~\ref{fig:1}c and d).
Figure~\ref{fig:1}e shows the physical realization of the optomechanical site in the microwave superconducting circuit platform, consisting of an LC circuit with a mechanically compliant vacuum-gap capacitor~\cite{teufel2011sideband}.
The Hamiltonian of such a lattice is described by
\begin{equation}
	\label{eq:hamiltonian_tot_nonD}
	\begin{split}
		\hat{H}/\hbar =&\sum_{i} \left( \omega_{\mathrm{c},i} \hat{a}^\dagger_i\hat{a}_i + \Omega_{\mathrm{m},i}\hat{b}^\dagger_i\hat{b}_i + g_{0,i} \hat{a}^\dagger_i\hat{a}_i (\hat{b}^\dagger_i+\hat{b}_i)  \right) \\
		+&\sum_{i \neq j} \left( J_{ij}\hat{a}^\dagger_i\hat{a}_j+J_{ji}\hat{a}^\dagger_j\hat{a}_i \right),
	\end{split}
\end{equation}
where $\hat{a}_i$ and $\hat{b}_i$ are the annihilation operators for the electromagnetic and mechanical modes at site $i$, respectively.
In the weak optomechanical coupling regime, the Hamiltonian of the microwave subsystem is diagonalized by collective microwave modes, described as $\hat{\alpha}_k = \sum_i \psi_i^k \hat{a}_i$, where $\psi_i^k$ is the normalized modeshape of collective microwave mode~$k$ at site $i$.
Using the collective mode basis, the total Hamiltonian is given by
\begin{equation}
	\label{eq:hamiltonian_tot}
	\begin{split}
		\hat{H}/\hbar = \sum_{k} \tilde{\omega}_\mathrm{c}^{k} \hat{\alpha}^\dagger_k\hat{\alpha}_k &+\sum_i \Omega_{\mathrm{m},i}\hat{b}^\dagger_i\hat{b}_i \\
		&+ \sum_{k,i} \left(g_{0,i} \cdot \eta_i^k\right) \cdot \hat{\alpha}^\dagger_k\hat{\alpha}_k (\hat{b}^\dagger_i+\hat{b}_i),
	\end{split}
\end{equation}
where $\tilde{\omega}_\mathrm{c}^{k}$ is the eigenfrequency of collective microwave mode $k$ and $\eta_i^k = |\psi_i^k|^2$ is the energy participation ratio of collective microwave mode~$k$ at site~$i$.
As schematically shown in Fig.~\ref{fig:1}b, each collective microwave mode is parametrically coupled to all the mechanical oscillators with the weight of the corresponding participation ratio.
Since each mechanical oscillator is locally coupled to the collective microwave modes, the optomechanical interaction can be used as a local probe to perform a microwave modeshape measurement.

When the coupling strengths are designed to be alternating along a 1D chain, the microwave subsystem corresponds to a bosonic SSH chain~\cite{asboth2016short,ozawa2019topological}. As shown in Fig.~\ref{fig:1}a, each unit cell consists of two optomechanical building blocks with an intra-cell electromagnetic coupling rate of $J$, and is connected with an inter-cell coupling rate of $J'$. Solving the two-band Bulk Hamiltonian of an infinite-size SSH chain yields upper and lower passbands (UPB and LPB). There are two topologically distinct phases, for which the transition occurs when the band gap closes at $J/J'=1$~(see Fig.~\ref{fig:1}g). For $J<J'$, the phase is \emph{topological}, with two \emph{zero-energy} modes emerging in the band gap when the SSH chain is truncated to be a finite system. The zero-energy modes are spatially localized at the edges of the chain, and are hybridized at both the edges for a disorder-free system. Figure~\ref{fig:1}f shows examples of the modeshapes for a few modes from the UPB and LPB, and the two zero-energy modes in the topological phase. The bulk-edge correspondence identifies the two phases by the winding number, which is topologically protected~\cite{asboth2016short,ozawa2019topological}. The winding number is a bulk property, defined as the number of times that the off-diagonal element $\rho(k)$ of the two-band bulk Hamiltonian winds around the origin of the complex plane, i.e., a chain with a non-zero winding number is in the topological phase, exhibiting the edge states~(see Fig.~\ref{fig:1}h and more details in SI).

\subsection*{Circuit optomechanical lattices}

\begin{figure*}[ht]
	\includegraphics[width=\textwidth]{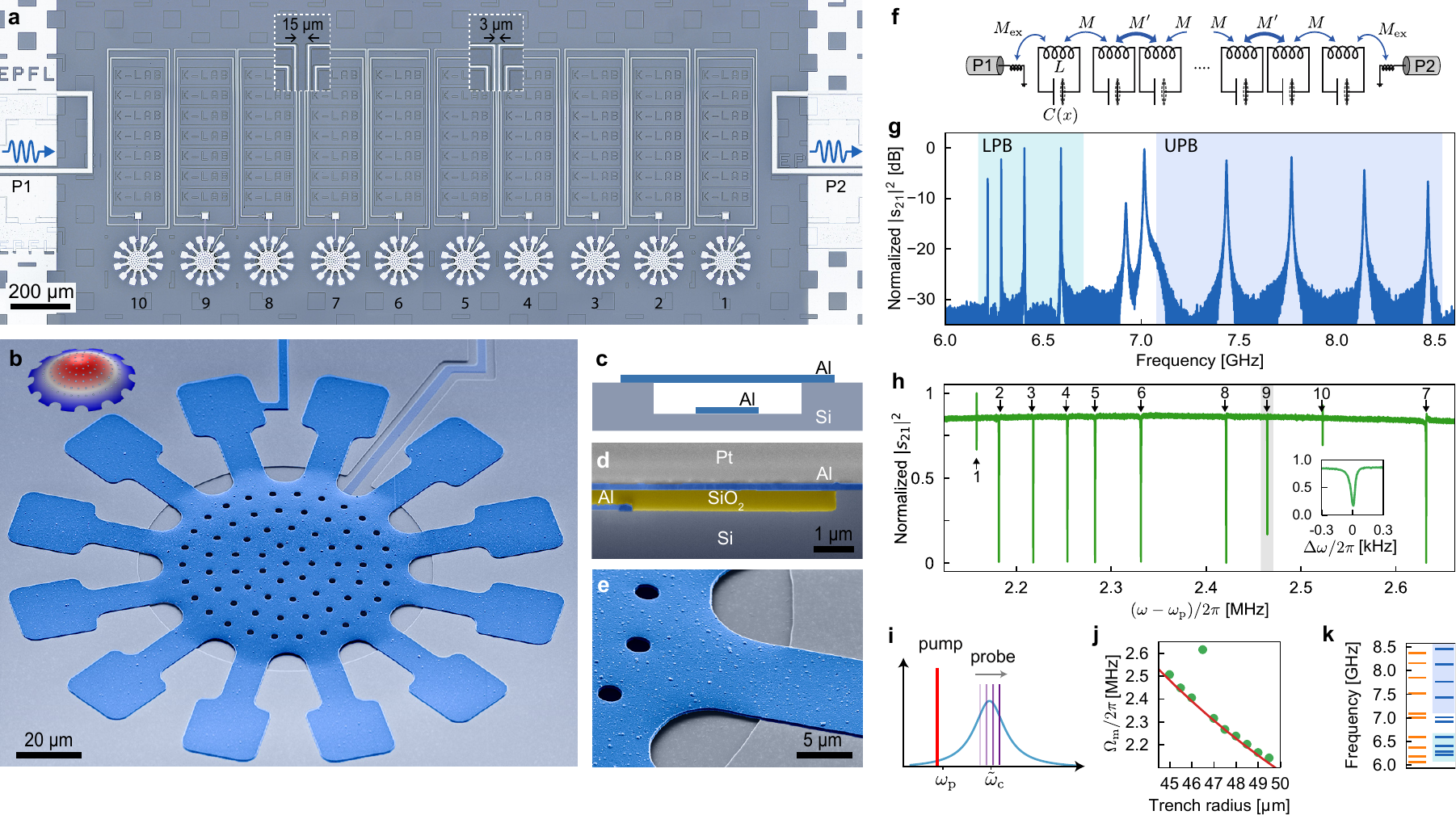}
	\caption{\footnotesize \linespread{1} \textbf{Superconducting circuit optomechanical chain realizing the 1D SSH model.} \textbf{a}, Microscope image of a 10-site circuit optomechanical chain. The mutual coupling is controlled by the spacing between adjacent spiral inductors (see insets). The chain is inductively coupled at both edges to co-planar waveguides. \textbf{b}, SEM of a mechanically compliant vacuum-gap capacitor. The inset shows the FEM simulation of the fundamental mechanical mode of the vibrating plate. \textbf{c}, Schematic cross-section of the capacitor. \textbf{d}, Focused ion beam cross-section of a capacitor before removing the SiO$_2$ sacrificial layer (Pt is used as the FIB protective layer). \textbf{e}, SEM of a suspended clamp over the trench. \textbf{f}, Equivalent circuit diagram of the SSH chain. \textbf{g}, Transmission spectrum of the device shown in \textbf{a}. Two topological edge modes are observed in the gap of the two bands (shaded regions), which are numerically calculated for the corresponding infinite SSH chain including parasitic couplings (see SI). \textbf{h}, Optomechanical induced transparency (OMIT) response of the highest frequency collective microwave mode. Each OMIT feature is associated to one of the drumhead capacitors along the chain, identified by the mechanical frequency. \textbf{i}, Frequency scheme for the OMIT measurement. \textbf{j}, Measured mechanical frequencies versus the trench radius (green dots). The solid line shows the inverse linear fit. One mechanical resonance is off due to an unusual deformation of the capacitor after the release. \textbf{k}, Microwave resonance frequencies of the device, design targets (orange), and measured values (blue). \label{fig:2}}
\end{figure*}

The main challenge of implementing such circuit optomechanical lattices is the low reproducibility of the gap size of vacuum-gap capacitors. The non-flat geometry of the movable capacitor plate~\cite{cicak2010low} prevents accurate control of the gap size and consequently the mechanical and microwave frequencies, as well as their coupling strength.

To overcome this challenge, we developed a nanofabrication process that significantly improves reproducibility and controllability over the gap size. We realize a flat aluminum membrane as the movable capacitor plate suspended over a trench (Figs.~\ref{fig:2}b, c, and d). In brief, we first etch a trench in a silicon substrate and cover it with a SiO$_2$ sacrificial layer. We then planarize the oxide layer to remove the topography and deposit the flat top plate of the capacitor. Finally, by removing the sacrificial layer, the top plate is suspended (see Figs.~\ref{fig:2}b and e). Cooling down of the device induces a tensile stress in the Al thin film, which guarantees the flatness and consequently the gap size to be controlled by the trench depth (see Methods and SI for details). This high-yield process allows us to control microwave and mechanical resonance frequencies with fluctuations of less than $0.5\%$ and 1\% respectively (see SI for details). Moreover the approach benefits from low mechanical dissipation $\Gamma_\mathrm{m}/2\pi \sim \mathcal{O}(1-10)$ $\mathrm{Hz}$ (see SI for the full characterizations).

To realize a circuit optomechanical 1D SSH chain in the topological phase (Fig.~\ref{fig:1}a), we fabricated a 10-site chain of mechanically compliant LC resonators with alternating mutual inductive coupling controlled by their physical distance (Figs.\ref{fig:2}a, and f).
The gap size is fabricated to be 245 nm, which results in a microwave frequency of $\omega_\mathrm{c}/2\pi = 7.12$~GHz and a single-photon optomechanical coupling rate of $g_0/2\pi = 10$~Hz for all sites. The microwave coupling rates are designed to be $J/2\pi =470$~MHz and  $J'/2\pi = 700$~MHz to achieve the non-trivial topological phase.

Figure~\ref{fig:2}g shows the microwave transmission spectrum of the chain, the UPB and LPB are highlighted by the shaded areas, and two topological edge modes can be observed in the middle of the band gap. The transmission response is in good agreement with the calculated eigenfrequencies of the desired design (Fig.~\ref{fig:2}k), which indicates reliable control over the system parameters in the fabrication process.
The linewidth of the modes varies from 7~MHz for the edge modes to 80~kHz for the first LPB mode depending on the collective modeshape.
The asymmetry of the band structure in the transmission spectrum originates from the small parasitic mutual inductive coupling between sites beyond the nearest neighbor (see SI) and does not change the topological properties of the chain~\cite{li2014topological}.

In order to identify each site by its mechanical frequency, we gradually increment the trench radius by $500$ nm along the chain to distinguish them in the further measurements~(see Figs.~\ref{fig:2}b inset).
The mechanical frequencies can be measured using optomechanically induced transparency (OMIT)~\cite{Weis2010} by applying a microwave pump red-detuned from a collective microwave mode while sweeping a weak probe tone across the resonance, as shown in Fig.~\ref{fig:2}i.
Figure~\ref{fig:2}h shows the OMIT response of the highest bulk mode. We observe ten OMIT features indicating mechanical frequencies, matching the inverse trench radius relationship (Fig.~\ref{fig:2}j).

\subsection*{Optomechanical modeshape measurement}
\begin{figure*}[ht]
	\includegraphics[width=\textwidth]{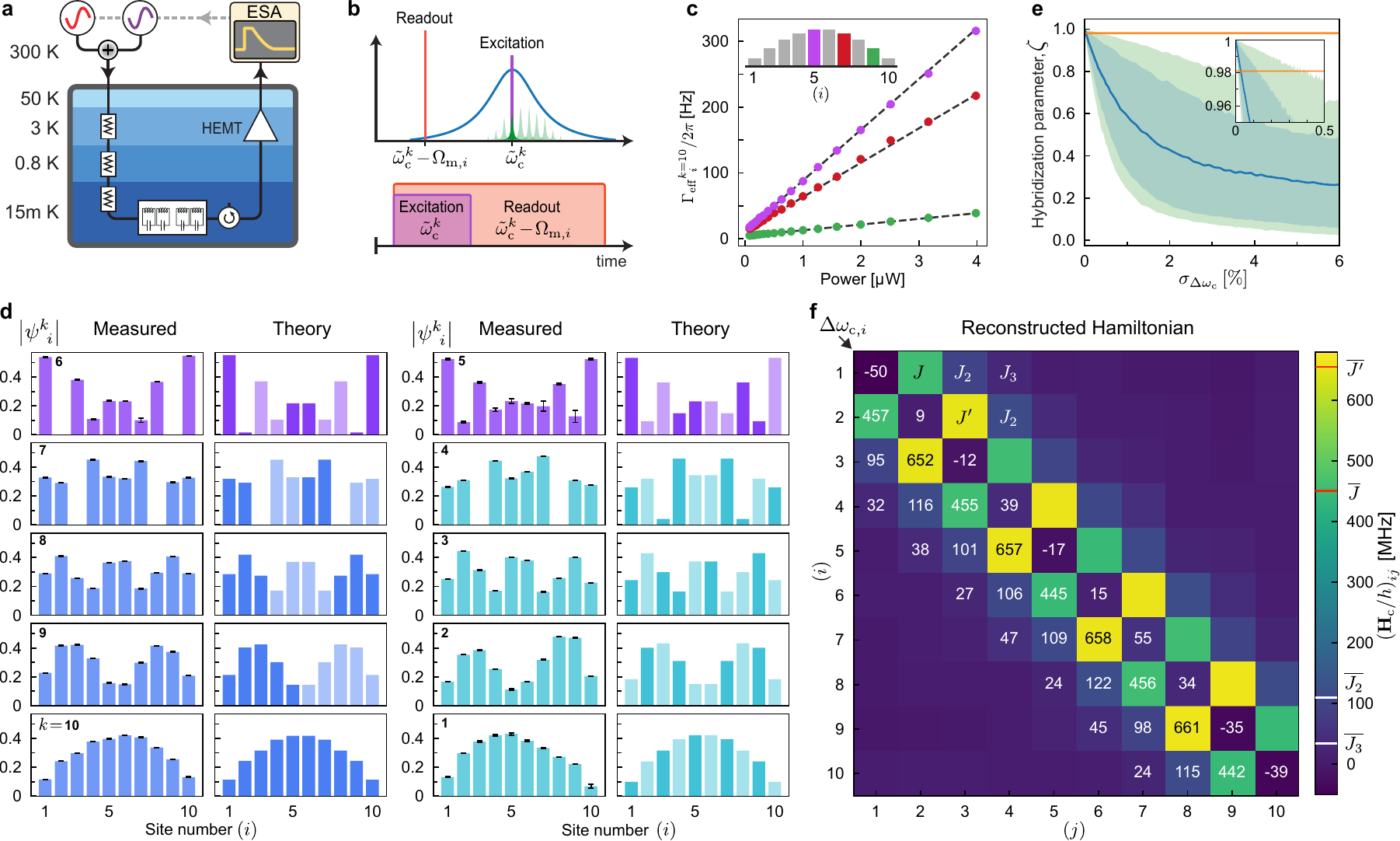}
	\caption{\footnotesize \linespread{1} \textbf{Optomechanical modeshape measurement and Hamiltonian reconstruction of a topological SSH chain.} \textbf{a}, Experimental setup for the modeshape measurement. \textbf{b}, Frequency and pulsing scheme: each mechanical mode is selectively excited by a resonant cavity drive while a readout pump on the lower sideband is on. \textbf{c}, Effective mechanical damping rate of the specific sites shown in the inset while driving the highest UPB mode. The slope of the linear fits depends on the collective modeshape. \textbf{d}, Measured and corresponding theoretical modeshapes of UPB, topological edge states, and LPB. Error bars are shown in black. In the theory plots, the sign of the phases is represented by different brightness. \textbf{e}, Stochastic analysis of the cavity frequency disorder effect on edge state hybridization, quantified by $\zeta$. The blue line shows expected value, blue and green shades reflect 70\% and 90\% certainty ranges, and the orange line shows the measured value. \textbf{f}, Reconstructed Hamiltonian matrix of the microwave subsystem in the rotating frame of the averaged cavity frequency. The diagonal elements show the cavity frequency disorder, and off-diagonal ones show the nearest-neighbor couplings ($J, J'$) as well as the parasitic couplings between distanced sites ($J_2, J_3$, etc.).}
	\label{fig:3}
\end{figure*}
Next, we exploit mechanical oscillators as embedded probes to directly and non-perturbatively measure the collective microwave modeshapes. We use the optomechanical damping effect~\cite{RMP_optomechanics} to deduce the modeshape information ($\eta_i^k$) from the mechanical oscillator's dynamics.
In the sideband resolved regime ($\kappa_\mathrm{tot}^k \ll \Omega_{\mathrm{m},i}$), the effective damping rate of mechanical mode $i$, in the presence of a pump on the lower sideband for collective microwave mode $k$, is given by
\begin{equation}
	\Gamma_{\mathrm{eff}, i}^{k}=\Gamma_{\mathrm{m}, i}+n_\mathrm{c}^{k} \ \frac{4 \left( {\eta_{\:i}^{k}}\cdot g_{0,i} \right)^2}{\kappa_\text{tot}^{k}},
	\label{eq:damping_rate}
\end{equation}
where $\Gamma_{\mathrm{m}, i}$ is the bare mechanical damping rate, $n_\mathrm{c}^k$ and $\kappa_\mathrm{tot}^k$ are the photon number and the linewidth of the collective mode, respectively.
In a time-domain protocol, we apply an excitation pulse on resonance to generate a beat note with a pump red detuned by $\Omega_{\mathrm{m},i}$ (Fig.~\ref{fig:3}b) and selectively excite mechanical mode~$i$ (the disorder in $\Omega_{\mathrm{m},i}$ from the designed value is smaller than the mechanical frequency increment to ensure correct identification of sites in the lattice). We then measure the ringdown signal of the optomechanical sideband in the presence of the red-detuned pump. The slope of $\Gamma_{\mathrm{eff},i}^k$ as a function of the pump power is proportional to $\left(\eta_i^k\right)^2$, as shown in Eq.~\ref{eq:damping_rate}. Figure~\ref{fig:3}c shows the ringdown data sets for a few mechanical modes in the chain measured on the highest UPB mode. By measuring the effective damping rate of every site in the chain, the full modeshape can be extracted for the corresponding microwave mode. Although it is challenging to independently obtain several parameters in Eq.~\ref{eq:damping_rate}, such as $g_{0,i}$ and $n_\mathrm{c}^k$, we can extract $\eta_i^k$ from all the slopes of $\Gamma_{\mathrm{eff},i}^{k}$ by using the normalization condition of the modeshapes, i.e., $\sum_k \eta_i^k = \sum_i \eta_i^k =1$ (see Methods for details). Figure~\ref{fig:3}e shows the full result of the modeshape characterization on the 10-site chain compared with the theoretical values with the design parameters.
The data is in excellent agreement with the theoretical predictions, demonstrating bulk modes in UPB, LPB, and topologically protected edge modes.

In prior experimental realizations of the SSH model, non-hybridized topological edge states were observed~\cite{kim2021quantum,st2017lasing} (localized on only one side of the chain). The edge state hybridization strongly depends on the disorder in the system parameters. Here, in contrast, we observe fully hybridized topological edge states, indicating minimal frequency disorder in the fabrication process. We perform numerical stochastic analysis to study the effect of disorder in the bare cavity frequency on the hybridization of the edge modes, quantified by $\zeta$~(see the definition in Methods). Figure~\ref{fig:3}e shows the expected value of $\zeta$ versus the standard deviation of the disorder $\sigma_{\Delta\omega}$, indicating less than 0.3\% disorder in the experimentally realized chain (see SI for details).

Finally, considering the knowledge of all the eigenvectors ($\psi_i^k$, modeshapes) and eigenvalues ($\tilde\omega_\mathrm{c}^k$, collective microwave frequencies) of the system, we can reconstruct the actual Hamiltonian of the microwave subsystem in the basis of the physical sites:
\begin{equation}
	\mathbf{H}_\mathrm{c}/\hbar = \mathbf{U}_\psi^\dagger
	\begin{bmatrix}
		\tilde{\omega}_\mathrm{c}^{1} &        & 0                              \\
		& \ddots &                                \\
		0                             &        & \tilde{\omega}_\mathrm{c}^{2N}
	\end{bmatrix}
	\mathbf{U}_\psi
\end{equation}
where $[\mathbf{U}_\psi]_{k,i} = \psi_i^k$ is a unitary matrix obtained from the measured participation ratios. Since our measurement protocol does not retrieve the phase of the eigenvectors, we infer it from the theoretically calculated ones. To more accurately reconstruct the Hamiltonian, we correct the unitary matrix by further imposing the orthogonality condition (see Methods for details).
Figure~\ref{fig:3}f shows the reconstructed Hamiltonian of the 1D SSH device. The diagonal elements represent the cavity frequency disorder of each site ($\sigma_{\Delta_\omega}$ = 0.5\% corresponds to a 2 nm gap size variation), while off-diagonal elements show the alternating microwave couplings as designed for the SSH model and parasitic second and third nearest neighbor couplings between distanced sites.

\subsection*{2D circuit optomechanical lattice}
\begin{figure*}[ht!]
	\includegraphics[width=\textwidth]{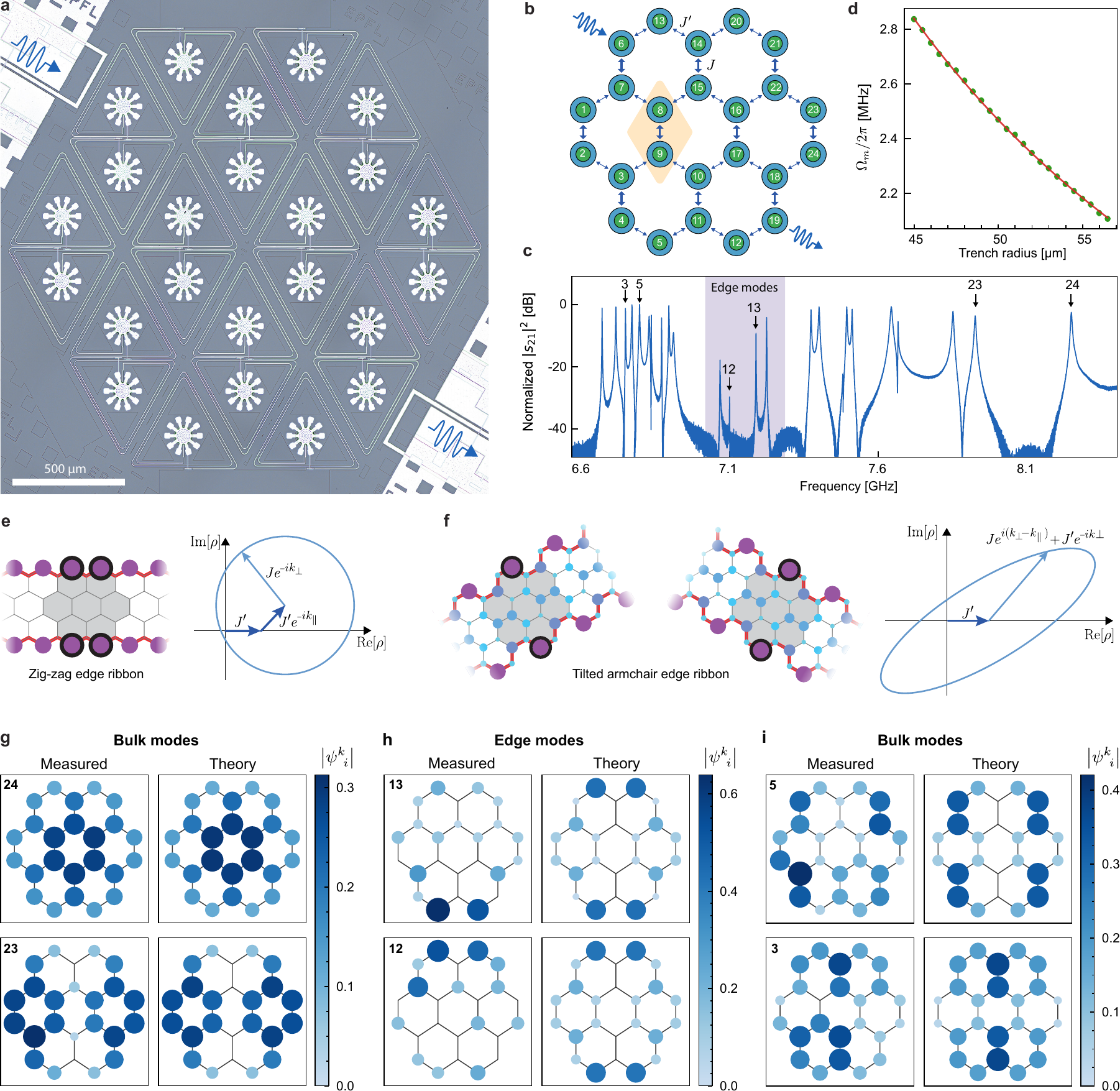}
	\caption{\footnotesize \linespread{1} \textbf{Two-dimensional superconducting circuit optomechanical honeycomb lattice realizing the strained graphene model.} \textbf{a}, Microscope image of a 24-site optomechanical honeycomb lattice (strained graphene flake) with alternating couplings. \textbf{b}, Mode diagram of the device shown in \textbf{a}. \textbf{c},~Transmission spectrum of the 2D lattice. Four edge modes are observed in the band gap. \textbf{d}, Measured mechanical frequencies versus trench radius and the inverse linear fit. \textbf{e,f},~24-site flake (gray-shaded region) as a truncated strained graphene ribbon in the zig-zag or tilted armchair orientations, respectively. The numerically simulated modeshapes of edge states are shown in amplitude, where the sites hosting the edge modes in the flake are identified by black circles. The off-diagonal element $\rho(k_\perp|k_\|)$ of the corresponding bulk Hamiltonian for a fixed wavenumber $k_\|$ is shown in the case of the topological phase, respectively. \textbf{g--i}, Examples of measured modeshapes of the 2D lattice in the upper and lower bulk modes, and edge modes, respectively. The observed edge modes are overlapping with the sites hosting the edge states in the analogous graphene ribbons in \textbf{e} and \textbf{f}. The modeshape information in amplitude is reflected on the color and area of each circle.
	}
	\label{fig:4}
\end{figure*}

The circuit optomechanical platform introduced here can be straightforwardly extended to 2D structures.
As a proof-of-concept experiment, we realize a 2D optomechanical honeycomb lattice. To demonstrate the coupling tunability, we alternate the mutual microwave couplings ($J$ and $J'$) along the vertical axis (see Fig.~\ref{fig:4}b) , which is known as the strained graphene model~\cite{pereira2009tight,naumis2017electronic}.
For strain-free graphene ($J=J'$), the band structure is gapless, and the upper and lower bands are connected to each other, forming Dirac cones. When applying strain in a certain orientation and decreasing the ratio of $J'/J$, a phase transition occurs at $J'/J=0.5$, and a band gap appears. For a finite-size graphene sheet, a set of edge modes can emerge depending on the structure of the edges and the strength of a strain~\cite{nakada1996edge,delplace2011zak}.

Figures~\ref{fig:4}a and b show a 24-site circuit optomechanical system and its mode diagram respectively, realizing a strained graphene flake, where the triangular spiral inductor in each site is inductively coupled to its three neighbors. The system is designed on the border of the phase transition $J'/J \simeq 0.5$ (see SI for details).
Figures~\ref{fig:4}c and d show all the microwave collective modes in the transmission spectrum and all the mechanical frequencies as a function of the trench radius, respectively.

Next, to gain insight into the properties of edge states and their topological origin, we consider our 24-site graphene flake as a truncated strained graphene ribbon, where the existence of edge states can be predicted using the bulk-edge correspondence~\cite{delplace2011zak}.
Figures \ref{fig:4}e and f, and the SI show all 6 possible orientations for the ribbon.   
Given a parallel wavenumber $k_\|$, which is a good quantum number for the ribbon Hamiltonian, the graphene ribbon can be reduced to a wavenumber-resolved 1D model in the perpendicular direction. In a similar manner to the two-band model for the standard 1D SSH chain, the off-diagonal element $\rho(k_\perp|k_\|)$ of the corresponding bulk Hamiltonian can be obtained by introducing a perpendicular wavenumber $k_\perp$. If $\rho(k_\perp|k_\|)$ winds around the origin in the complex plane, the ribbon with the $k_\|$ is in the nontrivial topological phase, possessing edge states. As shown in Figs.~\ref{fig:4} e and f, we find that only a ``zig-zag'' edge ribbon with $\rho(k_\perp|k_\|) = J'+ J'e^{-ik_\|} +J e^{-ik_\perp}$ and two ``tilted armchair'' edge ribbons with $\rho(k_\perp|k_\|) = J' +J e^{i(k_\perp-k_\|)} +J' e^{-ik_\perp}$ can possess edge states, while not for the other orientations. Both the two sites on the top and bottom edges of the flake overlap with edge modes seen in either the zig-zag or tilted armchair edge ribbons, enabling one to predict the existence of edge states on these 4 sites (see SI for details).

We perform a full modeshape measurement on the 2D lattice, revealing close agreement with the theoretical calculations. A few examples of the modeshapes are shown in Figs.~\ref{fig:4}e--h (the full results can be found in Methods). We find that there are four collective microwave modes whose modeshapes are localized in the two sites on either the top or bottom edges, as the topological analysis predicts.

\subsection*{Conclusion and outlook}

In summary, we realized optomechanical lattices demonstrating SSH model in 1D and strained graphene model in 2D with the exceptionally low disorder.
Moreover, we demonstrated how the optomechanical interaction can be exploited for direct measurements of the Hamiltonian, beyond the tight-binding approximation.
Looking forward, such optomechanical lattices offer a path to realize proposals exploring the rich physics in multimode optomechanics, ranging from quench~\cite{raeisi2020quench} and collective~\cite{heinrich2011collective} dynamics, to reservoir engineering~\cite{tomadin2012reservoir,yanay2020reservoir,zippilli2021dissipative}. Specifically, by using degenerate mechanical oscillators it is possible to create \emph{collective} long-range interactions and observe strong cooperative effects on mechanical motion~\cite{xuereb2012strong}. Finally, our system may enable the generation of highly entangled mechanical states~\cite{akram2012photon}, and viewed more broadly, it can be used to explore quantum correlations in topological optomechanical lattices~\cite{peano2015topological,zangeneh2020topological,ludwig2013quantum}.

\subsection*{Acknowledgements}
We thank the anonymous reviewer (V. Peano) for pointing out the connection to the strained graphene model. We sincerely thank O.~Yazyev and P.~Delplace for critical discussions with respect to the topological properties of strained graphene. We thank T. Sugiyama for fruitful discussions on the Hamiltonian reconstruction. This work was supported by the EU H2020 research and innovation programme under grant No. 101033361 (QuPhon), and from the European Research Council (ERC) grant No. 835329 (ExCOM-cCEO). This work was also supported by the Swiss National Science Foundation (SNSF) under grant No. NCCR-QSIT: 51NF40\_185902 and No. 204927. All devices were fabricated in the Center of MicroNanoTechnology (CMi) at EPFL.

\subsection*{Author contributions}

A.Y. conceived the experiment. 
A.Y. and A.B. designed and simulated devices.
S.K. provided the theoretical support with assistance of A.Y. and J.P. \
A.Y. and A.B. performed the numerical analysis.
A.Y. developed the fabrication process with assistance of M.C. and T.V.
M.C. and A.Y. fabricated the samples.
The measurement technique was implemented by A.Y., A.B., and S.K.
The data was collected by A.B. and S.K., with assistance of A.Y.
The data analysis was performed by A.B., A.Y., and S.K.  
The manuscript was written by A.Y., S.K., and A.B. with assistance of T.J.K. and all the other authors.
T.J.K. supervised all the efforts.

\subsection*{Competing interests}
The authors declare no competing interests.

\subsection*{Data availability}
The data used to produce the plots within this paper are available on Zenodo 

\href{https://doi.org/10.5281/zenodo.6987358}{(https://doi.org/10.5281/zenodo.6987358)}. 
All other data used in this study are available from the corresponding author on reasonable request.
\subsection*{Code availability}
The code used to produce the plots within this paper is available on Zenodo 

\href{https://doi.org/10.5281/zenodo.6987358}{(https://doi.org/10.5281/zenodo.6987358)}.

\let\oldaddcontentsline\addcontentsline
\renewcommand{\addcontentsline}[3]{}
\let\addcontentsline\oldaddcontentsline

\pagebreak
\subsection*{Methods}
\noindent
\textbf{Fabrication process}

\noindent
Fig.~\ref{fig:ext_fab} shows the nanofabrication process we developed for circuit optomechanics. We define a trench in the substrate containing the bottom plate of the capacitor. The trench then is be covered by a thick SiO$_2$ sacrificial layer, which inherits the same topography of the layer underneath. To remove this topography and obtain a flat surface, we use chemical mechanical polishing (CMP) to planarize the SiO$_2$ surface. We then etch back the sacrificial layer down to the substrate layer and deposit the top Al plate of the capacitor. Although after the release of structure by HF vapor etching of SiO$_2$ the drumhead will buckles up due to the compressive stress, at cryogenic temperatures the high tensile stress ensures the flatness of the top plate. This will guarantee the gap size to be precisely defined by the depth of the trench and the thickness of the bottom plate. The detailed description of each step can be find in SI.

\noindent
\textbf{Iterative normalization algorithm}

\noindent
Based on the theoretical expression, the participation ratio of collective microwave modes can be, in principle, obtained from their optomechanical damping rates, which are proportional to $(\eta_i^k)^2$. However, several parameters, such as single-photon optomechanical coupling rate $g_{0,i}$, cannot be characterized straightforwardly in our multimode optomechanical systems, which prevents us from obtaining the actual participation ratio just by dividing the damping rates by all the coefficients. Generally speaking, the unnormalized participation ratio of collective microwave mode $k$ at site $i$ is described as
\begin{equation}
	\widetilde{\eta_i^k} = C_i\: \eta_i^k \:D_k,
\end{equation}
where $C_i$ and $D_k$ are coefficients that depend on either site $i$ or collective microwave mode $k$, respectively.
Using the normalization conditions of the modeshapes, i.e.,  $\sum_k \eta_i^k = \sum_i \eta_i^k =1$, we can deduce the actual participation ratio by iteratively normalizing the unnormalized one along the row and column axes alternatively. Namely, we first define the initial participation ratio as 
\begin{equation}
	\widetilde{\eta_i^k}_{(0)} = \widetilde{\eta_i^k}.
\end{equation}
Then, we iterate normalization processes, defined as
\begin{equation}
	\label{eq:normi_eta}
	\widetilde{\eta_i^k}_{(n+1)}  = \frac{\widetilde{\eta_i^k}_{(n)}}{\sum_i\widetilde{\eta_i^k}_{(n)}} \quad (\forall \: k, \:\mathrm{if} \: n \:  \mathrm{is \: even})
\end{equation}
or
\begin{equation}
	\label{eq:normk_eta}
	\widetilde{\eta_i^k}_{(n+1)} = \frac{\widetilde{\eta_i^k}_{(n)}}{\sum_k \widetilde{\eta_i^k}_{(n)}} \quad (\forall \: i, \:\mathrm{if} \: n \: \mathrm{is \: odd})
\end{equation}
until the participation ratio is converged. We numerically confirm that the iterative normalization method allows us to obtain the actual participation ratio regardless of the amount of fluctuations in the coefficients.

\noindent 
\textbf{Orthogonalization of modeshapes}

\noindent
Our measurement scheme does not retrieve the sign of the modeshapes of the collective microwave modes to fully determine the unitary matrix $\mathbf{U}_\psi$, which is necessary for the Hamiltonian reconstruction.
By inferring the phase information from the theoretical modeshapes, we can construct the ``unitary'' matrix $\mathbf{\widetilde{U}}_\psi$ from $\eta_i^k$ as
\begin{equation}
	\left[\mathbf{\widetilde{U}}_\psi\right]_{k,i} = \quad + \sqrt{\eta_i^k} \quad \mathrm{or} \:\: -\sqrt{\eta_i^k}.
\end{equation}
While it satisfies the normalization conditions, the obtained matrix does not satisfy the orthogonalization condition due to the finite measurement error.
To more accurately reconstruct the Hamiltonian, we further correct $\mathbf{\widetilde{U}}_\psi$ to satisfy both the normalization and orthogonalization conditions.
We first numerically calculate the generator of $\mathbf{\widetilde{U}}_\psi$ as
\begin{equation}
	\mathbf{G} = \log\left(\mathbf{\widetilde{U}}_\psi \right).
\end{equation}
In general, the generator can be decomposed into a Hermitian matrix $\mathbf{H}$ and anti-Hermitian matrix $\mathbf{\bar{H}}$ as
\begin{equation}
	\mathbf{G} = \mathbf{H} + \mathbf{\bar{H}},
\end{equation}
where $\mathbf{H} = (\mathbf{G}+\mathbf{G}^\dag)/2$ and $\mathbf{\bar{H}} = (\mathbf{G}-\mathbf{G}^\dag)/2$.
Although the generator of a unitary matrix only contains an anti-Hermitian matrix, the experimentally obtained $\bm{G}$ contains the finite Hermitian component.
By neglecting the Hermitian component, the corrected unitary matrix can be obtained as
\begin{equation}
	\label{eq:correctU}
	\mathbf{U}_\psi = \exp\left(\mathbf{\bar{H}}\right),
\end{equation}
which is used for the Hamiltonian reconstraction.

\noindent
\textbf{Hybridization factor}

\noindent
The hybridization factor for the two edge states in the 1D SSH chain with $2N$ sites is defined as
\begin{equation}
	\zeta = \frac{1}{2} \left(\frac{\min\{\eta_1^N,\eta_{2N}^N\}}{\max\{ \eta_1^N,\eta_{2N}^N\}} + \frac{\min\{\eta_1^{N+1},\eta_{2N}^{N+1}\}}{\max\{\eta_1^{N+1},\eta_{2N}^{N+1}\}} \right)
\end{equation}
where $\eta_i^N$ and $\eta_i^{N+1}$ are the participation ratios of the topological edge modes. Here we use only the participation ratios of the topological modes ($k=N$ or $k+N+1$) at both the edges ($i = 1$ and $i=2N$) and define the factor such that it can achieve 1 only when the participation ratios at both the edges are equal, which is realized by the negligible frequency disorder. A detailed discussion on numerical analysis of hybridization factor in SSH chains can be find in SI.

\noindent
\textbf{Extended Data for 2D lattice}

The full OMIT response of the 24-site optomechanical system in the 2D honeycomb lattice is shown in Fig.~\ref{fig:SI_OMIT2D}a measured on the highest frequency collective microwave mode. By fitting each OMIT resonance, we extract all the mechanical frequencies, shown in Fig~\ref{fig:4}d. The minimum and maximum trench radius used in the device is highlighted in the figure. Fig.~\ref{fig:SI_OMIT2D}b shows the design target collective microwave frequencies versus the measured values.

\noindent
All the modeshapes of the 2D device obtained by the iterative normalization method are shown in Fig.~\ref{fig:SI_2D_modeshapes}.
For some of them, we don't see good agreements as we observed in the 1D case. This is because, given the much higher number of microwave modes, the effect of frequency disorder plays a more important role. Where the dots are completely missing it means that the experimental power participation ratio $\eta_i^k$ of site $i$ to the collective microwave mode $k$ was not high enough to produce a detectable signal.

Using the extracted modeshapes of the 2D device, we reconstruct the Hamiltonian of the microwave subsystem. Fig.~\ref{fig:SI_H_2D} shows the reconstructed Hamiltonian matrix in the rotating frame of the average microwave frequency and the ideal designed Hamiltonian including the second-nearest neighbor couplings. Due to the higher frequency disorder in the measured modeshapes of the 2D device, the unitary matrix and consequently Hamiltonian are slightly more disturbed compared to the 1D case, however, still, show well agreement with the design and theoretical predictions.


\renewcommand{\thefigure}{ED.\arabic{figure}}
\renewcommand{\theHfigure}{ED.\arabic{figure}}
\setcounter{figure}{0}

\begin{figure*}[t!]
	\includegraphics[scale=0.9]{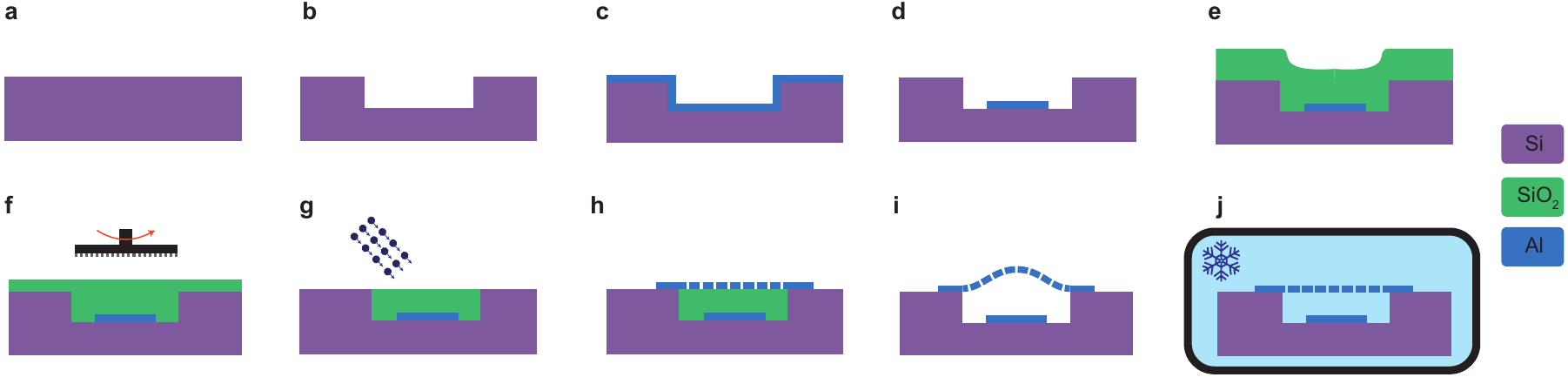}
	\caption{\textbf{Fabrication process} \textbf{a}, \textbf{b}, Etching a trench in a silicon wafer (325 nm). \textbf{c}, Aluminum deposition of the bottom plate (100 nm). \textbf{d}, Patterning of Al. \textbf{e}, SiO$_2$ sacrificial layer deposition (3 $\mu$m). \textbf{f}, CMP planarization. \textbf{g}, landing on the substrate using IBE etching. \textbf{h}, Top Al layer deposition and patterning (200 nm). \textbf{i}, Releasing the structure using HF vapor. Due to compressive stresses, the top plate will buckle up. \textbf{j}, At cryogenic temperatures, the drumhead shrinks and flattens. 
		\label{fig:ext_fab}}
\end{figure*}

\begin{figure*}[b!]
	\includegraphics[scale=0.9]{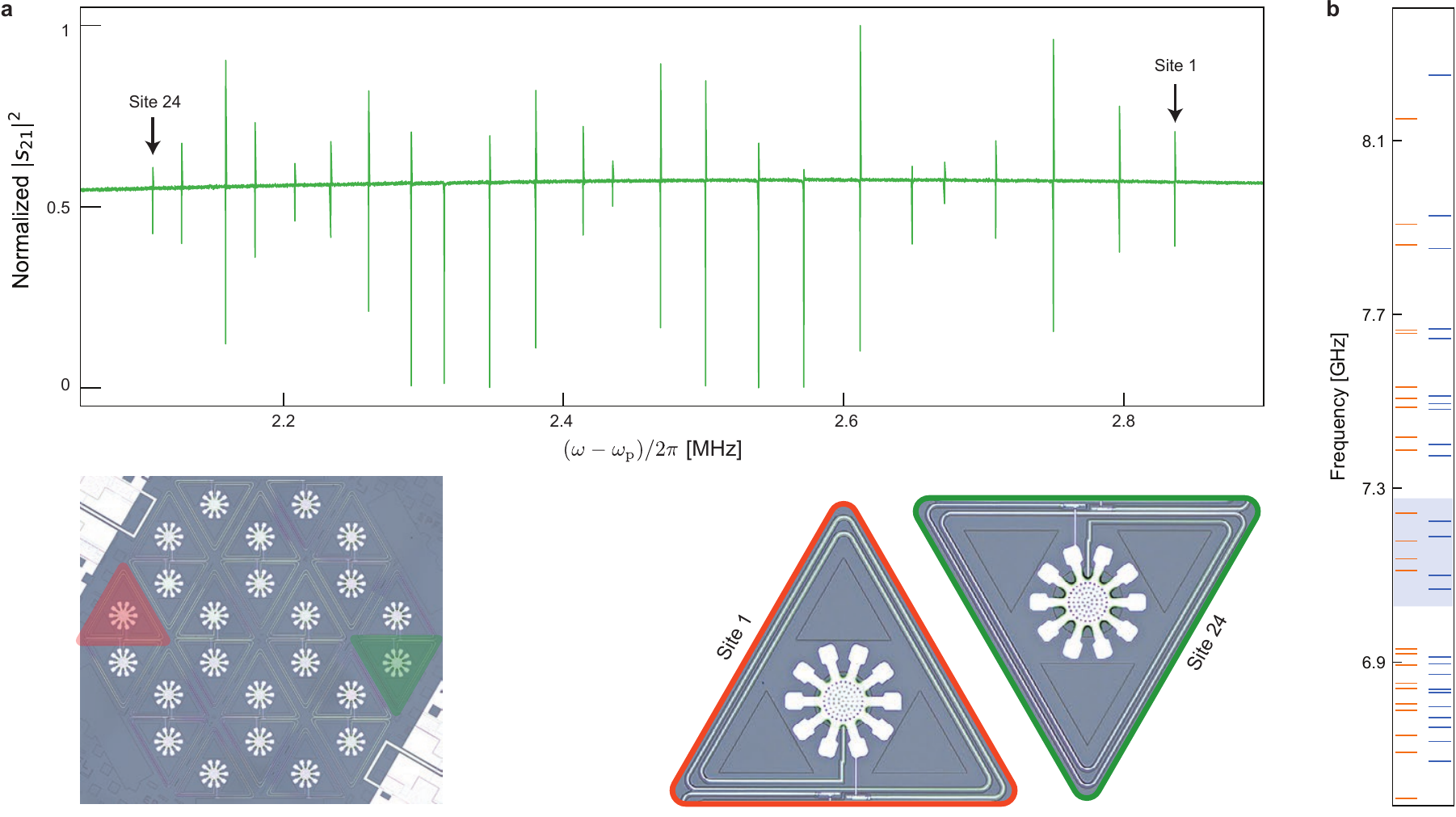}
	\caption{\textbf{Charecterization of 24-site 2D hanycomb lattice.} \textbf{a}, Optomechanically-induced transparency (OMIT) responce of the 2D device measured on the highest microwave bulk mode. Increasing the trench radius results in a slight shift of the mechanical frequencies. \textbf{b}, Microwave resonance frequencies of the device, design targets (orange), and measured values (blue). 
		\label{fig:SI_OMIT2D}}
\end{figure*}

\begin{figure*}[p!]
	\includegraphics[scale=0.9]{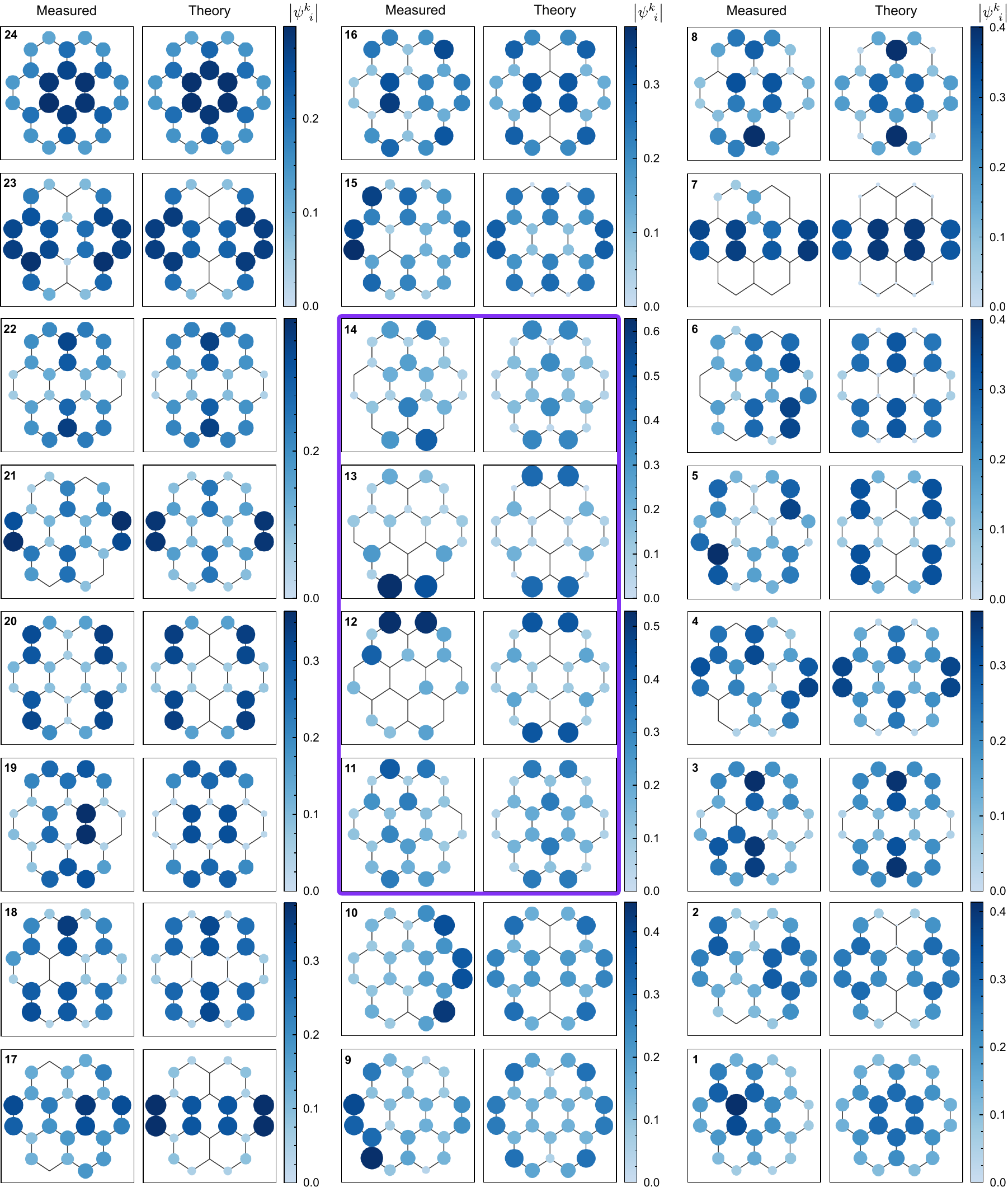}
	\caption{\textbf{Modeshapes of 24-site 2D honeycomb lattice.} The amplitude of the modeshape $|\psi^k_ i|$ is encoded in the area of the circles. Only for modes that share the same colorbar, the size and color of the circles can be compared. Highlighted in purple are the four edge modes. 
		\label{fig:SI_2D_modeshapes}}
\end{figure*}

\begin{figure*}[p!]
	\includegraphics[scale=0.9]{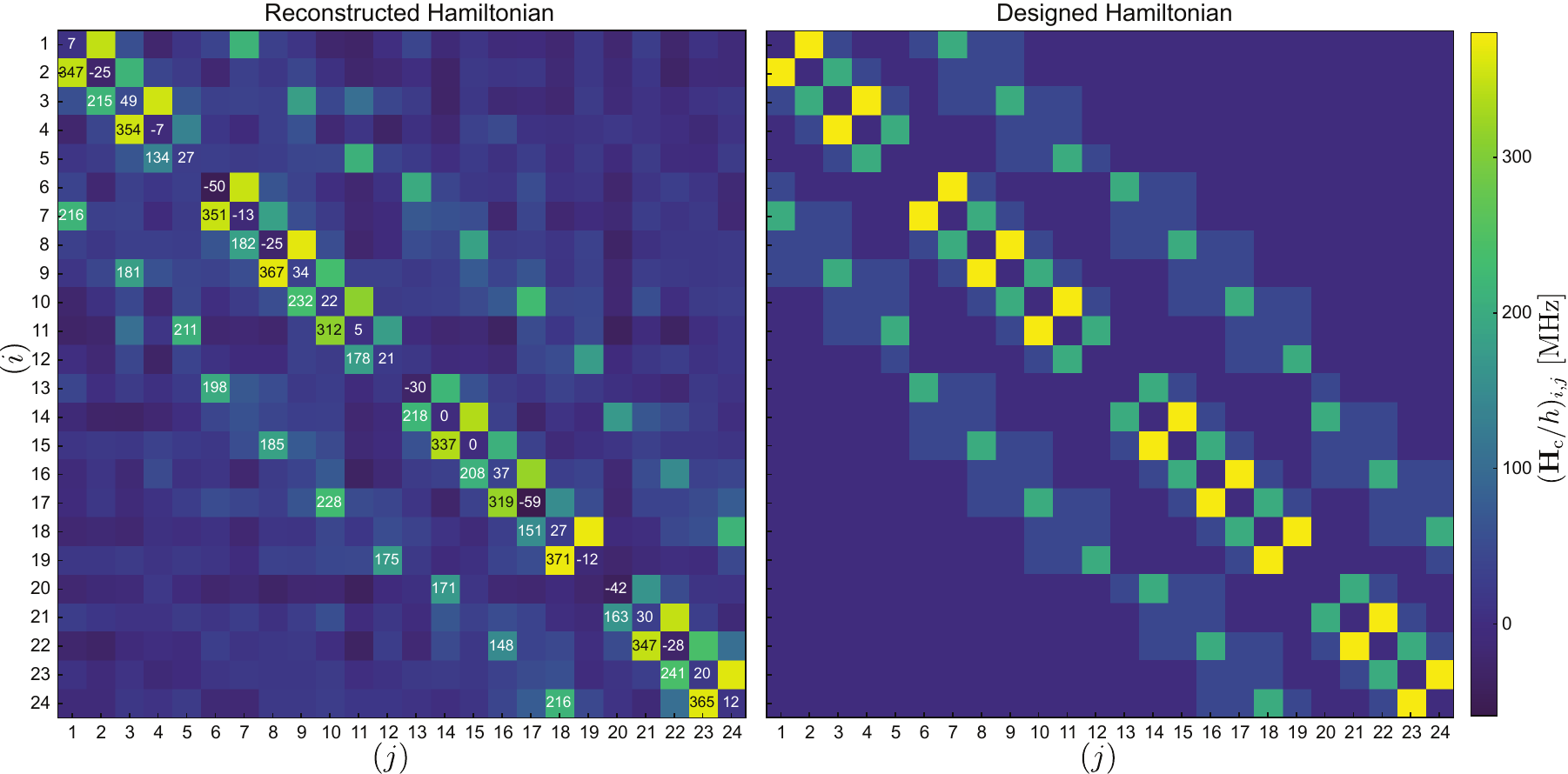}
	\caption{\textbf{Hamiltonian reconstruction of 24-site 2D honeycomb lattice.} The reconstructed Hamiltonian of 24 site 2D honeycomb device (left) and the designed Hamiltonian including second nearest-neighbor couplings (right). The diagonal elements represent the individual site's resonance frequency deviation from the average bare cavity frequency. \label{fig:SI_H_2D}}
\end{figure*}

\makeatletter
\close@column@grid
\clearpage
\onecolumngrid


\begin{center}
	\large{\textbf{Supplementary Information for: Topological lattices realized in superconducting circuit optomechanics}}
\end{center}
\begin{center}
	Amir Youssefi$^{*}$, Shingo Kono$^{*}$, Andrea Bancora$^{*}$, Mahdi Chegnizadeh, Jiahe Pan, Tatiana Vovk, and Tobias J. Kippenberg$^{\dagger}$\\[.1cm]
	{\itshape
		Laboratory of Photonics and Quantum Measurement, Swiss Federal Institute of Technology Lausanne (EPFL), Lausanne, Switzerland\\
	}
	$^\dagger$Electronic address: tobias.kippenberg@epfl.ch\\
\end{center}

\setcounter{equation}{0}
\setcounter{table}{0}
\renewcommand{\thefigure}{S\arabic{figure}}
\renewcommand{\theHfigure}{S\arabic{figure}}
\setcounter{figure}{0}
\setcounter{subsection}{0}
\setcounter{section}{0}

\tableofcontents


\section{Theory of multimode circuit optomechanics}
\subsection{Theoretical model}
This section provides the theoretical description of a multimode circuit optomechanical system. In our model, each electromechanical element consists of a single microwave mode optomechanically coupled to an individual mechanical mode. In addition, the microwave modes are electromagnetically coupled to each other. The Hamiltonian of the multimode system is in general given by
\begin{equation}
	\label{eq:hamiltonian_tot}
	\hat{H}/\hbar = \sum_{i} \left[ \omega_{\mathrm{c},i} \: \hat{a}^\dagger_i\hat{a}_i + \Omega_{\mathrm{m},i}\:\hat{b}^\dagger_i\hat{b}_i + g_{0,i}\: \hat{a}^\dagger_i\hat{a}_i (\hat{b}^\dagger_i+\hat{b}_i) \right] + \sum_{i \neq j}(J_{ij}\hat{a}^\dagger_i\hat{a}_j+J_{ji}\hat{a}^\dagger_j\hat{a}_i),
\end{equation}
where $\hat{a}_i$ and $\hat{b}_i$ are annihilation operators for microwave mode $i$ with a resonance frequency $\omega_{\mathrm{c},i}$ and mechanical mode $i$ with a resonance frequency $\Omega_{\mathrm{m},i}$, $g_{0,i}$ is the single-photon optomechanical coupling rate, and $J_{ij}$ is the coupling strength between microwave modes $i$ and $j$.

Assuming that the single-photon optomechanical coupling rates are \replaced{much smaller }{sufficiently smaller} than the microwave resonance frequencies and the electromagnetic coupling strengths, collective microwave modes can be well defined regardless of the coupling to the mechanical modes. Thus, we first focus on the Hamiltonian of the coupled microwave modes:
\begin{equation}
	\label{eq:hamiltonian_c}
	\hat{H}_\mathrm{c}/\hbar = \sum_{i} \omega_{\mathrm{c},i} \:\hat{a}^\dagger_i\hat{a}_i + \sum_{i \neq j}(J_{ij}\hat{a}^\dagger_i\hat{a}_j+J_{ji}\hat{a}^\dagger_j\hat{a}_i).
\end{equation}
In a matrix representation, the quadratic Hamiltonian $\hat{H}_\mathrm{c}$ can be described as
\begin{equation}
	\hat{H}_\mathrm{c} = \bm{\hat{a}}^\dag \mathbf{H}_\mathrm{c} \bm{\hat{a}},
\end{equation}
where
\begin{equation}
	\bm{\hat{a}} =
	\begin{bmatrix}
		\hat{a}_1 \\
		\hat{a}_2 \\
		\vdots    \\
		\hat{a}_i \\
		\vdots
	\end{bmatrix}, \quad
	\bm{\hat{a}}^\dag =
	\begin{bmatrix}
		\hat{a}_1^\dag \:\: \hat{a}_2^\dag \:\: \cdots \:\:\hat{a}_i^\dag \:\: \cdots
	\end{bmatrix}, \:\ \mathrm{and} \quad
	\mathbf{H}_\mathrm{c}/\hbar =
	\left[
	\begin{array}{ccccc}
		\ddots &                       &        &                       &        \\
		& \omega_{\mathrm{c},i} &        & J_{ji}                &        \\
		&                       & \ddots &                       &        \\
		& J_{ij}                &        & \omega_{\mathrm{c},j} &        \\
		&                       &        &                       & \ddots
	\end{array}
	\right].
\end{equation}

Since $\mathbf{H}_\mathrm{c}$ is an hermitian matrix, it can be diagonalized by a unitary matrix $\mathbf{U}_\psi$ as
\begin{equation}
	\label{eq:Dc}
	\mathbf{D}_\mathrm{c} = \mathbf{U}_\psi \mathbf{H}_\mathrm{c} \mathbf{U}_\psi^\dag,
\end{equation}
where $\mathbf{D}_\mathrm{c}$ is a diagonal matrix whose diagonal element $[\mathbf{D}_\mathrm{c}]_{k,k}$ corresponds to the eigenenergy $\hbar\tilde{\omega}_\mathrm{c}^k$ of the collective microwave mode~$k$.
On the basis of the collective microwave modes $\bm{\hat{\alpha}}=\mathbf{U}_\psi\bm{\hat{a}}$, Hamiltonian~(\ref{eq:hamiltonian_c}) is described as
\begin{equation}
	\label{eq:hamiltonian_hybrid}
	\begin{split}
		\hat{H}_\mathrm{c}/\hbar &= \bm{\hat{\alpha}}^\dag \left(\mathbf{D}_\mathrm{c} /\hbar \right)\bm{\hat{\alpha}} \\
		&=  \sum_k \tilde{\omega}_\mathrm{c}^k \hat{\alpha}^\dagger_k\hat{\alpha}_k.
	\end{split}
\end{equation}
Note that collective microwave mode~$k$ \replaced{can }{is} explicitly \added{be} represented using the bare microwave modes as
\begin{equation}
	\label{eq:alpha_a}
	\hat{\alpha}_k = \sum_i \psi_i^k \hat{a}_i,
\end{equation}
where $\psi_i^k =[\mathbf{U}_\psi]_{k,i}$ is the complex amplitude of the normalized mode function of  collective microwave mode~$k$ \added{at site ~$i$}. Thus, we define the participation ratio of bare mode~$i$ to collective mode~$k$ as
\begin{equation}
	\label{eq:eta}
	\eta_i^k = |\psi_i^k|^2.
\end{equation}
As discussed in the main text, our main goal is to characterize the participation ratio using the optomechanical damping effects in multimode optomechanics.

The property of the unitary transformation gives the normalization conditions of the mode functions:
\begin{equation}
	\label{eq:sumi_eta}
	\sum_i \eta_i^k = \sum_i |\psi_i^k|^2 = 1
\end{equation}
and
\begin{equation}
	\label{eq:sumk_eta}
	\sum_k \eta_i^k = \sum_k |\psi_i^k|^2 = 1,
\end{equation}
as well as the orthogonalization condition between different mode functions:
\begin{equation}
	\label{eq:orthogonal_eta}
	\sum_i {\psi_i^k}^* \psi_i^l = 0 \quad \left(\mathrm{if}\:\: l \neq k\right).
\end{equation}
The normalization condition of Eq.~(\ref{eq:sumi_eta}) describes the photon-number conservation in collective mode~$k$ when it is decomposed into the bare modes, while the normalization condition of Eq.~(\ref{eq:sumk_eta}) describes the photon-number conservation in bare mode~$i$ when it is decomposed into the collective modes.

Next, we describe the Hamiltonian of the multimode optomechanical system based on the collective microwave modes. Using the inverse unitary transformation $\bm{\hat{a}}=\mathbf{U}_\psi^\dag\bm{\hat{\alpha}}$, the bare microwave mode~$i$ is represented as
\begin{equation}
	\label{eq: a alpha}
	\hat{a}_i = \sum_k {\psi_i^k}^* \hat{\alpha}_k.
\end{equation}
With this relation, Hamiltonian~(\ref{eq:hamiltonian_tot}) is rewritten as
\begin{equation}
	\label{eq:hamiltonian_tot_hybrid}
	\hat{H}/\hbar = \sum_k \tilde{\omega}_\mathrm{c}^k\: \hat{\alpha}^\dagger_k\hat{\alpha}_k +\sum_i \Omega_{\mathrm{m},i}\:\hat{b}^\dagger_i\hat{b}_i + \sum_{k,l,i}\left[g_{0,i} {\psi_i^l}^* \psi_i^k \:\hat{\alpha}^\dagger_k\hat{\alpha}_l (\hat{b}^\dagger_i+\hat{b}_i) \right].
\end{equation}
Assuming that the frequency \replaced{spacing}{detuning} \added{$|\tilde{\omega}_k - \tilde{\omega}_l|$} \replaced{between}{among} the collective microwave modes are sufficiently larger than the single-photon optomechanical coupling rates, the rotating-wave approximation can be valid, and the non-energy \replaced{conserving}{conservation} term $\propto \hat{\alpha}^\dagger_k\hat{\alpha}_l$ (if $k\neq l$) \added{can be neglected}. Thus, the Hamiltonian can be approximated as
\begin{equation}
	\label{eq:hamiltonian_tot_hybrid_RWA}
	\begin{split}
		\hat{H}/\hbar &= \sum_k \tilde{\omega}_\mathrm{c}^k \:\hat{\alpha}^\dagger_k\hat{\alpha}_k +\sum_i \Omega_{\mathrm{m},i}\:\hat{b}^\dagger_i\hat{b}_i + \sum_{k,i} g_{0,i} {\psi_i^k}^* \psi_i^k \:\hat{\alpha}^\dagger_k\hat{\alpha}_k (\hat{b}^\dagger_i+\hat{b}_i) \\
		&= \sum_k \tilde{\omega}_\mathrm{c}^k\: \hat{\alpha}^\dagger_k\hat{\alpha}_k +\sum_i \Omega_{\mathrm{m},i}\:\hat{b}^\dagger_i\hat{b}_i + \sum_{k,i} \left(\eta_i^k g_{0,i}\right) \hat{\alpha}^\dagger_k\hat{\alpha}_k (\hat{b}^\dagger_i+\hat{b}_i),
	\end{split}
\end{equation}
\added{where ${\psi_i^k}^* \psi_i^k = \eta_i^k$.} In the collective-mode picture, mechanical mode~$i$ is optomechanically coupled to collective microwave mode~$k$ with the single-photon optomechanical coupling rate $g_{0,i}$ weighted by $\eta_i^k$, the participation ratio of bare microwave mode~$i$ to collective microwave mode~$k$.

In our multimode optomechanical system, it can be assumed that the collective microwave modes and the mechanical modes are well \replaced{separated}{isolated} from each other in frequency space and \added{the frequency spacing between the collective microwave modes is much larger than the mechanical frequencies}, enabling us to operate the multimode system as a single microwave mode coupled to a single mechanical mode.
Furthermore, our multimode system is locally connected to an input line on one side and an output line on the other side, resulting in collective microwave mode~$k$ being coupled to the input line with the external coupling rate $\kappa_1^k$ and the output line with the external coupling rate $\kappa_2^k$, respectively. Note that $\kappa_1^k$ and $\kappa_2^k$ depend on the mode function of collective mode~$k$.

Here, we consider the case when collective microwave mode~$k$ is driven by a coherent field with a frequency $ \omega_i^k$ from the input line to couple mechanical mode $i$ to collective microwave mode $k$.
The effective Hamiltonian in the rotating frame of $ \omega_i^k$ is given by
\begin{equation}
	\label{eq:hamiltonian_ki}
	\hat{H}_i^k/\hbar =  \Delta_i^k \hat{\alpha}^\dagger_k\hat{\alpha}_k + \Omega_{\mathrm{m},i}\hat{b}^\dagger_i\hat{b}_i + \left(\eta_i^kg_{0,i} \right) \hat{\alpha}^\dagger_k\hat{\alpha}_k (\hat{b}^\dagger_i+\hat{b}_i) + \sqrt{\kappa_1^k \dot n_\mathrm{d}^k}\left(\hat{\alpha}^\dagger_k + \hat{\alpha}_k\right),
\end{equation}
where $\Delta_i^k = \tilde{\omega}_\mathrm{c}^k - \omega_i^k$ is the detuning between the collective microwave frequency $\tilde{\omega}_\mathrm{c}^k$ and the drive frequency $\omega_i^k$ and $\dot n_\mathrm{d}^k$ is the photon flux of the cavity drive field.
With this Hamiltonian, the quantum Langevin equations of collective microwave mode $k$ and mechanical mode $i$ are given by
\begin{equation}
	\label{eq:langevin_ki}
	\begin{split}
		\dot{\hat{\alpha}}_k &= \left[-i\Delta_i^k -\frac{\kappa_\mathrm{tot}^k}{2} - i\left(\eta_i^k g_{0,i}\right) (\hat{b}^\dagger_i+\hat{b}_i) \right]\hat{\alpha}_k - i\sqrt{\kappa_1^k \dot n_\mathrm{d}^k} + \sqrt{\kappa_1^k}\hat{a}_{\mathrm{in},1} + \sqrt{\kappa_2^k}\hat{a}_{\mathrm{in},2} + \sqrt{\kappa_0^k}\hat{a}_{\mathrm{in},0} \\
		\dot{\hat{b}}_i &= \left(- i \Omega_{\mathrm{m},i} - \frac{\Gamma_{\mathrm{m},i}}{2}\right)\hat{b}_i - i\left(\eta_i^k g_{0,i}\right) \hat{\alpha}^\dagger_k\hat{\alpha}_k + \sqrt{\Gamma_{\mathrm{m},i}} \hat{b}_\mathrm{in},
	\end{split}
\end{equation}
where $\kappa_0^k$ and $\kappa_\mathrm{tot}^k=\kappa_0^k+\kappa_1^k +\kappa_2^k$ are the intrinsic loss rate and the total decay rate of the microwave mode, respectively. Moreover, $\Gamma_{\mathrm{m},i}$ is the intrinsic loss rate of the mechanical mode, and $\hat{a}_{\mathrm{in},0}$, $\hat{a}_{\mathrm{in},1}$, $\hat{a}_{\mathrm{in},2}$, are $\hat{b}_\mathrm{in}$ are input quantum \added{and thermal} noises from the corresponding baths.

Assuming that the coherent amplitude of the microwave mode is not affected by the optomechanical coupling, the \added{quantum} Langevin equation for the microwave mode can be divided into the classical part with $\langle \hat{\alpha}_k \rangle$ and the \replaced{quantum fluctuations }{quantum part with} $\delta\hat{\alpha}_k$. By using the linearizion $\hat{\alpha}_k = \langle \hat{\alpha}_k \rangle + \delta\hat{\alpha}_k$, we therefore have
\begin{equation}
	\label{eq:langevin_ki_linearized}
	\begin{split}
		\langle \dot{\hat{\alpha}}_k \rangle &= \left(-i\Delta_i^k -\frac{\kappa_\mathrm{tot}^k}{2}\right)\langle \hat{\alpha}_k \rangle - i\sqrt{\kappa_1^k \dot n_\mathrm{d}^k} \\
		\dot{\delta\hat{\alpha}}_k &= \left(-i\Delta_i^k -\frac{\kappa_\mathrm{tot}^k}{2}\right)\delta\hat{\alpha}_k - i\left(\eta_i^k g_{0,i}\right) \langle \hat{\alpha}_k \rangle (\hat{b}^\dagger_i+\hat{b}_i)  + \sqrt{\kappa_1^k}\hat{a}_{\mathrm{in},1} + \sqrt{\kappa_2^k}\hat{a}_{\mathrm{in},2} + \sqrt{\kappa_0^k}\hat{a}_{\mathrm{in},0} \\
		\dot{\hat{b}}_i &= \left(- i \Omega_{\mathrm{m},i} - \frac{\Gamma_{\mathrm{m},i}}{2}\right)\hat{b}_i - i\left(\eta_i^k g_{0,i}\right) \left(\langle \hat{\alpha}_k \rangle \hat{\alpha}^\dagger_k  +  \langle \hat{\alpha}_k \rangle^* \hat{\alpha}_k\right) + \sqrt{\Gamma_{\mathrm{m},i}} \hat{b}_\mathrm{in}.
	\end{split}
\end{equation}
Note that the bare optomechanical coupling terms are neglected. The time evolution of the mechanical mode is sufficiently slow for the classical amplitude of the microwave mode to be in the steady state. Using $\langle \dot{\hat{\alpha}}_k \rangle = 0$, the \replaced{mean}{classical} amplitude is therefore obtained as
\begin{equation}
	\label{eq:classical_amp}
	\langle \hat{\alpha}_k \rangle  = \frac{- i\sqrt{\kappa_1^k \dot n_\mathrm{d}^k}}{i\Delta_i^k+ \frac{\kappa_\mathrm{tot}^k}{2}}.
\end{equation}

\subsection{Extraction of participation ratio from optomechanical damping rate}
\label{subsec:eta_extraction}
As discussed in the main text, we characterize the participation ratio $\eta_i^k$ by measuring the optomechanical damping rate of mechanical mode $i$ coupled to collective mode $k$. From Eqs.~(\ref{eq:langevin_ki_linearized}), the optomechanical damping rate~\cite{SI_aspelmeyer2014cavity} is given by
\begin{equation}
	\label{eq:Gam_ki}
	\Gamma_{\mathrm{opt},i}^k = n^k_{\mathrm{c},i} \left(\eta_i^k g_{0,i} \right)^2 \left( \frac{\kappa_\mathrm{tot}^k}{\left(\Omega_{\mathrm{m},i} - \Delta_i^k\right)^2 + {\kappa_\mathrm{tot}^k}^2/4} - \frac{\kappa_\mathrm{tot}^k}{\left(\Omega_{\mathrm{m},i} + \Delta_i^k\right)^2 + {\kappa_\mathrm{tot}^k}^2/4}\right),
\end{equation}
where
\begin{equation*}
	n^k_{\mathrm{c},i} = |\langle \hat{\alpha}_k \rangle|^2 = \frac{\dot n_\mathrm{d}^k \kappa_1^k}{{\Delta_i^k}^2 + {\kappa_\mathrm{tot}^k}^2/4}
\end{equation*}
is the average photon number in collective mode $k$, where $\dot n_\mathrm{d}^k$ is the on-chip photon flux.
This yields the total effective damping rate of mechanical mode $i$:
\begin{equation}
	\label{eq:Gam_eff_ki}
	\Gamma_{\mathrm{eff},i}^k = \Gamma_{\mathrm{m},i}+ \Gamma_{\mathrm{opt},i}^k.
\end{equation}
To extract the optomechanical damping effect, we linearly change the drive power generated by a microwave source. Here, we define the transmittance between the microwave source \added{at room temperature} and the device as $R^k = \dot n_\mathrm{d}^k/\dot n_\mathrm{d}$, where $\dot n_\mathrm{d}$ is the photon flux at the output of the microwave source.
\added{The frequency dependence of the input wiring and components is weak and we assume that the transmittance depends only on $k$, but not on $i$.} \deleted{the frequency dependence of the transmittance can be negligible when the different mechanical modes are coupled to collective microwave mode $k$ by the drive field.}
Then, we obtain $\partial \Gamma_{\mathrm{eff},i}^k/\partial n_\mathrm{d}$, the slope of the total damping rate with respect to the photon flux $\dot n_\mathrm{d}$.

From Eq.~(\ref{eq:Gam_ki}), the slope of the total mechanical damping rate is analytically obtained as
\begin{equation}
	\label{eq:Gam_ki_nd}
	\frac{\partial \Gamma_{\mathrm{eff},i}^k}{\partial n_\mathrm{d}} =\frac{\kappa_1^k R^k \left(\eta_i^k g_{0,i} \right)^2}{{\Delta_i^k}^2 + {\kappa_\mathrm{tot}^k}^2/4} \left( \frac{\kappa_\mathrm{tot}^k}{\left(\Omega_{\mathrm{m},i} - \Delta_i^k\right)^2 + {\kappa_\mathrm{tot}^k}^2/4} - \frac{\kappa_\mathrm{tot}^k}{\left(\Omega_{\mathrm{m},i} + \Delta_i^k\right)^2 + {\kappa_\mathrm{tot}^k}^2/4}\right).
\end{equation}
From this equation, we experimentally obtain the unnormalized participation ratio \added{$\left ( \sum_{k~\mathrm{or}~i} \widetilde{\eta_i^k} \neq 1 \right ) $} as
\begin{equation}
	\label{eq:unnorm_eta}
	\widetilde{\eta_i^k} = \sqrt{ \frac{\partial \Gamma_{\mathrm{eff},i}^k}{\partial n_\mathrm{d}} \left({\Delta_i^k}^2+ {\kappa_\mathrm{tot}^k}^2/4 \right) \left( \frac{\kappa_\mathrm{tot}^k}{\left(\Omega_{\mathrm{m},i} - \Delta_i^k\right)^2 + {\kappa_\mathrm{tot}^k}^2/4} - \frac{\kappa_\mathrm{tot}^k}{\left(\Omega_{\mathrm{m},i} + \Delta_i^k\right)^2 + {\kappa_\mathrm{tot}^k}^2/4}\right)^{-1}}.
\end{equation}
By definitions in Eqs.~(\ref{eq:Gam_ki_nd}) and (\ref{eq:unnorm_eta}), the relation between the normalized and unnormalized participation ratios is described as
\begin{equation}
	\label{eq:eta_eta}
	\widetilde{\eta_i^k}  = g_{0,i} \: \eta_i^k \sqrt{\kappa_1^k R^k}.
\end{equation}
In principle, the \deleted{normalized} participation ratio $\eta_i^k$ can be determined by \deleted{using} \added{normalizing Eq.~(\ref{eq:eta_eta}) with} $g_{0,i}$, $\kappa_1^k$ and $R^k$ that would be obtained independently.
However, the local optomechanical coupling $g_{0,i}$ can not be straightforwardly measured in such a multimode optomechanical system. Furthermore, it is also not trivial to determine $\kappa_1^k$ and $R^k$ using our setup in the transmission configuration, \added{since it is difficult to differentiate the input contribution to the output one.}

Nevertheless, we can determine $\eta_i^k$ from experimentally-obtainable $\widetilde{\eta_i^k}$ based on the normalization conditions of the unitary transformation~[Eqs.~(\ref{eq:sumi_eta}) and (\ref{eq:sumk_eta})].
Here, we apply an iterative normalization method for $\widetilde{\eta_i^k} $ as follows.
We first define the initial unnormalized participation ratio as
\begin{equation}
	\label{eq:eta0}
	\widetilde{\eta_i^k}_{(0)} = \widetilde{\eta_i^k} .
\end{equation}
Then, we iterate the normalization process in the row and column axes alternatively.
Namely, the unnormalized participation ratio $\widetilde{\eta_i^k}_{(n)}$ \added{at step $n$} is updated in each normalization step to $\widetilde{\eta_i^k}_{(n+1)} $ as
\begin{equation}
	\label{eq:normi_eta}
	\widetilde{\eta_i^k}_{(n+1)}  = \frac{\widetilde{\eta_i^k}_{(n)}}{\sum_i\widetilde{\eta_i^k}_{(n)}} \quad (\forall \: k, \:\mathrm{if} \: n \:  \mathrm{is \: even})
\end{equation}
or
\begin{equation}
	\label{eq:normk_eta}
	\widetilde{\eta_i^k}_{(n+1)} = \frac{\widetilde{\eta_i^k}_{(n)}}{\sum_k \widetilde{\eta_i^k}_{(n)}} \quad (\forall \: i, \:\mathrm{if} \: n \: \mathrm{is \: odd}).
\end{equation}
In the iterative normalization method, the unnormalized participation ratio at step $n$ can be always described as
\begin{equation}
	\label{eq:unnorm}
	\widetilde{\eta_i^k}_{(n)} = C_{i\:(n)}\: \eta_i^k \:D_{k\:(n)},
\end{equation}
where $C_{i\:(n)} $ and $D_{k\:(n)}$ are coefficients depending only on either $i$ or $k$, \added{respectively}.
Importantly, each normalization process updates only \deleted{these} \added{the} coefficients \added{$C_{i\:(n)} $ and $D_{k\:(n)}$} without any changes in $\eta_i^k$.
\added{This can be easily confirmed from Eqs.~(\ref{eq:normi_eta}) and (\ref{eq:normk_eta}).}

\begin{figure}[h]
	\centering
	\includegraphics[scale=1.2]{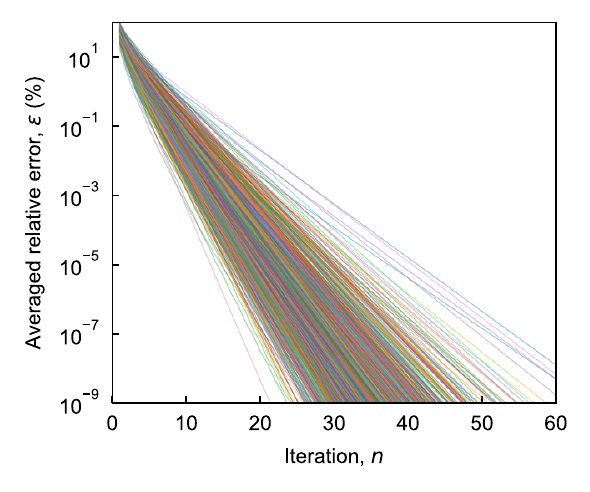}
	\caption{\added{\textbf{Numerical justification of iterative normalization method}. The averaged relative error of the unnormalized participation ratio compared with the original participation ratio as a function of step $n$ of the iterative normalization. Each different-colored line corresponds to a different dataset. The system size is chosen to be $N=10$.}}
	\label{fig:Iteration}
\end{figure}

Here, we numerically confirm that both $C_{i\:(n)} $ and $D_{k\:(n)}$ converge to 1, \added{i.e., $\widetilde{\eta_i^k}_{(n)} \rightarrow \eta_i^k$} with an sufficiently large number of the iteration.
\added{For this purpose, we prepare a random $N \times N$ unitary matrix $\mathbf{U}$ and obtain the participation ratio as $\eta_i^k=|[\mathbf{U}]_{k,i}|^2$.
	In addition, we prepare random coefficients $0 < C_i < 1$ and $0 < D_k < 1$ and artificially generate the unnormalized participation ratio as $\widetilde{\eta_i^k} = C_i\eta_i^k D_k$ to simulate the unnormalized participation ratio.
	The task is to deduce $\eta_i^k$ from $\widetilde{\eta_i^k}$.
	We apply the iterative normalization method to the unnormalized participation ratio.
	As an evaluation function, we define the averaged relative error of the unnormalized participation ratio at $n$ step compared with the original one as}
\begin{equation}
	\label{eq:error_iteration}
	\varepsilon = \frac{\sum_{i,k}\left|\widetilde{\eta_i^k}_{(n)}-\eta_i^k\right|/\eta_i^k}{N^2}.
\end{equation}
\added{In Fig.~\ref{fig:Iteration}, we plot the averaged relative error $\varepsilon$ at each iteration step for 1000 different data sets. 
	From these results, we confirm that the unnormalized participation ratio converges to the original one, i.e.\ $\widetilde{\eta_i^k}_{(n)} \rightarrow \eta_i^k$ when the number of the iteration is sufficiently large.}

\deleted{Thus, even without knowledge of the coefficients, the initial participation ratio converges to the normalized one, i.e.\ $\widetilde{\eta_i^k}_{(n)} \rightarrow \eta_i^k$ when $n \rightarrow \infty$.
	Empirically, $n=50$ is sufficient for $\widetilde{\eta_i^k}$ to converge to $\eta_i^k$.}

\section{\added{Topological theory of 1D and 2D multimode systems}}

\subsection{\added{SSH model for a 1D optomechanical chain}}\label{sec:1DSSH}
To demonstrate a topological property, we realize the 1D SSH model in our multimode optomechanical system.
Here, we report the analysis of J. K. Asb{\'o}th, \textit{et al.} (2011)~\cite{SI_asboth2016short} and discuss the topological aspects of the SSH model.

\subsubsection{\added{Band structure}}
\begin{figure*}[h!]
	\includegraphics[scale=1]{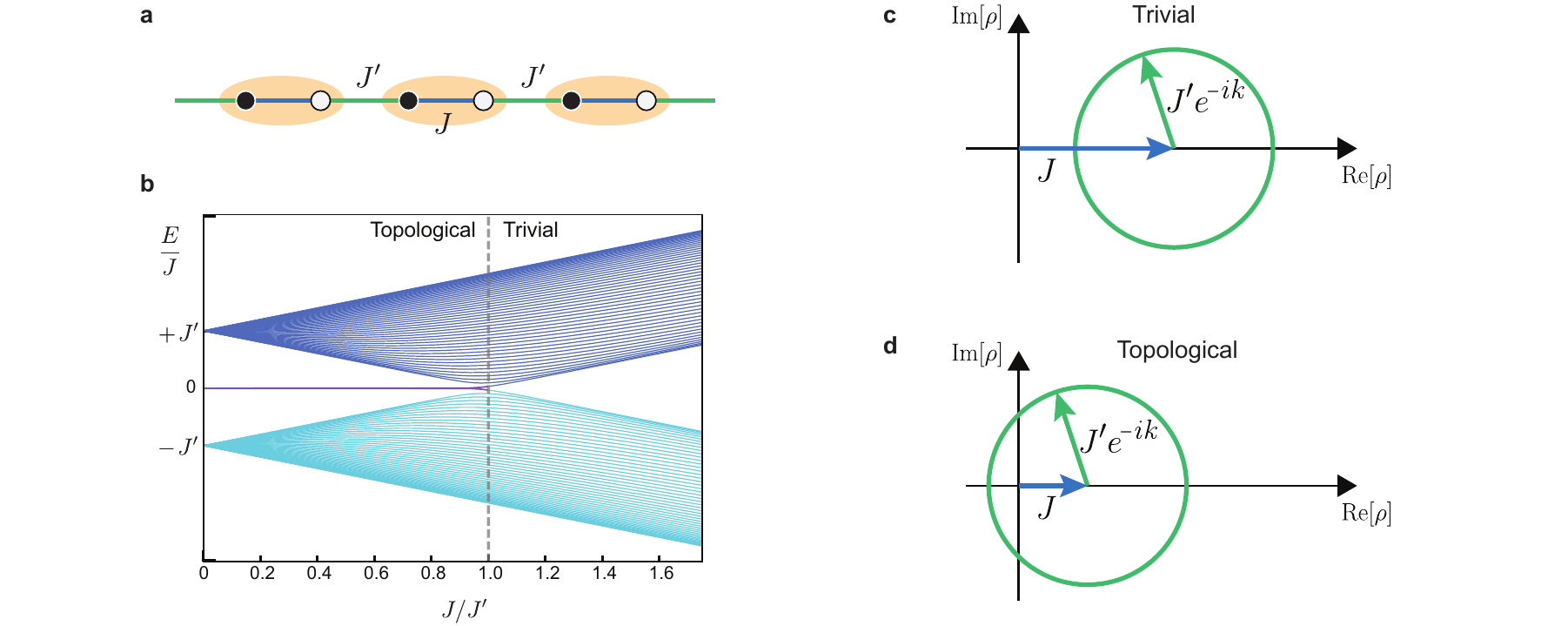}
	\caption{
		\textbf{1D SSH model}.
		\textbf{a},~Schematic of a 1D SSH chain. The black and white circles describe sites $A$ and $B$, respectively. 
		\textbf{b},~Band structure as a function of the coupling ratio $J/J'$. 
		\textbf{c,d},~off-diagonal element $\rho(k)$ in the bulk Hamiltonian in the trivial and topological phase, respectively.
	}
	\label{fig:SI_1DSSH}
\end{figure*}

The SSH model is constructed by a 1D chain of unit cells individually consisting of two sites ($A$ and $B$).
The two sites in a unit cell are coupled to each other with a coupling strength of $J$, while a unit cell is connected to adjacent cells with a strength of $J'$, as shown in Fig.~\ref{fig:SI_1DSSH}a.
To analyze the band structure, we consider a 1D SSH chain with a total cell number of $N$ and impose a periodic boundary condition on the chain.
The Hamiltonian of such a chain is given by
\begin{equation}
	\hat{H} = \sum_n \left(J|n, B \rangle \langle n, A| + J'|n+1, A \rangle \langle n, B|+ \mathrm{h.c.}\right),
\end{equation}
where $|n, A\rangle$ and $|n, B\rangle$ denote a state vector at site $A$ and $B$ of unit cell $n$, respectively.
For convenience, the Hamiltonian is described by a tensor product of the intra-cell and inter-cell subsystems, i.e.,
\begin{equation}
	\hat{H} = \sum_n \left(J|n\rangle \langle n|\otimes \hat{\sigma}_+ + J'|n+1 \rangle \langle n|\otimes \hat{\sigma}_- + \mathrm{h.c.}\right),
\end{equation}
where $|n\rangle$ is a state vector of the inter-cell subsystem and $\hat{\sigma}_+ = |B\rangle \langle A|$ ($\hat{\sigma}_- = |A\rangle \langle B|$) is a ladder operator describing the intra-cell coupling.
Based on Bloch's theorem, we can diagonalize the Hamiltonian of the inter-cell subsystem using a wavenumber basis, which is defined as
\begin{equation}
	\label{eq:wavenumberbasis}
	|k\rangle=\frac{1}{\sqrt{N}}\sum_{n} e^{ik n}|n\rangle,
\end{equation}
where $k=2\pi \nu/N$ ($\nu=1,2,\cdots,N$) is the wavenumber.
Thus, the Hamiltonian can be rewritten as
\begin{equation}
	\hat H = \sum_{k} |k\rangle \langle k|\otimes \left[J\hat{\sigma}_x + J' \left(\hat{\sigma}_+ e^{ik} + \hat{\sigma}_- e^{-ik}\right)\right].
\end{equation}
Then, the band structure of the 1D chain is obtained as the eigenenergy of the so called 'bulk Hamiltonian' related to the intra-cell subsystem:
\begin{align}
	\hat{H}(k) & = \langle k | \hat{H}|k\rangle                                                     \\
	& = J\hat{\sigma}_x + J' \left(\hat{\sigma}_+ e^{ik} + \hat{\sigma}_- e^{-ik}\right) \\
	& =
	\left(
	\begin{array}{cc}
		0          & \rho(k) \\
		\rho^*(k) & 0
	\end{array}
	\right),
\end{align}
where $\hat{\sigma}_x$ is the Pauli-$x$ matrix of the intra-cell system.
Here, the off-diagonal element $\rho(k)$ of the bulk Hamiltonian is defined as
\begin{equation}
	\rho(k) = J+J'e^{-ik} \equiv |\rho(k)|e^{-i\phi(k)}.
\end{equation}
From the $2 \times 2$ matrix, the eigenenergy for each wavenumber is obtained as
\begin{equation}
	E(k) = \pm |\rho(k)| = \pm\sqrt{J^2 + {J'}^2 +2JJ'\cos k}
\end{equation}
with eigenvectors (Bloch wave function) of
\begin{equation}
	\label{eq:eigenvec1Dssh}
	|u_{k, \pm}\rangle =\frac{1}{\sqrt{2}}
	\left(
	\begin{array}{c}
		e^{-i\phi(k)} \\
		\pm1
	\end{array}
	\right).
\end{equation}
In the thermodynamic limit of $N\rightarrow\infty$, the band structure shows upper and lower bands when $J \neq J'$, while the two bands are closed at the phase transition point of $J=J'$ (see the Fig.~\ref{fig:SI_1DSSH}b).

\subsubsection{\added{Bulk-edge correspondence}} \label{sec:1DZak}
The bulk-edge correspondence~\cite{SI_delplace2011zak} reveals the existence of edge states of a truncated 1D SSH chain from the bulk properties of the corresponding infinite chain.
More specifically, we can predict the existence from the Zak phase, which is calculated using the eigenvectors of Eq.~(\ref{eq:eigenvec1Dssh}) as
\begin{equation}
	\mathcal{Z} = i \oint d k \langle u_{k, \pm}| \partial_{k} |u_{k, \pm}\rangle = \frac{1}{2} \oint dk \:\partial_{k} \phi(k),
\end{equation}
where the integration is performed over the first Brillouin zone. 
Note that the Zak phase is a bulk property, which can be defined in the infinite size limit.
The chiral symmetry confirms that the Zak phase is discretized to be $\pi$ multiplied by an integer. As we will see in the following section, when $\mathcal{Z} = 0$, the chain is in the trivial phase, where the truncated chain does not show edge states. On the other hand, when $\mathcal{Z} = \pi$, the chain is in the nontrivial topological phase, showing two edge states when the chain is truncated to be a finite system. For the 1D SSH model, $\mathcal{Z} = 0$ when $J>J'$ while $\mathcal{Z} = \pi$ when $J<J'$. 

Furthermore, the Zak phase can be graphically obtained from the winding number of $\rho(k)$.
As shown in Figs.~\ref{fig:SI_1DSSH}c and d, the off-diagonal element $\rho(k)$ of the bulk Hamiltonian for the trivial and topological cases is described by a closed curve in the complex plane, respectively. The Zak phase is obtained as the product of $\pi$ and the winding number of the closed curve around the origin. Namely, the existence of edges states is simply determined by whether the closed curve winds around the origin or not for the 1D SSH model. 

\subsubsection{\added{Finite-size effect}} \label{sec:1D_finite}
Here, we report the analysis presented and introduced by Delplace, \textit{et al.} (2011)~\cite{SI_delplace2011zak}, explaining how the Zak phase can predict the existence of edge states in the 1D SSH model. Then, we discuss the finite-size effects on the prediction. Importantly, the following analysis can be extended to a generalized 1D SSH model, where a 1D chain is constructed by a chain of 2-site unit cells with arbitrary inter-cell couplings that do not break the translational symmetry.
Furthermore, the analysis can be valid as long as the inter-call couplings do not induce the Pauli-$z$ component in the bulk Hamiltonian. This is the case for our 1D SSH model including the second and third nearest-neighbor couplings, as well as a wavenumber-resolved 1D model reduced from a strained graphene ribbon.

When an open boundary condition is imposed on a 1D SSH chain, an eigenstate of the full Hamiltonian can be described by a superposition of two degenerate plane waves.
Using Bloch's theorem, a state vector of a plane wave with an eigenenergy of $E(k)=\pm|\rho(k)|$ is described by
\begin{equation}
	|\psi_k,\pm \rangle = \frac{1}{\sqrt{2N}}\sum_n e^{ikn}
	\left(
	\begin{array}{c}
		e^{-i\phi(k)} \\
		\pm1
	\end{array}
	\right)
	(|n, A \rangle \:\: |n, B \rangle).
\end{equation}
Note that there are two degenerate plane waves, except for the maximum and minimum absolute values of eigenenergy.  
Thus, a trial eigenstate of bulk modes is given by a superposition of $|\psi_k,\pm \rangle$ and $|\psi_{k'},\pm \rangle$ with the constraint of the degeneracy, i.e., $E(k) = E(k')$.

To satisfy the open boundary condition, the trial eigenstate should vanish at the nearest site outside of the chain, that is, at site $B$ for $n = 0$ and at site $A$ for $n = N + 1$. These conditions determine a possible wavenumber for bulk modes. 
The condition at site $B$ for $n=0$ is straightforwardly satisfied by superposing the two plane waves such that $(|\psi_k,\pm \rangle - |\psi_{k'},\pm \rangle)/\sqrt{2}$. Then, the other condition at site A for $n=N+1$ imposes 
\begin{equation}
	\label{1DSSHopen0}
	\left[k(N+1) - \phi(k)\right] - \left[k'(N+1) - \phi(k')\right] = 2\pi j, 
\end{equation}
where $j$ is an integer.
By defining  $\bar{k} = \frac{k-k'}{2}$ and $\bar{\phi}\left(\bar{k}\right) = \frac{\phi(k)-\phi(k')}{2}$, Eq.~(\ref{1DSSHopen0}) is rewritten as 
\begin{equation}
	\label{1DSSHopen}
	(N+1)\bar{k} - \pi j= \bar{\phi}\left(\bar{k}\right).
\end{equation}
Thus, counting the number of the solutions, which is the number of the possible bulk modes, corresponds to counting the number of intersections between $(N+1)\bar{k} - \pi j$ and $\bar{\phi}\left(\bar{k}\right)$ in the range of $0<\bar{k}<\pi$. 

\begin{figure*}[h!]
	\includegraphics[scale=1]{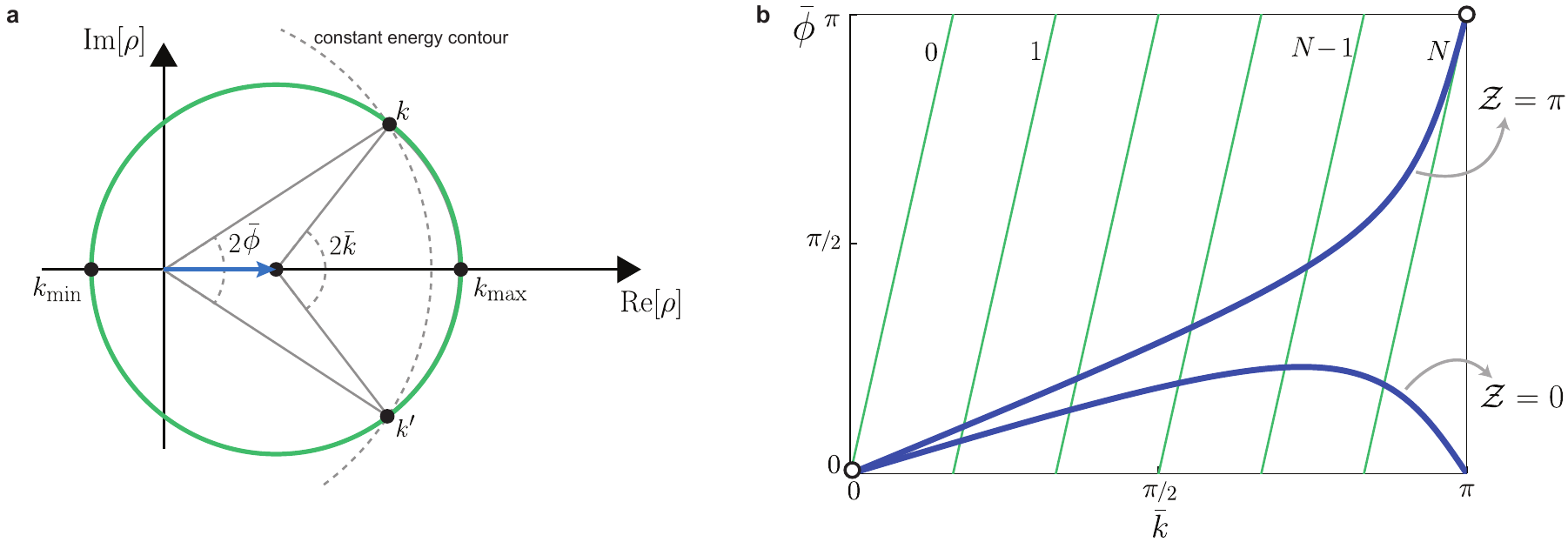}
	\caption{
		\textbf{finite-size effect in 1D SSH model}.
		\textbf{a},~off-diagonal element $\rho(k)$ in the complex plane. Its value corresponding to the maximum and minimum absolute values of the eigenenergy are shown at wavenumbers $k_\mathrm{max}$ and $k_\mathrm{min}$. The wavenumber $k$ and $k'$ that satisfy the degeneracy constraint are found by the two intersections between $\rho(k)$ and the constant energy contour.
		\textbf{b},~$(N+1)\bar{k} - \pi j$ (green lines for $j=0,1,\cdots N$) and $\bar{\phi}\left(\bar{k}\right)$ (blue lines for the cases in the trivial and topological phases, respectively) as a function of $\bar{k}$.
	}
	\label{fig:SI_1DSSHfinite}
\end{figure*}

For simplicity, we assume that the phase references are determined by the plane wave with the highest absolute value of the eigenenergy such that $\bar{k}\rightarrow0$ and $\bar{\phi}\left(\bar{k}\right) \rightarrow 0$ when $k\rightarrow k_\mathrm{max}$, where $k_\mathrm{max}$ is the wavenumber associated with the maximum absolute value of the eigenenergy (see Fig.~\ref{fig:SI_1DSSHfinite}a). 
Moreover, without loss of generality, we can define $\bar{k}$ to be a positive number.
With these conditions, $\bar{k}$ depends on only $k$ due to the constraint of $E(k) = E(k')$. Then, we systematically find solutions from the wavenumber with the highest absolute value of the eigenenergy $\left(\bar{k}=0\right)$ to one with the lowest absolute value of the eigenenergy $\left(\bar{k}=\pi\right)$ by plotting $(N+1)\bar{k} - \pi j$ and $\bar{\phi}\left(\bar{k}\right)$, as shown with an example of the standard 1D SSH case in Fig.~\ref{fig:SI_1DSSHfinite}b. While $\bar{\phi}(0) = 0$ by definition, $\bar{\phi}(\pi)$ depends on the Zak phase, i.e., while $\bar{\phi}(\pi)=0$ if $\mathcal{Z} = 0$, $\bar{\phi}(\pi)=\pi$ if $\mathcal{Z} = 1$ (see Fig.~\ref{fig:SI_1DSSHfinite}a for the topological case).
As shown in Fig.~\ref{fig:SI_1DSSHfinite}b, we can always find $N$ solutions in the range of $0<\bar{k}<\pi$ when in the trivial phase ($\mathcal{Z} = 0$), confirming that all the solutions are found in the bulk modes and there are no edges states. On the other hand, when in the nontrivial topological phase ($\mathcal{Z} = \pi$), the function of $(N+1)\bar{k} - \pi j$ for $j=1,2,\cdots N-1$ have an intersection with $\bar{\phi}\left(\bar{k}\right)$, but not for $j=N$ in the range of $0<\bar{k}<\pi$, for sufficiently large $N$. Thus, there are only $N-1$ solutions for bulk modes, implying that there is one another mode that is not described by the plane waves, corresponding to an edge state~\cite{SI_delplace2011zak}. Since we can independently apply the same analysis to both the upper and lower bands, there are two edge states in total when $\mathcal{Z} = \pi$.

When the system size is large enough, the Zak phase calculation accurately predicts the existence of edge states. However, when the system becomes smaller, the number of solutions of Eq.~(\ref{1DSSHopen}) may change. More precisely, even though the condition of $\mathcal{Z} = \pi$ is satisfied, we can find an intersection between $(N+1)\bar{k} - \pi j$ for $j=N$ and $\bar{\phi}\left(\bar{k}\right)$ in the range of $0<\bar{k}<\pi$ when the slope of $\bar{\phi}\left(\bar{k}\right)$ at $\bar{k}$ is steeper than $N+1$.
Therefore, the slope condition for the existence of edge states is mathematically described as
\begin{equation}
	\label{1DSSHslope0}
	\left|\frac{\partial \bar{\phi}(\bar{k})}{\partial \bar{k}}\vline_{\:\bar{k}=\pi}\right|<N+1,
\end{equation}
which is actually satisfied in the example shown in Fig.~\ref{fig:SI_1DSSHfinite}b for the case in the topological phase.
Furthermore, Eq.~(\ref{1DSSHslope0}) can be simplified as
\begin{equation}
	\label{1DSSHslope}
	\left|\frac{\partial \phi(k)}{\partial k}\vline_{\:k=k_\mathrm{min}}\right|<N+1, 
\end{equation}
where we use $\frac{\partial \phi(k)}{\partial k} = - \frac{\partial \phi(k')}{\partial k'}$ for $k,k' \rightarrow k_\mathrm{min}$. Here, $k_\mathrm{min}$ is the wavenumber associated with the minimum absolute value of the eigenenergy. Therefore, in order to possess edge states, a 1D SHH chain needs to satisfy two conditions: the condition for the Zak phase ($\mathcal{Z} = \pi$) and the slope condition for $\phi(k)$ at $k=k_\mathrm{min}$, described by Eq.~(\ref{1DSSHslope}).
Note that the slope condition is automatically satisfied in the limit of infinite size ($N \rightarrow \infty$), implying the Zak phase calculation accurately predicts the existence of edge states for a large-size system.

\subsubsection{\added{Effect of parasitic couplings}}
\label{subsec: SSHmodel_beyond}

In most physical implementations of the SSH model in lattices, parasitic couplings between distanced sites are unavoidable and deviate the actual response of the devices from the ideal model. In our case, the parasitic couplings arise from the mutual inductance between distanced spiral inductors of LC circuits. The parasitic coupling rate decreased with $J_\mathrm{par.}\propto\frac{1}{l^3}$, where $l$ is the effective distance between two spirals. 
The effect of these higher-order couplings on the band structure, modeshapes, and topological properties of the SSH chains has been studied in several works~\cite{SI_li2014topological, SI_perez2018ssh, SI_perez2019interplay}. Here we summarize these results and calculate the band structure of the designed devices in the presence of the parasitic coupling.

\begin{figure*}[h!]
	\includegraphics[scale=1]{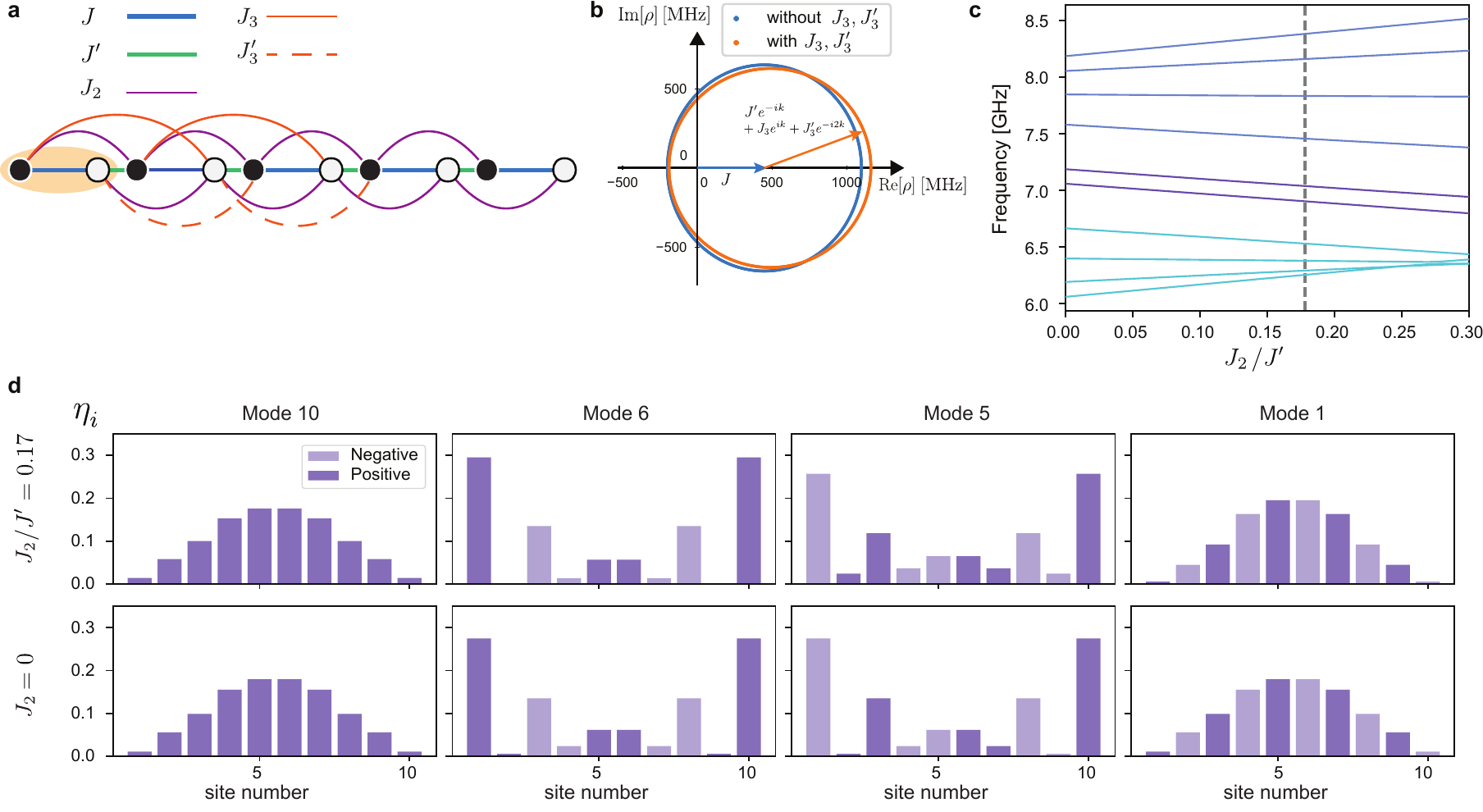}
	\caption{
		\textbf{Effect of parasitic couplings on the 1D SSH model}.
		\textbf{a}, Mode diagram of a 1D SSH chain including second and third nearest-neighbor couplings. The black and white circles describe sites $A$ and $B$, respectively. \textbf{b}, Off-diagonal element $\rho(k)$ in the complex plane with and without taking into account the third nearest-neighbor couplings. We use the system parameters of the 1D device presented in the main text. \textbf{c}, The energy levels (eigen frequencies) as a function of the relative second nearest-neighbor coupling strength for the 1D SSH chain. The dashed line shows the value of the second nearest-neighbor coupling of the actual devices discussed in the main text. Color coding denotes the LPB, edge states and UPB. \textbf{d}, Comparison of the modeshapes between an ideal 10-site SSH chain and a chain with the second nearest-neighbor couplings, showing a minor change in the modeshapes.
	}
	\label{fig:SI_1DSSH_parasitic_couplings}
\end{figure*}

Figure \ref{fig:SI_1DSSH_parasitic_couplings}a shows the mode diagram including the second and third nearest-neighbor couplings $J_2$, $J_3$ and $J_3'$.
In our 1D optomechanical system, the second and third nearest-neighbor couplings are found to be approximately $J_2/2\pi = 100$~MHz, $J_3'/2\pi = 37$~MHz and $J_3/2\pi = 27$~MHz respectively, while the alternating nearest-neighbor couplings are found to be $J/2\pi = 470$~MHz and $J'/2\pi = 700$~MHz. Importantly, we can safely assume that the two different second nearest-neighbor couplings are strictly identical due to the geometry of our 1D system. 
The Hamiltonian of a 1D SSH chain including such parasitic couplings is given by
\begin{equation}
	\begin{split}
		\hat{H} = \sum_n &[J|n, B \rangle \langle n, A| + J'|n+1, A \rangle \langle n, B|\\
		&J_2|n+1, A \rangle \langle n, A| + J_2|n+1, B \rangle \langle n, B|\\
		&J_3|n+1, B \rangle \langle n, A| + J'_3|n+2, A \rangle \langle n, B|\\
		&+ \mathrm{h.c.}].
	\end{split}
\end{equation}
Since the parasitic couplings does not distort the translational symmetry, we can apply Bloch's theorem using the wavenumber basis defined as Eq.~(\ref{eq:wavenumberbasis}), and obtain the bulk Hamiltonian as
\begin{align}
	\hat{H}(k) & =
	\left(
	\begin{array}{cc}
		\epsilon(k)         & \rho(k) \\
		\rho^*(k) & \epsilon(k)
	\end{array}
	\right),
\end{align}
where
\begin{equation}
	\rho(k) = J+J'e^{-ik} +J_3e^{ik} + J_3'e^{-i2k} \equiv |\rho(k)|e^{-i\phi(k)}
\end{equation}
and
\begin{equation}
	\epsilon(k) = J_2\cos(k).
\end{equation}
The second nearest-neighbor couplings modify the diagonal elements of the bulk Hamiltonian, not inducing a component of the Pauli-$z$ matrix but only the identity matrix. This is because the two different second nearest-neighbor couplings, which, in general, may differ, are identical in our 1D system.  As a result, we can diagonalize the Bulk Hamiltonian with eigenvectors in the same form as Eq.~(\ref{eq:eigenvec1Dssh}) and apply the same analysis as the ideal 1D SSH model in order to predict the existence of edges states. On the other hand, the third nearest-neighbor couplings modify the off-diagonal element $\rho(k)$.

In Fig.~\ref{fig:SI_1DSSH_parasitic_couplings}b, we plot $\rho(k)$ in the complex plane, using the experimentally obtained coupling strengths with and without taking into account the third nearest-neighbor couplings. Since the third nearest-neighbor couplings are negligible in our 1D optomechanical system, the winding number is found to be 1 for both the cases. As long as it shows a well-defined band gap ($J_2<J'/2$), the 1D SSH chain with the non-trivial winding number possess edges modes~\cite{SI_perez2018ssh}. This is the case for the 1D optomechanical chain presented in the main text.

Furthermore, we numerically study the effect of parasitic couplings on the eigenenergies and modeshapes in the SSH model. For simplicity, we here neglect the third nearest-neighbor couplings. 
Figure \ref{fig:SI_1DSSH_parasitic_couplings}c shows the energy levels as a function of the relative second nearest-neighbor coupling strength for our 1D SSH chain. The second nearest-neighbor coupling breaks the symmetry of the band structure resulting in a wider UPB and narrower LPB~\cite{SI_perez2018ssh}, while the edge states splitting and the band gap is barely affected. 
Figure \ref{fig:SI_1DSSH_parasitic_couplings}e shows  the modeshapes of an ideal 10-site SSH chain and a chain with the second nearest-neighbor couplings, respectively. The second nearest neighbor couplings of our 1D device do not significantly change the modeshapes.

\subsection{\added{Strained graphene model}}
To demonstrate the extendability of our scheme to a 2D structure, we realize the strained graphene model in our optomechanical system, as shown in the main text. Here, we explain the basic concepts of the graphene model~\cite{SI_ni2008uniaxial,SI_naumis2017electronic,SI_pereira2009tight}, including the band structure of an infinite-size system, the bulk-edge correspondence, and the numerically-simulated results for a finite-size system. These will give us a good intuition about the modeshapes of our 24-site 2D optomechanical system.

\subsubsection{\added{Band structure}} \label{sec:graphene_band}
\begin{figure*}[h!]
	\includegraphics[scale=1]{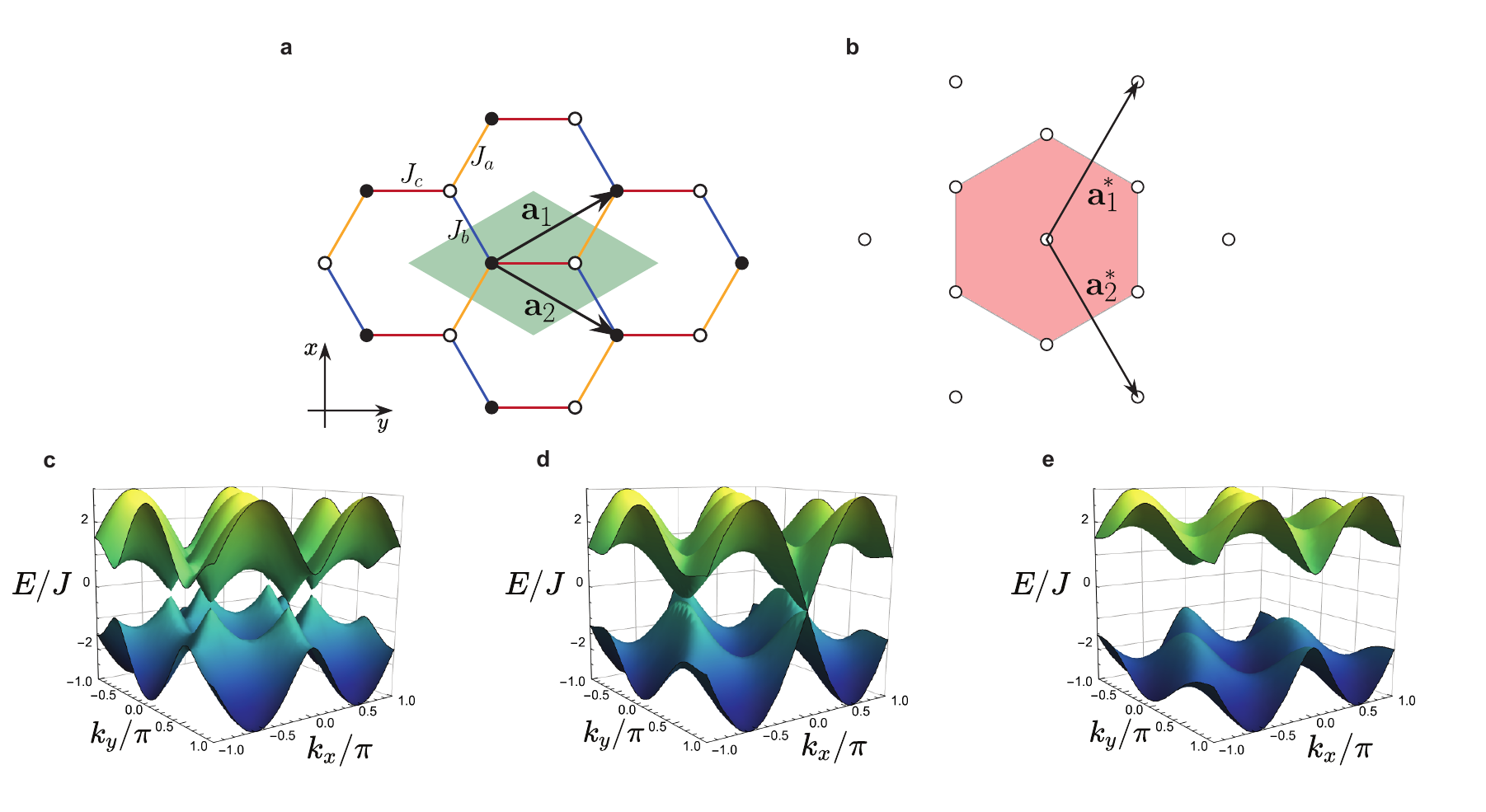}
	\caption{\textbf{Honeycomb lattice for a graphene of infinite size.} 
		\textbf{a},~Multimode system in a honeycomb lattice with a unit cell (green region) and two lattice vectors ($\textbf{a}$ and $\textbf{b}$). Three coupling strengths ($J_a$, $J_b$, and $J_c$) in different orientations are shown with lines of different colors, respectively. The black and white circles describe sites $A$ and $B$, respectively. 
		\textbf{b},~Reciprocal honeycomb lattice with two reciprocal lattice vectors ($\textbf{a*}$ and $\textbf{b*}$)., where the first Brillouin zone is highlighted. 
		\textbf{c--e},~Band structure of a strain-free and strained graphenes for $J'/J = 1$, $J'/J = 0.5$, and $J'/J = 0.25$, respectively.
	}
	\label{fig:SI_2D_theory}
\end{figure*}

To label every site in a honeycomb lattice, we define two lattice vectors, as denoted with $\textbf{a}_1$ and $\textbf{a}_2$ in Fig.\ref{fig:SI_2D_theory}a, where the coordinate is chosen to have lattice vectors defined as $\textbf{a}_1,\textbf{a}_1=(\pm\frac{\sqrt{3}}{2},\frac{3}{2})$.
In addition, one unit cell, enclosed in the green shaded area, consists of two sites.
As shown with lines at different angles, connecting sites in Fig.~\ref{fig:SI_2D_theory}a, three unique couplings ($J_a$, $J_b$, and $J_c$) exist in the honeycomb lattice.

The Hamiltonian of such a 2D multimode system can be formally written as
\begin{equation}
	\hat H = \sum_{\textbf{i}-\textbf{j}=\textbf{a}_1}J_a|\textbf{i}, A \rangle \langle \textbf{j}, B| + \sum_{\textbf{i}-\textbf{j}=\textbf{a}_2}J_b|\textbf{i}, A \rangle \langle \textbf{j}, B| +\sum_{\textbf{i}}J_c|\textbf{i}, B \rangle \langle \textbf{i}, A| + \mathrm{h.c.},
\end{equation}
where $|\textbf{i}, A \rangle$ and $|\textbf{i}, B \rangle$ denotes a state vector at site $A$ and $B$ of unit cell $\textbf{i}=m\textbf{a}_1+n\textbf{a}_2$, respectively.

In a similar way to the 1D SSH model, we first consider the Hamiltonian of an infinite-size honeycomb lattice with periodic conditions in the two translational-symmetry directions (thermodynamical limit of $N\rightarrow\infty$). Using Bloch's theorem, we can diagonalize the Hamiltonian of the inter-cell subsystem by using a wavenumber basis, which is defined as
\begin{equation}
	|\textbf{k}\rangle=\frac{1}{\sqrt{N}}\sum_\textbf{i}e^{i\textbf{i}\cdot \textbf{k}}|\textbf{i}\rangle,
\end{equation}
where \textbf{i} denotes a lattice point and \textbf{k} is the wave vector, shown with red in Fig.\ref{fig:SI_2D_theory}b, together with the reciprocal lattice. The two reciprocal lattice vectors are $\textbf{a}_1^*,\textbf{a}_2^*=2\pi (\pm\frac{\sqrt{3}}{3}, \frac{1}{3})$ in the coordinate we choose, following the convention $\mathbf{i} \cdot \textbf{j}^* = 2\pi\delta_{ij}\ (i, j \in \{\mathbf{a, b}\})$.
Thus, the wavenumber-resolved intra-cell Hamiltonian, called bulk Hamiltonian, is obtained as
\begin{align}
	\hat H(\textbf{k})&=\langle \textbf{k}|\hat H|\textbf{k}\rangle \\
	& = \left(
	\begin{array}{cc}
		0 & \rho(\mathbf{k})\\
		\rho^*(\mathbf{k}) & 0
	\end{array}
	\right),
	\label{H_k_2D}
\end{align}
where
\begin{equation}
	\rho(\mathbf{k}) = J_c + J_ae^{-i\textbf{a}\cdot \textbf{k}}+J_be^{-i\textbf{b}\cdot \textbf{k}} \equiv |\rho(\mathbf{k})|e^{-i\phi(\mathbf{k})}.
	\label{phi_definition}
\end{equation}
By diagonalizing the bulk Hamiltonian, the band structure can be obtained as $E(\textbf{k}) = \pm |\rho(\mathbf{k})|$. The upper and lower bands for strain-free graphene of infinite size ($J_a=J_b=J_c$) are shown in Fig.~\ref{fig:SI_2D_theory}c, where the upper and lower bands are connected at the Dirac points. A strain applied to the lattice will introduce non-equal couplings for different angles, depending on the orientation of the strain. 
For the threshold $J_i + J_j = J_k\ (i,j,k\in\{a,b,c\})$, pairs of Dirac points will merge (see Fig.~\ref{fig:SI_2D_theory}d), while the upper and lower bands become gapped for further anisotropy (see Fig.~\ref{fig:SI_2D_theory}e). For the model presented in the main text, two of the couplings are equal ($J_a = J_b = J$) while the other is different from them ($J_c=J'$). In this case, the infinite 2D honeycomb lattice shows a phase transition at $J'/J = 0.5$, where the band structure will be gapped for $J'/J<0.5$.

\subsubsection{\added{Graphene ribbon}} \label{sec:graphene_cut}
The 2D optomechanical lattice presented in the main text can be interpreted as a truncated graphene ribbons with different orientations of the boundaries, as shown in Fig.~\ref{fig:SI_strained_graphane_ribbon}a. It is well known that the existence of edge states depends on the edge structure of a graphene ribbon~\cite{SI_nakada1996edge, SI_kohmoto2007zero,SI_wang2016giant,SI_hatsugai2009bulk,SI_rechtsman2013topological,SI_plotnik2014observation} and can be predicted from the bulk structure of the corresponding graphene model of infinite size. This is known as the bulk-edge correspondence. One of the examples of such edge states appears on the so-called zig-zag edges of a graphene ribbon~\cite{SI_nakada1996edge} and has been experimentally observed in 2D materials~\cite{SI_wang2016giant,SI_plotnik2014observation} as well as photonic~\cite{SI_rechtsman2013topological} and microwave~\cite{SI_bellec2013topological,SI_bellec2014manipulation} structures. To predict the existence of edge states for our 2D optomechanical lattice, we follow the method formalized by Delplace, \textit{et al.} (2011)~\cite{SI_delplace2011zak}. Here, we explain how a graphene ribbon with different orientations of boundaries can be described to predict the existence of edge states using the bulk-edge correspondence.

A graphene ribbon is modeled by a multimode system in a honeycomb lattice, where the periodic boundary condition is imposed on the ribbon direction while the open boundary condition is imposed on the width direction that is parallel to a lattice vector defining the ribbon width. The translational symmetry in the ribbon direction ensures that the wavenumber in the ribbon direction is well defined. Therefore, given a certain wavenumber in the ribbon direction, the 2D lattice can be reduced to a 1D chain in the width direction, enabling us to predict the existence of edge states of the ribbon by using a similar analysis to the 1D SSH model~\cite{SI_delplace2011zak}. Note that a different wavenumber in the ribbon direction gives a different result on the prediction of the existence of edge states.

The choice of a unit cell and two lattice vectors is naturally determined by the boundary structure of a graphene ribbon to be considered. This is crucial to predicting the edge states on the boundary using the bulk-edge correspondence. Since our 2D optomechanical system can be considered to have either so-called zig-zag or armchair edges, we describe the Hamiltonian with two different choices of a unit cell and two lattice vectors.

\begin{figure*}[h!]
	\includegraphics[scale=1]{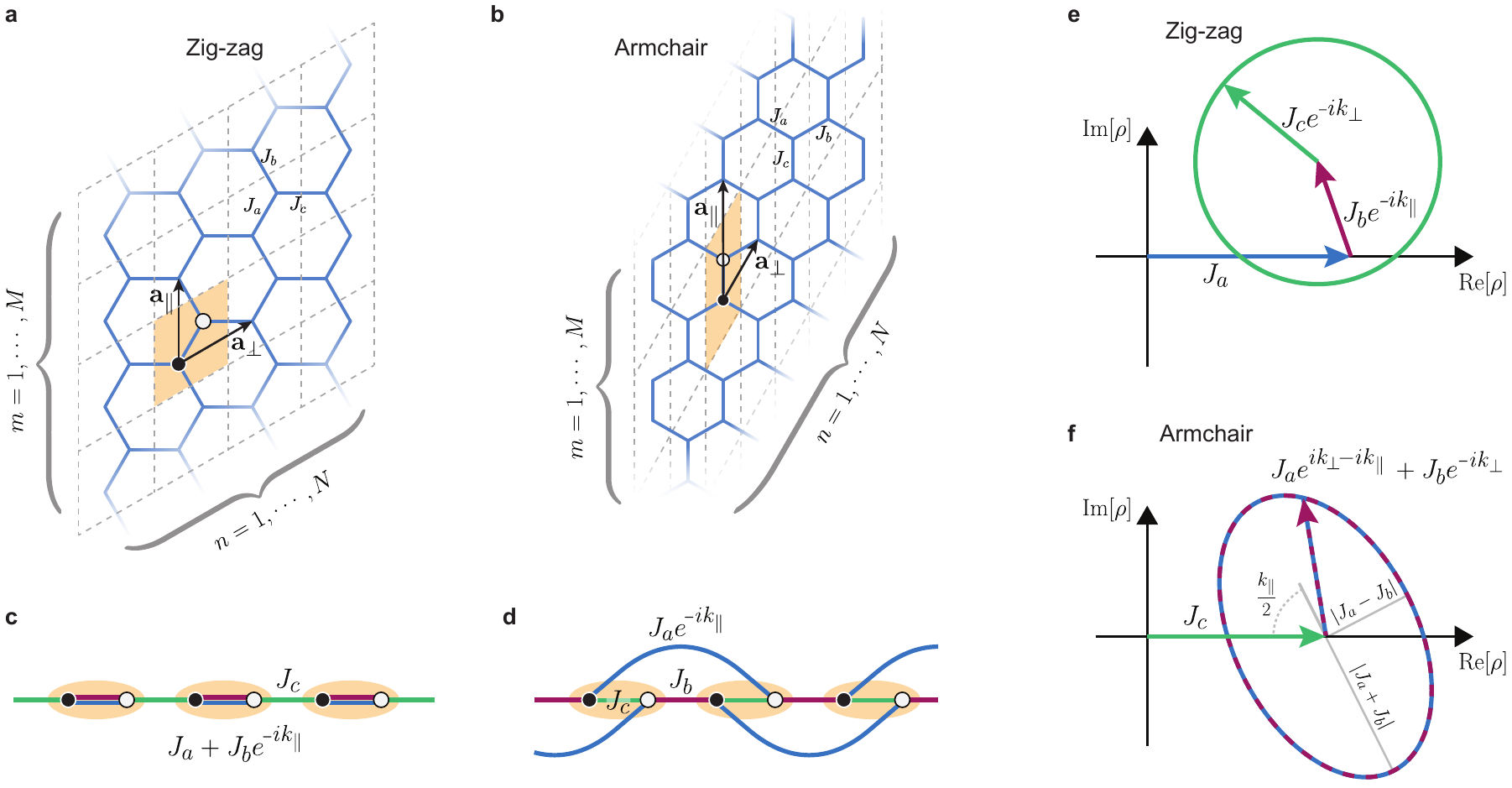}
	\caption{
		\textbf{Graphene ribbon.}
		\textbf{a},~Zig-zag edge graphene ribbon. \textbf{b}~Armchair edge graphene ribbon. \textbf{c,d},~Schematic of wavenumber-resolved generalized 1D SSH chains for zig-zag and armchair edges, respectively. The black and white circles describe sites $A$ and $B$, respectively. \textbf{e,f},~$\rho(k_\perp|k_\|)$ as a function of $k_\perp$ in the complex plane for zig-zag and armchair edges, respectively. 
	}
	\label{fig:SI_graphane_ribbon}
\end{figure*}

The zig-zag edges of a graphene ribbon can be described by using a unit cell and two lattice vectors of $\textbf{a}_{\|}$ and $\textbf{a}_\perp$ that are shown in Fig.~\ref{fig:SI_graphane_ribbon}a. The direction of $\textbf{a}_\|$ is parallel to the ribbon direction, while the size of $\textbf{a}_\perp$ defines the width of the ribbon. More precisely, a graphene ribbon with zig-zag edges can be defined by imposing an open boundary condition on the basis of $\textbf{a}_\perp$ while a periodic boundary condition is imposed on the basis of $\textbf{a}_\|$.
The Hamiltonian of such a graphene ribbon is given by 
\begin{equation}
	\label{H2Dzigzag}
	\hat H = \sum_{n,m}\left[J_a\left|m,n\right\rangle \langle m,n|\otimes \hat{\sigma}_- + J_b|n,m+1 \rangle \langle m,n |\otimes \hat{\sigma}_- + J_c  | m,n+1  \rangle \langle m,n |\otimes \hat{\sigma}_- + \mathrm{h.c.}\right],
\end{equation}
where $|m,n\rangle$ for $n=1,2,\cdots N$ and $m=1,2,\cdots M$ is a state vector at unit cell $(m,n)$ of the lattice.
By using the translation symmetry in the ribbon direction ($\textbf{a}_\|$), the Hamiltonian is partially diagonalized as 
\begin{equation}
	\hat H = \sum_{k_\|} |k_\| \rangle \langle k_\| |\otimes \sum_m \left[(J_a+J_b e^{-ik_\|}) | n \rangle \langle n|\hat{\sigma}_- + J_c |n+1 \rangle \langle n|\hat{\sigma}_- + \mathrm{h.c.}\right],
\end{equation}
where the wavenumber basis in the ribbon direction is defined as 
\begin{equation}
	|k_\|\rangle=\frac{1}{\sqrt{M}}\sum_{m} e^{ik_\| m}|m\rangle
\end{equation}
with the wavenumber $k_\|=2\pi \mu/M$ ($\mu=1,2,\cdots,M$).
By projecting the full Hamiltonian onto a certain wavenumber subspace with $k_\|$, we obtain the Hamiltonian for a wavenumber-resolved 1D chain:
\begin{equation}
	\hat H(k_\|) = \langle k_\| |\hat{H}|k_\| \rangle  = \sum_n \left[(J_a+J_b e^{-ik_\|}) |n \rangle \langle n|\hat{\sigma}_- + J_c |n+1 \rangle \langle n|\hat{\sigma}_- + \mathrm{h.c.}\right].
\end{equation}
As schematically shown in Fig.~\ref{fig:SI_graphane_ribbon}c, the 1D chain is considered as a generalized 1D SSH chain, where unit cells, individually consisting of two sites, are connected to each other in a chain. Here, an intra-cell coupling coefficient is a complex number and depends on wavenumber $k_\|$.

To predict the existence of edges states in the reduced 1D chain with the given wavenumber of $k_\|$, or edge states on the boundaries of the graphene ribbon, we here use a similar analysis to the 1D SSH model, explained in Sec.~\ref{sec:1DSSH}.
By imposing a periodic boundary condition on the chain, we can diagonalize the Hamiltonian using a well-defined wavenumber basis, that is given by  
\begin{equation}
	|k_\perp \rangle=\frac{1}{\sqrt{N}}\sum_{n} e^{ik_\perp n}|n\rangle,
\end{equation}
where $k_\perp =2\pi \nu/N$ ($\nu=1,2,\cdots,N$).
Then, we have the bulk Hamiltonian:
\begin{align}
	\hat{H}(k_\perp|k_\|) & = \langle k_\perp | \hat{H}(k_\|)|k_\perp\rangle  \\
	& =
	\left(
	\begin{array}{cc}
		0          & \rho(k_\perp|k_\|)  \\
		\rho^*(k_\perp|k_\|)  & 0
	\end{array}
	\right),
	\label{Hbulk_2D}
\end{align}
where
\begin{equation}
	\rho(k_\perp|k_\|) = J_a+J_b e^{-ik_\|} +J_c e^{-ik_\perp} \equiv \left|\rho(k_\perp|k_\|)\right|e^{-i\phi(k_\perp|k_\|)}.
\end{equation}
Using the wavenumber-resolved bulk Hamiltonian in the thermodynamic limit ($N\rightarrow \infty$), we can predict the existence of edge states by the following discussions. Note that the thermodynamic limit in the ribbon direction ($M \rightarrow \infty$) is not necessarily required as long as the periodic boundary condition is imposed. For a finite-length ribbon, the wavenumber $k_\|$ is discretized depending on the size $M$.

With a graphical approach, we plot the closed curve of the off-diagonal element $\rho(k_\perp|k_\|)$ of the bulk Hamiltonian for a given $k_\|$ and varying $k_\perp$, in the complex plane, as shown in Fig.~\ref{fig:SI_graphane_ribbon}e.
When the winding number of the closed curve around the origin is one, the wavenumber-resolved 1D chain is in the nontrivial topological phase, supporting two edge states on the zig-zag edges for the given $k_\|$.   
On the other hand, when the winding number is zero, the chain is in the trivial phase, showing no edge states.

Alternatively, we can also predict the existence of edge states by calculating the wavenumber-resolved Zak phase, given by
\begin{equation}
	\mathcal{Z}(k_\|) = i \oint d k_\perp \langle u_{k_\perp|k_\|, \pm}| \partial_{k_\perp} |u_{k_\perp|k_\|, \pm}\rangle = \frac{1}{2} \oint dk_\perp \:\partial_{k_\perp} \phi(k_\perp|k_\|),
\end{equation}
where $ |u_{k_\perp|k_\|, \pm}\rangle=1/\sqrt{2}(e^{-i\phi(k_\perp|k_\|)}, \:\: \pm1)^T$ is an eigenvector (Bloch wave function), diagonalizing the two-band bulk Hamiltonian of Eq.~(\ref{Hbulk_2D}).
The wavenumber-resolved generalized 1D SSH chain is in the topological phase when $\mathcal{Z}(k_\|) = \pi$, while the chain is in the trivial phase when $\mathcal{Z}(k_\|) = 0$.

\begin{figure*}[h!]
	\includegraphics[scale=0.8]{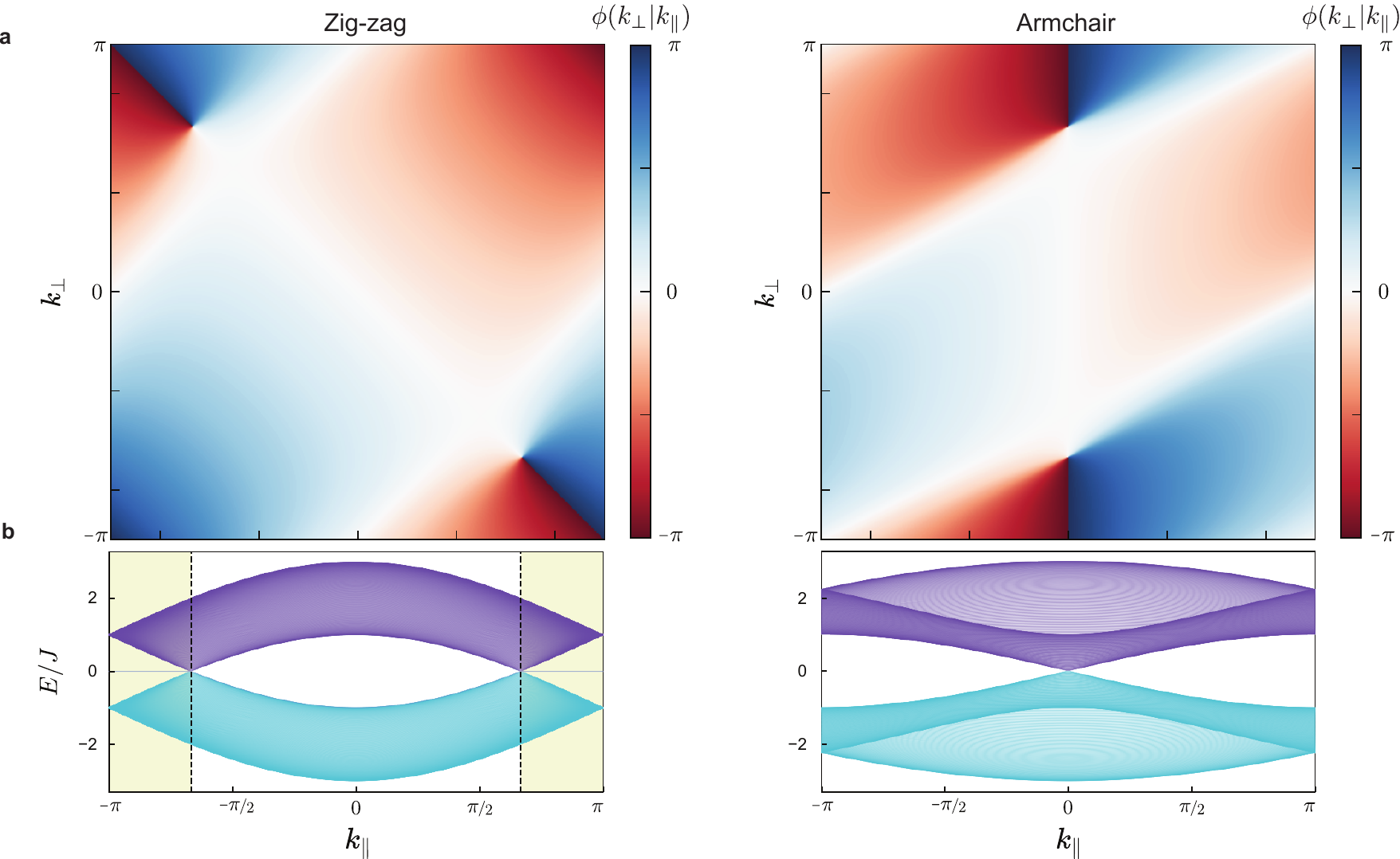}
	\caption{
		\textbf{Strain-free graphene ribbons.}
		\textbf{a},~Phase $\phi(k_\perp|k_\|)$ for zig-zag and armchair edge graphene ribbons, respectively. The discontinuities of the phase are located at the boundary between the red region ($\phi=-\pi$) and the blue region ($\phi=\pi$).
		\textbf{b},~Band structure as a function of $k_\|$ for zig-zag and armchair edge graphene ribbons with an approximately 100-cell width, respectively. The regions of $k_\|$ satisfying $\mathcal{Z}=\pm\pi$ are highlighted. 
	}
	\label{fig:SI_normal_graphane_ribbon}
\end{figure*}

In the same manner, we can predict the existence of edge states in a graphene ribbon with armchair edges. 
An armchair-edge graphene ribbon is defined by a unit cell and two lattice vectors that are shown in Fig.~\ref{fig:SI_graphane_ribbon}b.
According to the connectivity among the unit cells, the full Hamiltonian of the graphene ribbon is given by
\begin{equation}
	\label{H2Darmchair}
	\hat H = \sum_{n,m}\left(J_c\left| m,n \right\rangle \langle  m,n |\otimes \hat{\sigma}_-  + J_b| m,n+1 \rangle \langle m,n |\otimes \hat{\sigma}_- + J_a| m+1,n-1  \rangle \langle m,n |\otimes \hat{\sigma}_- + \mathrm{h.c.}\right).
\end{equation}
Importantly, the connectivity is different from the Hamiltonian for the zig-zag edges [see Eq.~(\ref{H2Dzigzag})], which results in a different prediction of the existence of edges states.
By diagonalizing the full Hamiltonian with the wavenumber basis in the ribbon direction ($k_\|$), we have the Hamiltonian of a wavenumber-resolved generalized 1D SSH chain:
\begin{equation}
	\hat H(k_\|) = \langle k_\| |\hat{H}|k_\| \rangle  = \sum_n \left(J_c |n \rangle \langle n|\otimes\hat{\sigma}_-  +J_b |n \rangle \langle n+1|\otimes\hat{\sigma}_- + J_a e^{-ik_\|} |n-1 \rangle \langle n|\otimes\hat{\sigma}_- + \mathrm{h.c.}\right).
\end{equation}
As schematically shown in Fig.~\ref{fig:SI_graphane_ribbon}d, the wavenumber-resolved 1D chain can be considered as a generalized 1D SSH chain.
By introducing the wavenumber basis in the width direction ($k_\perp$), the bulk Hamiltonian is obtained as Eq.~(\ref{Hbulk_2D}), where the off-diagonal element is modified as
\begin{equation}
	\begin{split}
		\rho(k_\perp|k_\|) &= J_c +J_b e^{-ik_\perp} +J_a e^{-ik_\|+ik_\perp}  \\
		&= J_c + e^{-ik_\|/2} \left[\left(J_a+J_b\right) \cos(k_\perp-k_\|/2)+ i \left(J_a-J_b\right) \sin(k_\perp-k_\|/2)\right].
	\end{split}
\end{equation}
In this case, the closed curve of $\rho(k_\perp|k_\|)$ is an ellipse in the complex plane, as shown in Fig.~\ref{fig:SI_graphane_ribbon}f. However, we can apply the same analysis to the armchair-edge graphene ribbon in order to predict the existence of edge states, as in the case of the zig-zag edges. Namely, if the closed curve winds up the origin, the armchair-edge graphene ribbon shows edge states for the given $k_\|$. 

As examples, we analyze strain-free graphene ribbons with zig-zag and armchair edges, for which $J_a=J_b=J_c$. In Fig.~\ref{fig:SI_normal_graphane_ribbon}a, the phase $\phi(k_\perp|k_\|)$ for the zig-zag and armchair edge graphene ribbons are shown as a function of $k_\|$ and $k_\perp$, respectively, as introduced by Delplace \cite{SI_delplace2011zak}. The phase is restricted to the interval $[-\pi,\pi]$ to be a single-valued function. The two singularity points, corresponding to the Dirac points, are shown in each plot. With a fixed $k_\|$, the phase of $\phi(k_\perp|k_\|)$ along $k_\perp$ gives the wavenumber-resolved Zak phase to predict the existence of edge states. More simply, the discontinuities of $\phi(k_\perp|k_\|)$ are useful to determine the discretized Zak phase ($\mathcal{Z} = 0$ or $\pi$) since a path along $k_\perp$ necessarily goes across a discontinuity to give a non-zero Zak phase.

In Fig.~\ref{fig:SI_normal_graphane_ribbon}b, we numerically calculate the band structure of graphene ribbons with zig-zag and armchair edges for different $k_\|$, respectively. The ribbon width is set to be on the order of 100 to avoid the finite-size effect. Since the ribbon width is sufficiently wide, edge states are found to be zero-energy states in the band gap. The regimes of $\mathcal{Z}(k_\|) = \pi$ are highlighted in Fig~\ref{fig:SI_normal_graphane_ribbon}b. As the Zak phase predicts, there are edge states in the zig-zag edge ribbon for $|k_\||>2\pi/3$, while not in the armchair edge ribbon for all $k_\|$. 

\begin{figure*}[h!]
	\includegraphics[scale=1]{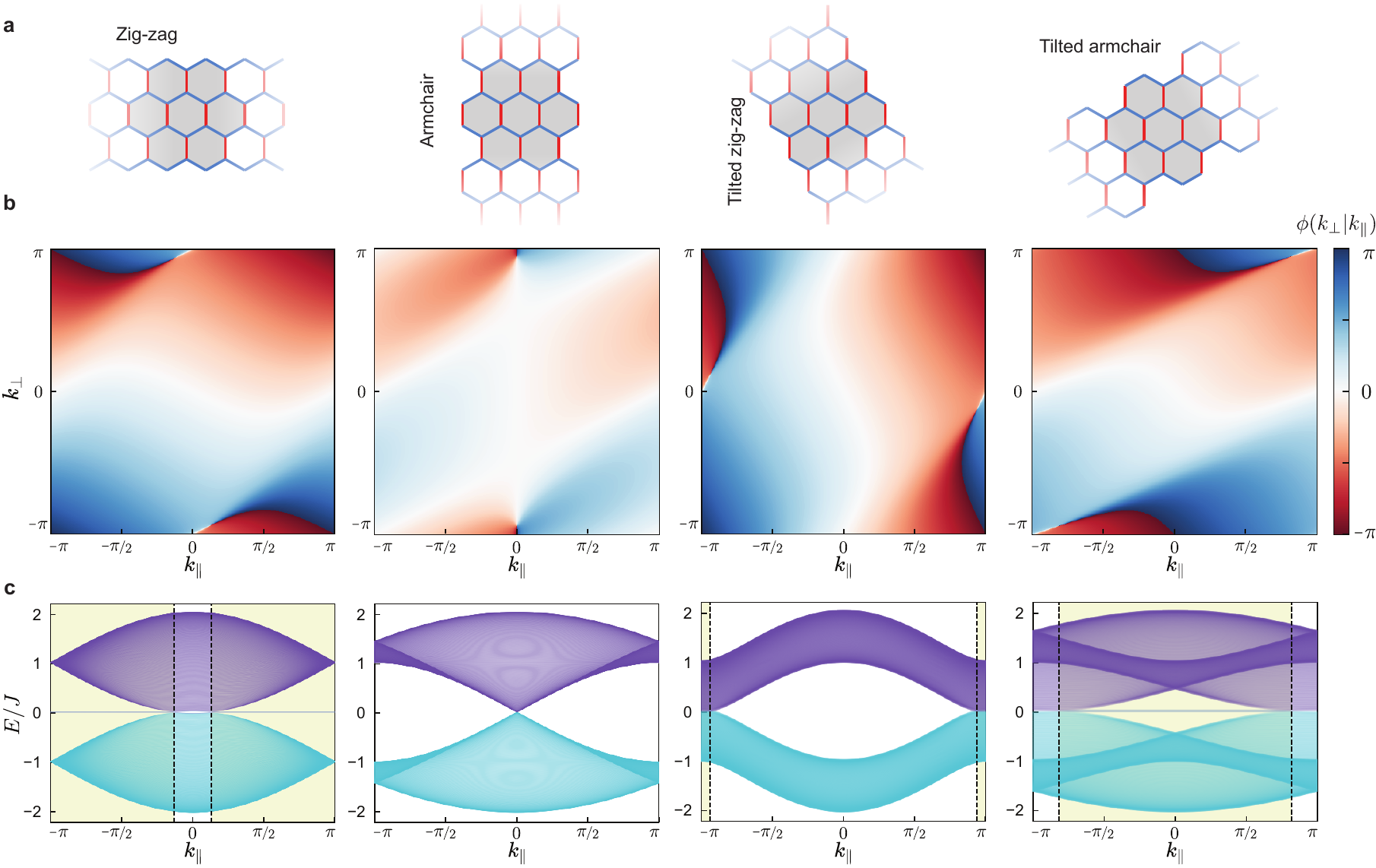}
	\caption{
		\textbf{Strained graphene ribbons.}
		\textbf{a},~Four different interpretations of our 2D lattice as a truncated strained graphene ribbon: zig-zag, armchair, tilted armchair, and tilted zig-zag edge.
		\textbf{b},~The phase $\phi(k_\perp|k_\|)$ for the four different orientations. The discontinuities of the phase are located at the boundary between the blue region ($\phi=-\pi$) and the red region ($\phi=\pi$).
		\textbf{c},~Band structure as a function of $k_\|$ for the four different graphene ribbons with an approximately 100-cell width, respectively. The region of $k_\|$ satisfying $\mathcal{Z}=\pm\pi$ are highlighted. 
	}
	\label{fig:SI_strained_graphane_ribbon}
\end{figure*}

\subsubsection{\added{Strained graphene ribbon}}\label{sec:strained_graphene_cut}
In our 2D optomechanical lattice, only one of the couplings in three different orientations is different, where the coupling strength is denoted by $J$ while the other two coupling strengths are denoted by $J'$. This corresponds to applying a strain to a graphene ribbon in a direction parallel or perpendicular to the direction of coupling $J$. As shown in Fig.~\ref{fig:SI_strained_graphane_ribbon}a, our 2D lattice is interpreted as a truncated strained graphene ribbon with zig-zag or armchair edges. More precisely, there are four different possible orientations for the ribbon: zig-zag edges perpendicular to the direction of coupling $J$ (zig-zag ribbon), armchair edges parallel to the direction of coupling $J$ (armchair ribbon), zig-zag edges tilted with respect to the direction of coupling $J$ (tilted zig-zag ribbon), and armchair edges tilted with respect to the direction of coupling $J$ (tilted armchair). Note that due to the reflection symmetry, there are two orientations of strained graphene ribbons with tilted zig-zag and tilted armchair edges, respectively, that are not shown in Fig.~\ref{fig:SI_strained_graphane_ribbon}a to avoid redundancy.

We plot the phase $\phi(k_\perp|k_\|)$ for the four different orientations of the ribbon. To describe a zig-zag strained graphene ribbon, we set $J_c=J$ and $J_a=J_b=J'$ in the Hamiltonian of Eq.~(\ref{H2Dzigzag}), while we set $J=J_a$ and $J_b=J_c=J'$ (or $J=J_b$ and $J_a=J_c=J'$) for a tilted zig-zag ribbon. To describe an armchair strained graphene ribbon, we set $J_c=J$ and $J_a=J_b=J'$ in the Hamiltonian of Eq.~(\ref{H2Darmchair}), while we set $J_a=J$ and $J_b=J_c=J'$ (or $J=J_b$ and $J_a=J_c=J'$) for a tilted armchair ribbon. We note that due to the reflection symmetry, the phase for the two tilted zig-zag ribbons can be converted by replacing $J$ and $J'$, resulting in identical phase trajectories in the complex plane. The same argument is valid for the two tilted armchair orientations.

For our specific setting with $J'/J = 0.51$, the phase $\phi(k_\perp|k_\|)$ is plotted as a function of $k_\|$ and $k_\perp$ in Fig.~\ref{fig:SI_strained_graphane_ribbon}b. The two Dirac points almost merge since the coupling ratio is close to the transition point of 0.5. In the same way as the cases without a strain, we can determine for which $k_{\|}$ edge states exist from the phase $\phi(k_\perp|k_\|)$ in the path along $k_\perp$. To confirm that the prediction of the existence of edge states from the Zak phase is correct, we numerically simulate the band structure of strained graphene ribbons of approximately 100-cell width, as shown in Fig.~\ref{fig:SI_strained_graphane_ribbon}c, where the regions of $\mathcal{Z} = \pi$ are highlighted. Edge states appear as zero-energy modes in the band gap~\cite{SI_kohmoto2007zero,SI_delplace2011zak}. The analytical calculation of the Zak phase accurately predicts the existence of edge states. Importantly, for our specific setting with $J'/J=0.51$, the zig-zag ribbon and the tilted armchair ribbon show edge states for almost all $k_\|$.

\subsubsection{\added{Finite-width effect on graphene ribbon}} \label{sec:finite_strained_graphene_ribbon}
\begin{figure}[h!]
	\includegraphics[scale=1]{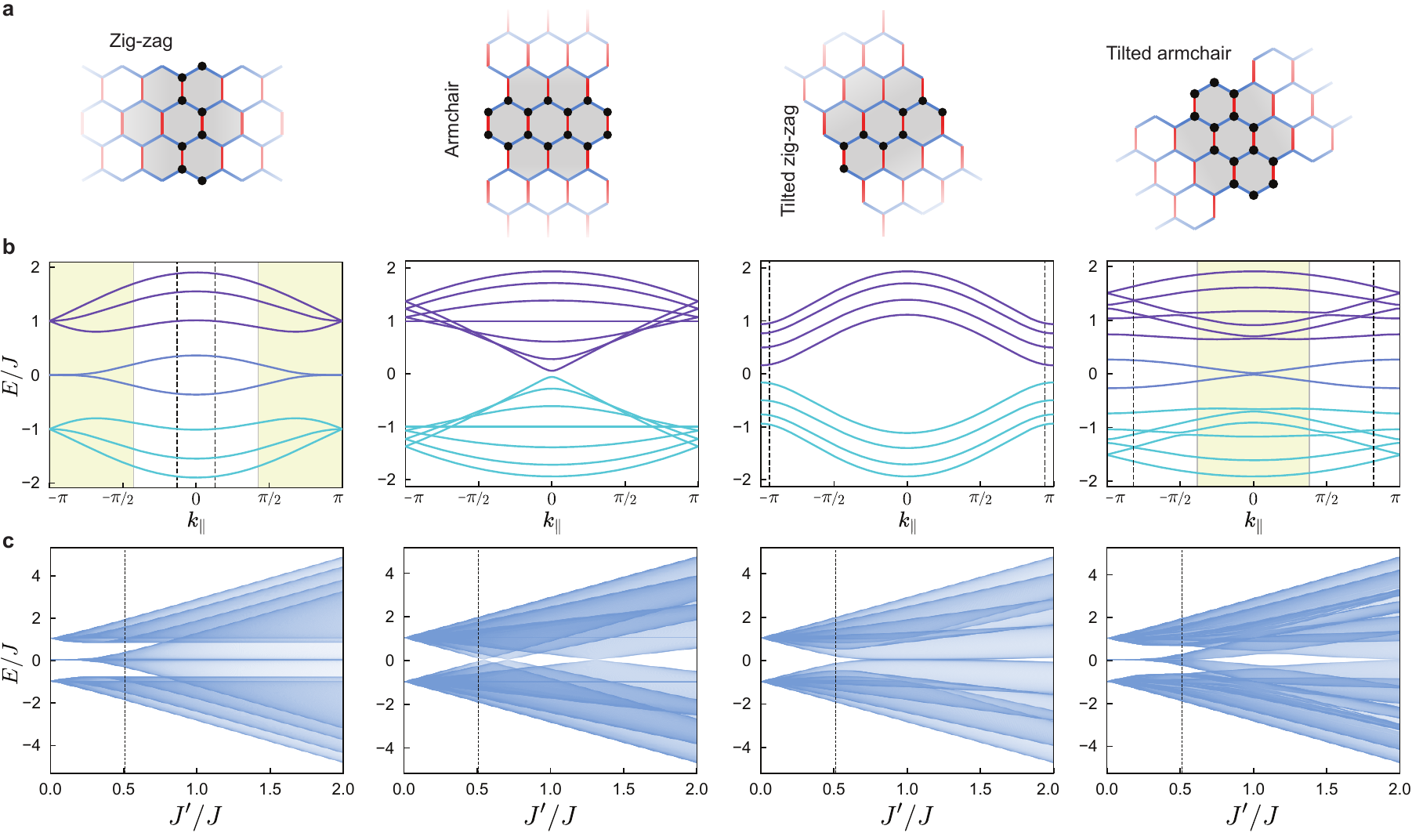}
	\caption{\textbf{Strained graphene ribbons of finite width.}
		\textbf{a},~Four different interpretations of our 2D lattice as a truncated strained graphene ribbon: zig-zag, armchair, tilted zig-zag, and tilted armchair edges. The black dots denote the unit cell used for the calculation of the band structure based on Bloch's theorem. 
		\textbf{b},~Band structure as a function of $k_\|$ for the four different orientations. The coupling ratio is set to be $J'/J=0.51$ and the width is the same as our 24-site lattice. The number of the energy levels corresponds to the number of sites in the unit cell (black dots in \textbf{a}). The regions of $k_\|$ satisfying both $\mathcal{Z}=\pi$ and the slope condition are highlighted while the transition points calculated only from the Zak phase are shown with the black dashed lines. 
		\textbf{c},~Band structure of the corresponding graphene ribbons as a function of $J'/J$. The dashed line points out $J'/J=0.51$, our design value.}
	\label{fig:SI_strained_graphane_ribbon_finite}
\end{figure}

Depending on the boundary structures, the width of the graphene ribbon is found to be $N=4$ for the zig-zag edges and $N=7$ for the armchair edges (see Fig.~\ref{fig:SI_strained_graphane_ribbon_finite}a). As discussed in Sec.~\ref{sec:1DSSH}, the finite-size effect modifies the prediction of the existence of edge states. Since the wavenumber-resolved 1D chain reduced from a graphene ribbon can be considered as a generalized 1D SSH chain, we can apply the same discussions as for the standard 1D SSH model to predict the existence of edge states of a graphene ribbon in the presence of the finite-width effect.
Namely, edge states appear when both $\mathcal{Z}=\pi$ and the slope condition are satisfied simultaneously. Here, the slope condition for the wavenumber-resolved 1D chain is described as
\begin{equation}
	\label{ribbon_slope}
	\left|\frac{\partial \phi(k_\perp|k_\|)}{\partial k_\perp}\vline_{\:k_\perp=k_{\perp,\mathrm{min}}}\right|<N+1, 
\end{equation}
where $k_{\perp,\mathrm{min}}$ is the wavenumber associated with the minimum absolute value of the eigenenergy for the given $k_\|$. 

As shown in Fig.~\ref{fig:SI_strained_graphane_ribbon_finite}b, we numerically simulate the energy levels as a function of $k_\|$ for the four different ribbons with the same width as our 24-site lattice.
We find that for specific ranges of $k_\|$, there are zero-energy modes, corresponding to edge states, in the band gap, only for the zig-zag and tilted armchair edge ribbons. This is consistent with the prediction. However, due to the finite width of the ribbons, the edge states are hybridized, showing an energy splitting. For further study, the band structure as a function of the coupling ratio of $J'/J$ is shown in Fig.~\ref{fig:SI_strained_graphane_ribbon_finite}c. As expected, zero-energy states appear when $J'/J \rightarrow 0$ only for the case with the zig-zag edges and the tilted armchair edges. 

Using the Zak phase condition ($\mathcal{Z}=\pi$) and the slope condition, we can predict the existence of the edges states. In Fig.~\ref{fig:SI_strained_graphane_ribbon_finite}b, the regions of $k_\|$ that satisfy both the conditions are highlighted, while the transition points that are predicted from the Zak phase calculation are shown with the black dashed lines. We find that the region of $k_\|$ showing edges states are decreased from the case with an infinite width due to the finite-width effect. Importantly, edge states exist only in either the zig-zag edge ribbon or the tilted armchair edge ribbon.

\subsection{\added{24-site honeycomb lattice}}
\subsubsection{\added{Band structure and modeshapes}}
\begin{figure}[h]
	\includegraphics[scale=1]{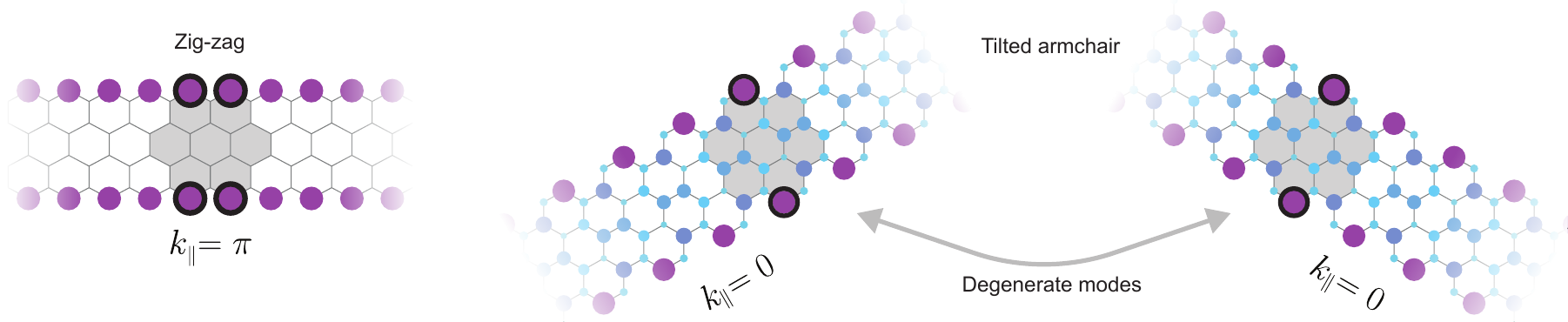}
	\caption{\textbf{Edge states of graphene ribbons overlapping the flake structure.} Examples of numerically simulated modeshapes of edge states in zig-zag and tilted armchair edge graphene ribbons, overlapping both the two sites on the top and bottom edges of the 24-site flake. The titled armchair orientation has a reflection symmetry resulting in two degenerate mode shapes.}
	\label{fig:SI_2D_edge_overlap}
\end{figure}

The 24-site multimode optomechanical system in a honeycomb lattice presented in the main text is designed to be close to the transition point for strained graphene ($J'/J=0.5$). From the discussions for strained graphene ribbons, we could expect that edge states appear on both the two sites of the top and bottom edges of the flake. This is because the two sites overlap the edge states seen in the zig-zag and tilted armchair edge ribbons with the same width as our 24-site lattice, as shown in Fig.~\ref{fig:SI_2D_edge_overlap} (see Sec.~\ref{sec:finite_strained_graphene_ribbon}). However, the prediction can be precise only when a periodic boundary condition is imposed in the ribbon direction. Our 24-site lattice could be obtained by truncating the corresponding strained graphene ribbon and imposing an open boundary condition in the ribbon direction. Therefore, it is not trivial and beyond the scope of this work and still and open question how to precisely predict the existence of edge states in such a small flake of strained graphene, since the finite-size effects in both the two lattice directions could be mixed up. 

\begin{figure}[h]
	\includegraphics[scale=1]{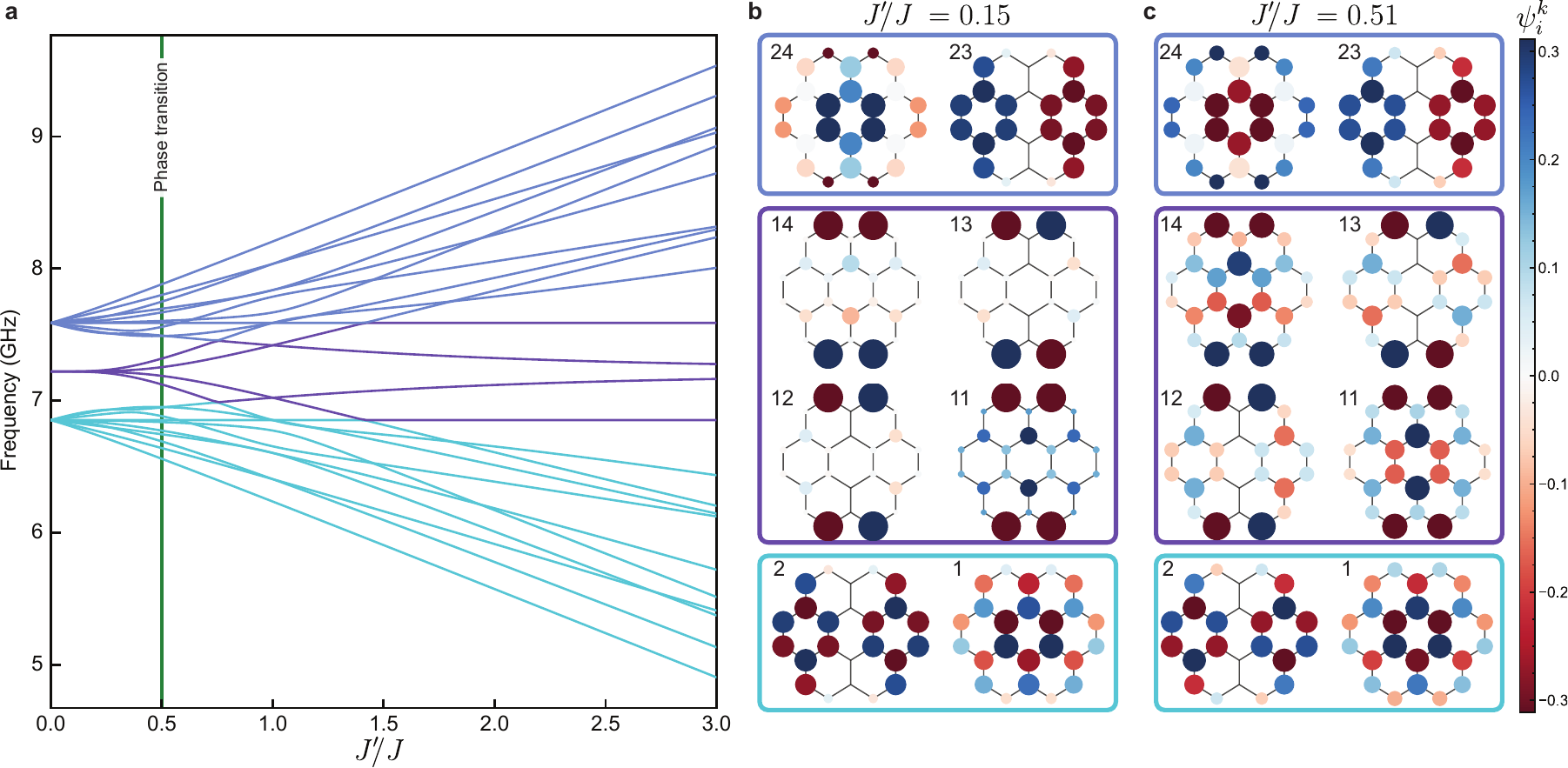}
	\caption{\textbf{24-site honeycomb lattice.} \textbf{a}, Energy levels of the 24-site honeycomb lattice as a function of $J'/J$. The green line shows the phase transition for a graphene flake of infinite size. \textbf{b,c}, Examples of the mode shapes in two of the upper bulk modes, all the edge states, and two of the lower bulk modes for $J/'J=0.15$ and $J'/J=0.51$ (design values), respectively.}
	\label{fig:SI_2D_design}
\end{figure}

Nevertheless, we can numerically simulate the band structure and the modeshapes of the 24-site multimode system. In Fig.~\ref{fig:SI_2D_design}a, we show the energy levels of the 24-site system as a function of $J'/J$. In the small coupling ratio limit ($J'/J\rightarrow0$), four zero-energy modes appear in the band gap. Figures~\ref{fig:SI_2D_design}b and c show several examples of modeshapes of two of the upper bulk modes,  all the zero-energy states, and two of the lower bulk modes for $J/'J=0.15$ and $J'/J=0.51$ (design values), respectively. As expected, the four zero-energy modes are localized at the four sites of the top and bottom edges. This is consistent with the topological prediction of the existence of edge states from strained graphene ribbons, discussed in Sec.~\ref{sec:finite_strained_graphene_ribbon}.

\subsubsection{\added{Effect of parasitic couplings}}
\label{sec:2D_SNNC}
\begin{figure*}[h!]
	\includegraphics[scale=1]{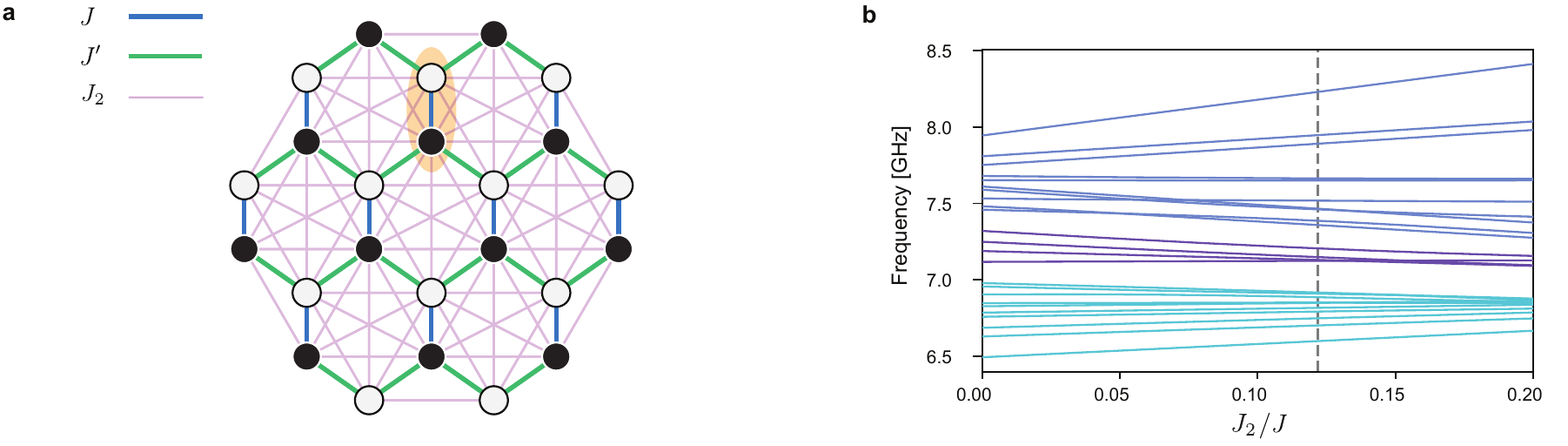}
	\caption{
		\textbf{Effect of parasitic coupling on the 2D honeycomb lattice}.
		\textbf{a}, Mode diagram of the 24-site 2D honeycomb lattice including second nearest-neighbor couplings. \textbf{b}, Energy levels as a function of the relative second nearest-neighbor coupling rate for the 2D case. The dashed line shows the values of the actual device discussed in the main text. Color coding denotes the LPB, edge states and UPB.
	}
	\label{fig:SI_2DSSH_parasitic_couplings}
\end{figure*}

The ideal nearest neighbor coupling in the honeycomb lattice is realized by mutual inductance between adjacent triangular spirals with shared edges in the circuit shown in Fig.\ref{fig:4} of the main text. Beyond this, there are parasitic mutual inductances between distanced spirals in the realized lattice. Based on the electromagnetic FEM simulations, we expect that parasitic mutual inductance between triangles which have a shared vertex (as shown in Fig.~\ref{fig:SI_2DSSH_parasitic_couplings}a) resulting in approximately identical parasitic coupling rates, while the parasitic coupling is an order of magnitude smaller than the nearest neighbor coupling rate. 
Figure \ref{fig:SI_2DSSH_parasitic_couplings}b shows the numerically calculated energy levels including the parasitic couplings for the 24-site graphene flake versus the relative second nearest-neighbor coupling rate. This can explain the asymmetry in the band structure of the 24-site flake which was experimentally measured and presented in the main text.

\section{Design and simulation}

\subsection{Circuit theory of the dimerized SSH arrays}
\label{section:circuit_theory}
\begin{figure*}[h]
	\includegraphics[scale=1]{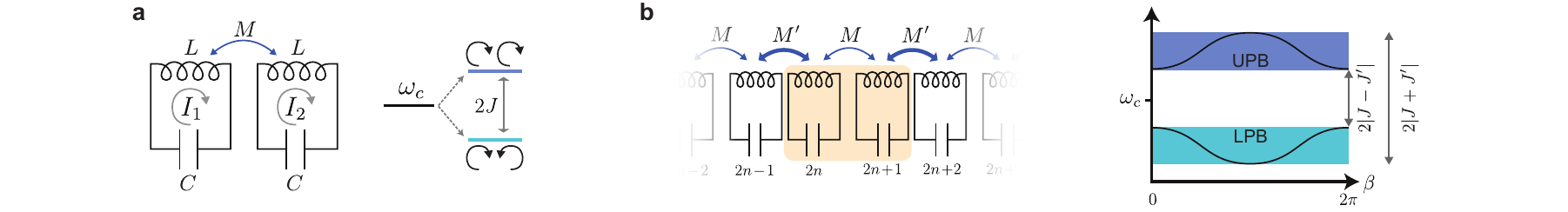}
	\caption{ \textbf{Circuit theory of the dimerized arrays}. \textbf{a}, Two coupled LC circuits. The inductive coupling results in energy splitting. The higher frequency mode supports symmetric currents. \textbf{b}, Infinite chain of dimerized circuits. The energy band structure consists of two passbands (UPB and LPB). \label{fig:SI_circ}}
\end{figure*}
Here we briefly discuss the circuit theory of microwave SSH chains and link the circuit elements of such a system to the Hamiltonian parameters. We first consider two coupled identical LC circuits as shown in Fig.~\ref{fig:SI_circ}a. The system can be described by electrical currents, $I_1(t)$ and $I_2(t)$. Kirchhoff's equations result in:
\begin{equation}
	L \ddot{I_1}(t) - M \ddot{I_2}(t) + \frac{1}{C} I_1(t) = 0 \ \ , \ \ L \ddot{I_2}(t) - M \ddot{I_1}(t) + \frac{1}{C} I_2(t) = 0 \,
\end{equation}
Which can be transferred to the frequency domain:
\begin{equation}
	(\omega_c^2 - \omega^2) L I_1 = -M \omega^2 I_2 \ \ , \ \ (\omega_c^2 - \omega^2) L I_2 = -M \omega^2 I_1 \,
	\label{eq:KVL_dimer}
\end{equation}
Where $\omega_c = \sqrt{\frac{1}{LC}}$.
Equation \ref{eq:KVL_dimer} results in two eigenfrequencies
\begin{equation}
	\omega_c \sqrt{\frac{1}{1\mp\frac{M}{L}}} \simeq \omega_c (1 \pm \frac{M}{2L})
	\label{eq:splitting}
\end{equation}
corresponding to two eigen modes of $I_1 = \pm I_2$ respectively. Based on the energy splitting, the energy coupling rate can be extracted as $J = \omega_c \frac{M}{2L}$. It worth to mention that the symmetric mode ($I_1=I_2$) corresponds to the higher resonance frequency and the asymmetric mode to the lower one.

Now we describe the infinite array of coupled dimer LC circuits with the staggered couplings of $M$ and $M'$ as shown in Fig.~\ref{fig:SI_circ}b. For each dimer (sites $2n$ and $2n+1$) we can derive the frequency domain circuit equations:
\begin{equation}
	\begin{split}
		& (\omega_c^2 - \omega_2) I_{2n} + \frac{M}{L} \omega^2 I_{2n+1} +  \frac{M'}{L} \omega^2 I_{2(n-1)+1}  = 0\\
		& (\omega_c^2 - \omega_2) I_{2n+1} + \frac{M'}{L} \omega^2 I_{2(n+1)} +  \frac{M}{L} \omega^2 I_{2n} = 0 \ \ .
		\label{eq:KVL_SSH}
	\end{split}
\end{equation}
Considering Bloch's theorem, the propagating mode in such array can be described by a harmonic function:
\begin{equation}
	I_{2n} = \added{I_0} e^{i \beta 2n} \ \ \ , \ \ \ I_{2n+1} = \added{I_0'} \lambda e^{i \beta 2n}
\end{equation}
Where $\beta \in \mathbb{R}$ is the wave number of the propagating mode,  $\lambda \in \mathbb{C}$ is defining the phase and amplitude difference between circuits in a dimer and $I_0 \in \mathbb{R}$ is one of the current's amplitudes.
Therefore equation \ref{eq:KVL_SSH} will be simplified to:
\begin{equation}
	\begin{split}
		& (\omega_c^2-\omega^2) + \lambda \frac{M+M' e^{-i\beta}}{L} \omega^2 = 0 \\
		& \lambda (\omega_c^2-\omega^2) + \frac{M+M' e^{i\beta}}{L} \omega^2 = 0 \ \ .
		\label{eq:bloch_SSH}
	\end{split}
\end{equation}
Solving for $\omega$ from equations \ref{eq:bloch_SSH} will result in the band structure of the infinite chain:
\begin{equation}
	\omega = \frac{\omega_c}{\sqrt{1 \pm \frac{\sqrt{M^2 + M'^2 + 2 M M' \cos(\beta)}}{L}}} \ .
	\label{eq:SSH_bands}
\end{equation}
As shown in Fig.~\ref{fig:SI_circ}b, the energy spectrum of the infinite chain consists of two passbands. In case of small mutual coupling ($\frac{M}{L},\frac{M'}{L} \ll 1$) The upper passband (UPB) and lower passband boundaries are:
\begin{equation}
	\begin{split}
		& \omega^\mathrm{UPB}_{\pm} \ = \ \omega_c + |J \pm J'| \\
		& \omega^\mathrm{LPB}_{\pm} \ = \ \omega_c - |J \mp J'|
		\label{eq:UPB_LPB}
	\end{split}
\end{equation}
Where $J = \omega_c \frac{M}{2L}$, $J' = \omega_c \frac{M'}{2L}$ are the mutual coupling rates in the chain. In the finite chains, when $J < J'$, two edge states arise in the middle of the band gap due to the truncated boundary condition of the chain. This can be intuitively understood in the extreme case of $J \ll J'$, when the first and last LC circuits are fully isolated from the rest of the chain, which consists of strongly coupled circuits.

\subsection{Design principles}
Here we review design rules and relations between the system parameters and the geometrical parameters of the circuits. First we consider a single building block microwave LC resonator. The capacitance of a vacuum-gap capacitor can be approximated by $C_\mathrm{VG} = \frac{\epsilon_0 A_\mathrm{eff}}{d}$, where $A_\mathrm{eff}$ is the effective area of the parallel plate and $d$ is the gap size. \added{Assuming the transverse dimensions of the conductors can be neglected}, the self-inductance or mutual inductance of a spiral inductors can be precisely calculated using Neumann's formula:
\begin{equation}
	L_{A,B} =  \frac{\mu_0}{4\pi} \oint_A \oint_B \frac{d\mathbf{r}_A\cdot d\mathbf{r}_B}{|\mathbf{r}_A-\mathbf{r}_B|}
\end{equation}
Where the path integral is carried out over the geometrical curve of the spiral. In the case of self inductance $A=B$ and and in case of mutual inductance $A$ and $B$ refer to the shape functions of the two spirals.
This results in two simple approximate scaling rules: The self inductance of a spiral scales with $L\propto N_s^2 r_\mathrm{eff}$ where $N_s$ is the number of turns and $r_\mathrm{eff}$ is the effective radius of spiral. The mutual inductance between two spirals $A$ and $B$ with the distance of $l$ scales with $M \propto N_{s,A} N_{s,B} r_\mathrm{eff,A}^2 r_\mathrm{eff,B}^2 /l^3 $ in far distances ($l\gg r_{\mathrm{eff},A} , r_{\mathrm{eff},B} $).
To increase microwave coupling rates ($J=\omega_c \frac{M}{2L}$) in a design, $M$ needs to increase while $L$ is fixed. This can be achieved by increasing the aspect ratio of spirals to have longer adjacent wires in neighbor sites, like the spirals used in the 10 site 1D chain.

The mechanical frequency of the first fundamental mode of a drumhead resonator can be approximated as $\Omega_\mathrm{m} \simeq \frac{2.4}{R}\sqrt{\frac{\sigma^\mathrm{Al}}{\rho^\mathrm{Al}}}$ where $R$ is the radius of the drumhead (radius of the circular trench as fixed mechanical boundary condition) and $\sigma^\mathrm{Al}$, $\rho^\mathrm{Al}$ are the stress and the density of the Al thin film. In the device, we \replaced{gradually }{slightly} change the trench radius of different sites \added{by $ 500~\mathrm{nm}$ ($1 \%$)} to distinguish them \added{in the mode-shape measurement experiment by their mechanical frequency. It is important to mention that the mechanical frequency disorder in the fabrication process should be smaller than the mechanical frequency shift introduced by incrementing trench radius in order to keep them in correct designed order. This can be verified by comparing the measured mechanical frequencies from OMIT response and the theoretical $\Omega_{\mathrm{m},i}\propto\frac{1}{R_i}$ relation from the design, as shown in Fig. 2j and Fig. 4e in the main text.} We note that this slight variation on the trench radius does not have a significant effect on the electrical boundary conditions of the \replaced{capacitor }{circuit}, \added{given that the bottom plate of the capacitor only sees $\approx 20 \% $ of the center top plate's area} and \added{thus } does not perturb the microwave resonance frequencies of the LC circuits, $\omega_\mathrm{c} = 1/\sqrt{LC}$.

To probe the system, the two ends of the chain are inductively coupled to coplanar waveguides using two short-circuited inductive loop couplers. The external coupling rate to the outermost sites is designed to be comparably smaller than internal microwave couplings in the chain to avoid deviation from the ideal SSH model (See Fig. \ref{fig:SI_loop_sweep}).

\subsection{Electromagnetic simulations}
\begin{figure*}[h!]
	\includegraphics[scale=1]{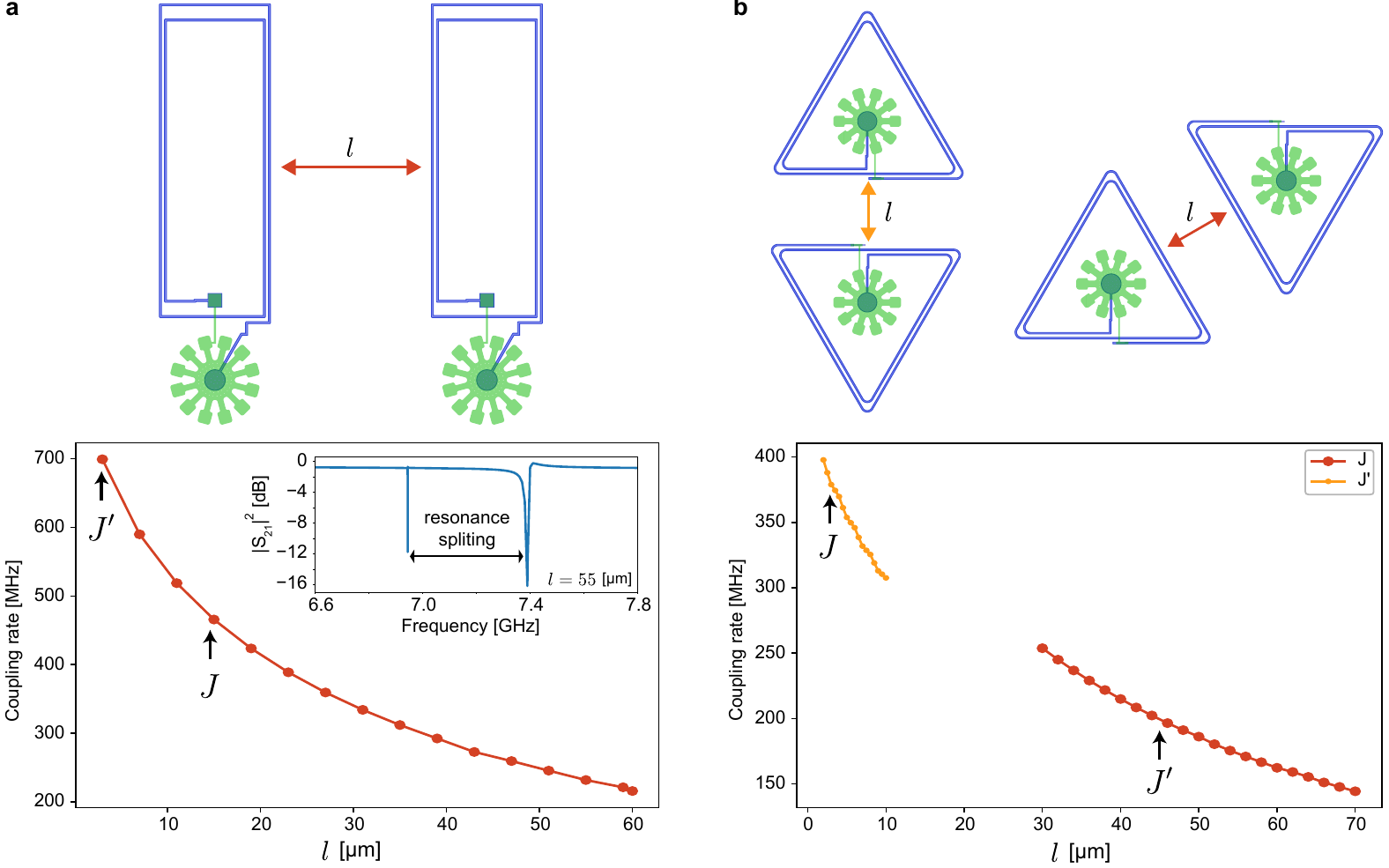}
	\caption{\textbf{Mutual inductive coupling rate as a function of the distance between two identical sites} \textbf{a}, Unit cells for 1D. \textbf{b}, Unit cells for 2D in different directions. The black arrows show the designed value. The coupling rates are extracted from splitting of hybridized modes observed in the FEM simulation with Sonnet as shown in the inset. \label{fig:SI_sim_sweep}}
\end{figure*}

All the electromagnetic simulations were performed using $\text{Sonnet}^\text{®}$. We first simulated the single LC resonator and varied the geometric inductor parameters and bottom capacitor plate diameter (which changes the capacitance) to get the desired resonance frequency around 7 GHz. As detailed in Sec. \ref{section:circuit_theory}, when two harmonic modes get coupled together, two hybridized modes appear (Fig. \ref{fig:SI_sim_sweep} \textbf{a} inset).
The frequency splitting between such two modes is exactly double the coupling rate between them as can be easily seen from Eq. \eqref{eq:splitting}.
In our case, the coupling rate between two neighbouring modes is inductive, and its strength can be changed by varying the distance between them. We ran parametrised $\text{Sonnet}^\text{®}$ simulations by varying the distance between them, and for each run we extracted the coupling rate from the resulting frequency splitting. As the distance between the two circuits increases, the coupling rate decreases, as it is expected. (See Fig. \ref{fig:SI_sim_sweep})

From these simulations we choose the distances that would result in the coupling strengths that we desired. For the 1D case (Fig. \ref{fig:SI_sim_sweep} \textbf{a}) there is one direction of coupling. In the graph the chosen values of the inter-cell coupling $J'$ and intra-cell coupling $J$ can be seen. From these same simulations unwanted second and third order coupling rates can also be extracted, which we used to compute the expected modeshapes.
For the 2D case (Fig. \ref{fig:SI_sim_sweep} \textbf{b}) we ran two different distance sweeps because of the slightly asymmetric shape of the spiral inductor, due to the needed crossover point to connect it to the capacitor plates.

\begin{figure*}[h]
	\includegraphics[scale=1]{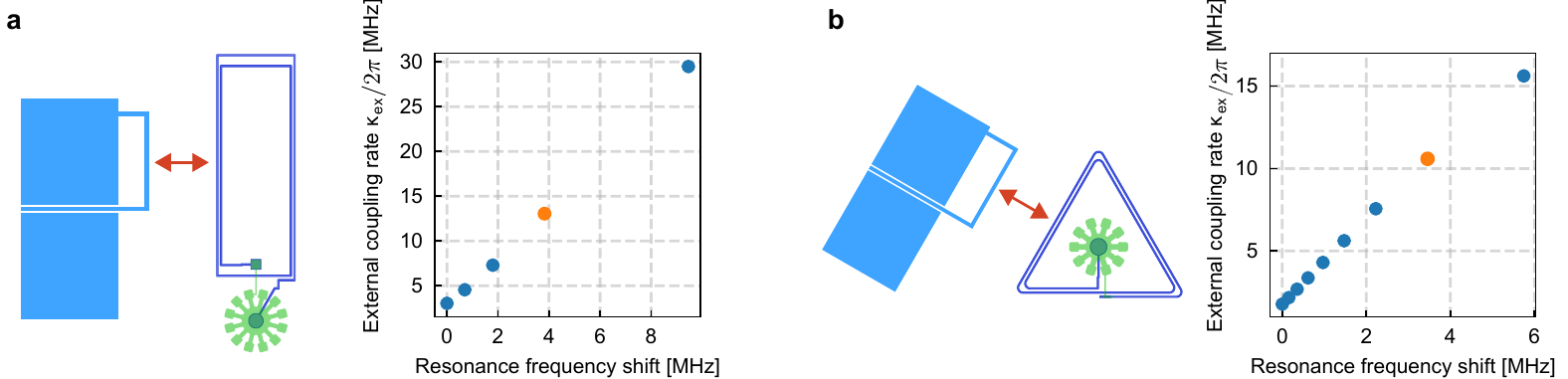}
	\caption{\textbf{Coupling to the input-output waveguide as a function of distance.} Sweeping the distance of a single site to the loop coupler for the 1D (\textbf{a}) and 2D (\textbf{b}) design, the external coupling rates as well as the resonance frequency shift due to probing the system are extracted. The orange dot corresponds to the distance chosen in the final designs. \label{fig:SI_loop_sweep}}
\end{figure*}

Another important aspect is how much the input/output loop couplers are shifting the resonance frequency of the outermost sites. If the loop coupler is too close to the circuit, we get a resonance frequency shift, which will only affect the edges of the array, seriously impairing the hybridization of the modes. To study this effect, after having optimized the dimensions of the loop itself, we swept the distance to a single LC site.
In Fig.\ref{fig:SI_loop_sweep} the external coupling rate - resonance shift tradeoff is shown.
We chose a distance that could provide a sufficiently high coupling rate, but for which the resonance frequency shift was much lower than the coupling rates in the system $\Delta\omega_\mathrm{c} / 2\pi \ll J,J'$/.

\begin{figure*}[h]
	\includegraphics[scale=1]{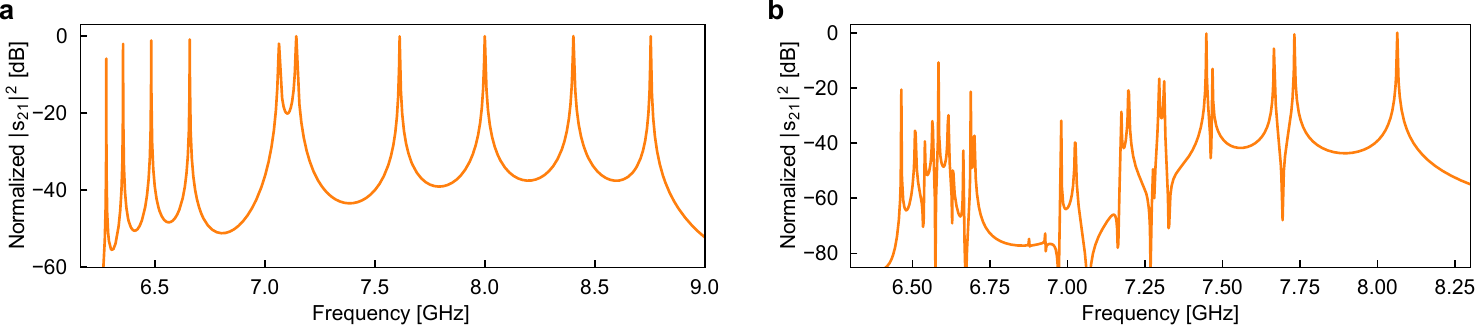}
	\caption{\textbf{ $\text{Sonnet}^\text{®}$ simulation of the full designs}. (\textbf{a}) 1D array. (\textbf{b}) 2D lattice.\label{fig:SI_Sonnet_S21}}
\end{figure*}

Finally, the whole device can be simulated in $\text{Sonnet}^\text{®}$. The result can be seen in Fig. \ref{fig:SI_Sonnet_S21}.  For the 1D we directly simulated the mask designed used in the fabrication process. For the 2D, given the size of the structure, we substituted the vacuum-gap capacitors with ideal capacitor components. This should not change the coupling between the sites in an appreciable way because the electromagnetic field of the vacuum-gap capacitor is confined between the plates, and the magnetic field responsible for the coupling of neighbouring sites is mostly confined around the spiral lines.

\section{Nanofabrication techniques for circuit optomechanics}
\subsection{Challenges and limitations of the conventional nanofabrication process}
Since 2010, when the conventional nanofabrication process of making superconducting vacuum-gap capacitors was introduced by Cicak, \textit{et al.}~\cite{SI_cicak2010low} the design and process did not have substantial change, while it was used to implement outstanding quantum experiments in optomechanics. The main steps of the conventional process (Fig.\ref{fig:old_PF}a) consist of deposition and definition the bottom plate of the capacitor, deposition of a sacrificial layer covering the bottom layer, deposition and definition of the top capacitor plate, and finally releasing the device by removing (isotropic etching) the sacrificial layer. Following the same principle, several research groups realized circuit optomechanical systems using various set of sacrificial materials on different substrates such as Si$_3$N$_4$ on sapphire at NIST~\cite{SI_teufel2011circuit}, polymer on Si at Caltech~\cite{SI_suh2014mechanically}, SiO$_2$ on quartz at Aalto~\cite{SI_pirkkalainen2015squeezing}, and aSi on sapphire at EPFL~\cite{SI_toth2017dissipative}. Due to the deposition induced compressive stress in the superconducting thin film (mostly Al), the drumhead capacitor is buckling up after the release, which increases the gap size up to a few micro meters. Cooling down such devices, induces tensile stress in the Al film due to the high thermal expansion rate difference with the substrate. Under the tensile stress, the drumhead shrinks and  buckles in the opposite direction, resulting in a small, but not accurately predictable, nor reproducible, gap size in the order of $\sim 50$ nm. This prevents precisely controlling microwave and mechanical properties of the system at low temperatures and reduces reproducibility of the design given the high probability of deformations and collapses after the release~\cite{SI_toth2018dissipation}.

\subsection{\replaced{Improved, high-yield, and high-accuracy }{reproducible} nanofabrication process}
Here we present a novel nanofabrication process (Fig.\ref{fig:PF}) to overcome such challenges with a significant improvement on control and yield. We define a trench in the substrate containing the bottom plate of the capacitor. The trench then is be covered by a thick SiO$_2$ sacrificial layer, which inherits the same topography of the layer underneath. To remove this topography and obtain a flat surface, we use chemical mechanical polishing (CMP) to planarize the SiO$_2$ surface. We then etch back the sacrificial layer down to the substrate layer and deposit the top Al plate of the capacitor. Although after the release of structure by HF vapor etching of SiO$_2$ the drumhead will buckles up due to the compressive stress, at cryogenic temperatures the high tensile stress ensures the flatness of the top plate. This will guarantee the gap size to be precisely defined by the depth of the trench and the thickness of the bottom plate. We describe every step of the process in detail here:
\begin{figure}[h]
	\centering
	\includegraphics[width=\textwidth]{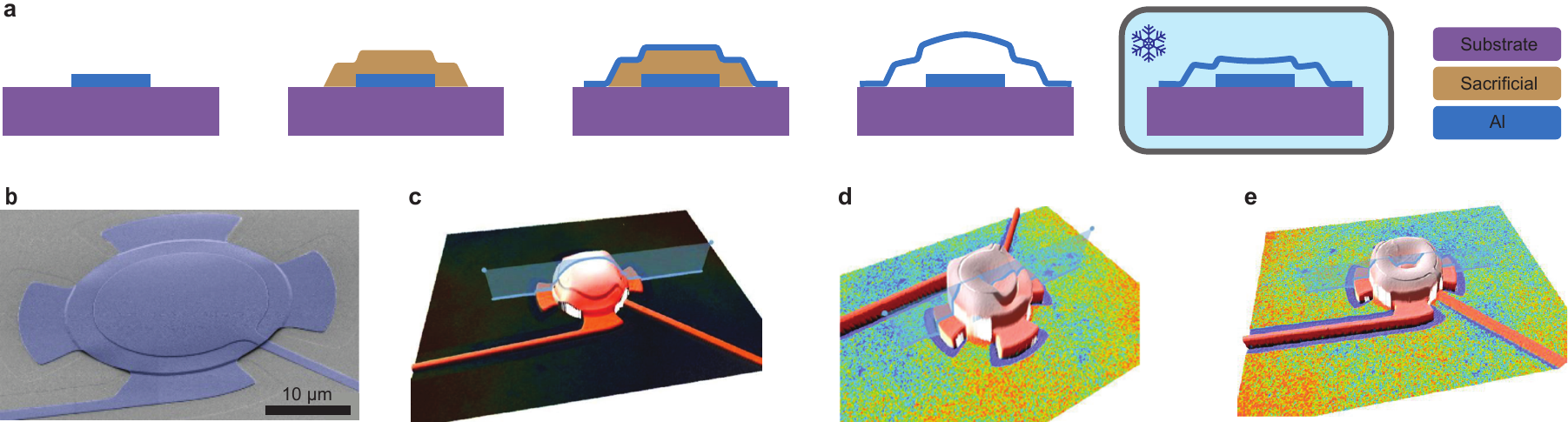}
	\caption{\textbf{Challenges of the conventional fabrication process for circuit optomechanics.} \textbf{a}, The \replaced{original }{traditional} nanofabrication process used to make mechanically compliant vacuum-gap capacitors: a sacrificial layer (aSi or Si$_3$N$_4$) used to support the top layer. After the release, the capacitor buckles up more than 1 $\mu$m due to the compressive stress in deposited superconducting metal. At low temperatures, the drumhead shrinks resulting in a small gap size of $\sim50$ nm. \textbf{b}, SEM image of a drumhead fabricated with the conventional process at EPFL\cite{SI_Toth2017}. \textbf{c}, Laser profilometry of a successfully released drumhead. \textbf{d,e}, Laser profilometry of a deformed and a collapsed drumhead correspondingly.}
	\label{fig:old_PF}
\end{figure}
\begin{figure}[h]
	\centering
	\includegraphics[width=\textwidth]{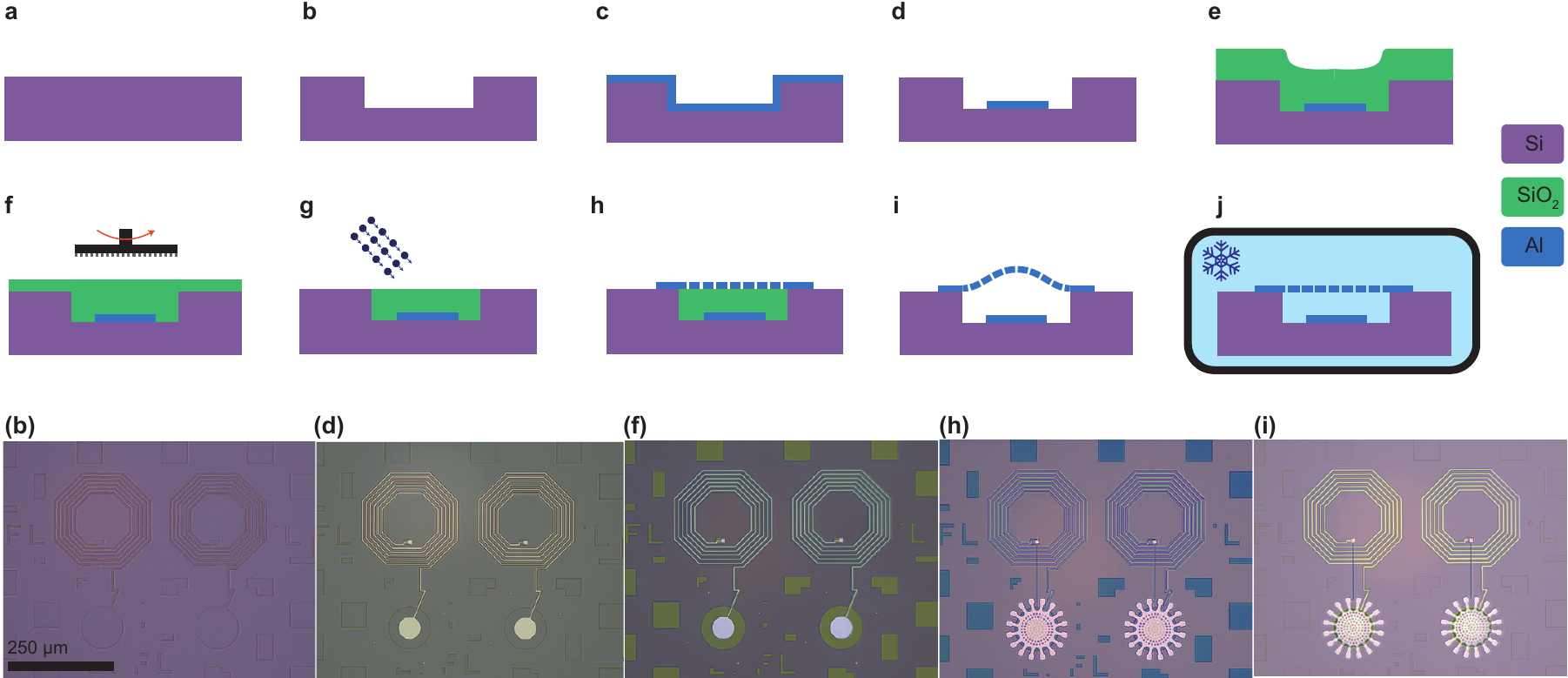}
	\caption{\textbf{The new reproducible nanofabrication process for circuit optomechanics.} \textbf{a}, \textbf{b}, Etching a trench in a silicon wafer (325 nm). \textbf{c}, Aluminum deposition of the bottom plate (100 nm). \textbf{d}, Patterning of Al. \textbf{e}, SiO$_2$ sacrificial layer deposition (3 $\mu$m). \textbf{f}, CMP planarization. \textbf{g}, landing on the substrate using IBE etching. \textbf{h}, Top Al layer deposition and patterning (200 nm). \textbf{i}, Releasing the structure using HF vapor. Due to compressive stresses, the top plate will buckle up. \textbf{j}, At cryogenic temperatures, the drumhead shrinks and flattens. The optical micrographs shows selected steps of the process flow.}
	\label{fig:PF}
\end{figure}
We use high-resistivity silicon wafers (Fig. \ref{fig:PF}a). First, a 325 nm trench is etched in the silicon substrate to define the bottom part of the circuit (Fig. \ref{fig:PF}b). Silicon etching is done by deep reactive ion etching (DRIE) with $\rm{C_4F_8}$ gas (Adixen AMS200). The next step is the deposition of 100 nm aluminum by electron beam evaporation (Alliance-Concept EVA 760, Fig. \ref{fig:PF}c). We pattern and etch Al using wet etchant ($\rm{H_3PO_4}\; 85\% + \rm{CH_3COOH }\; 100\% + \rm{HNO_3 }\; 70\%$ 83:5.5:5.5, Fig. \ref{fig:PF}d). Afterwards, we deposit 2 $\mu$m silicon oxide using low thermal oxide deposition (LTO) (Fig. \ref{fig:PF}d). To remove the surface topography, we use CMP (ALPSITEC MECAPOL E 460) to planarize the surface (Fig. \ref{fig:PF}f) and reduce the surface topography to less than 10 nm (an example of the polishing process is shown in Fig. \ref{fig:CMP}). We made dummy trenches on all the empty space of the wafer to increase the uniformity in the CMP process. After planarization we etch back the sacrificial layer by ion beam etching (IBE,Veeco Nexus IBE350) to land on the silicon substrate (Fig. \ref{fig:PF}g). Before deposition of the top plate Al, we make an opening in the oxide to ensure the galvanic connection between top and bottom layers of the circuit in the spiral inductor and capacitor. This is done by DRIE etching of the SiO$_2$ (SPTS APS) with $\rm{CHF_3}$. Prior to the opening etch, the photo-resist is re-flowed to make slanted sidewalls to smoothly connect two layers. Then we use a 1:1 DRIE etch of SiO$_2$ (SPTS APS with $\rm{CHF_3}$ etchant), where the resist pattern is transferred to the oxide (Fig. \ref{fig:Cross}b). The top Al layer (200 nm) is evaporated afterwards to form the top plate of the vacuum-gap capacitor. After dicing the wafer into chips, we finally release the structure using Hydrofluoric (HF) acid vapor (SPTS uEtch) which is an isotropic etch process dedicated for MEMS structuring and does not attack Al. The holes on the drumhead are there to facilitate the release process. All patterning steps are done by direct mask-less optical lithography (Heidelberg MLA 150) using 1$\mu$m photo resist (AZ ECI 3007).

\begin{figure}[h]
	\centering
	\includegraphics[width=\textwidth]{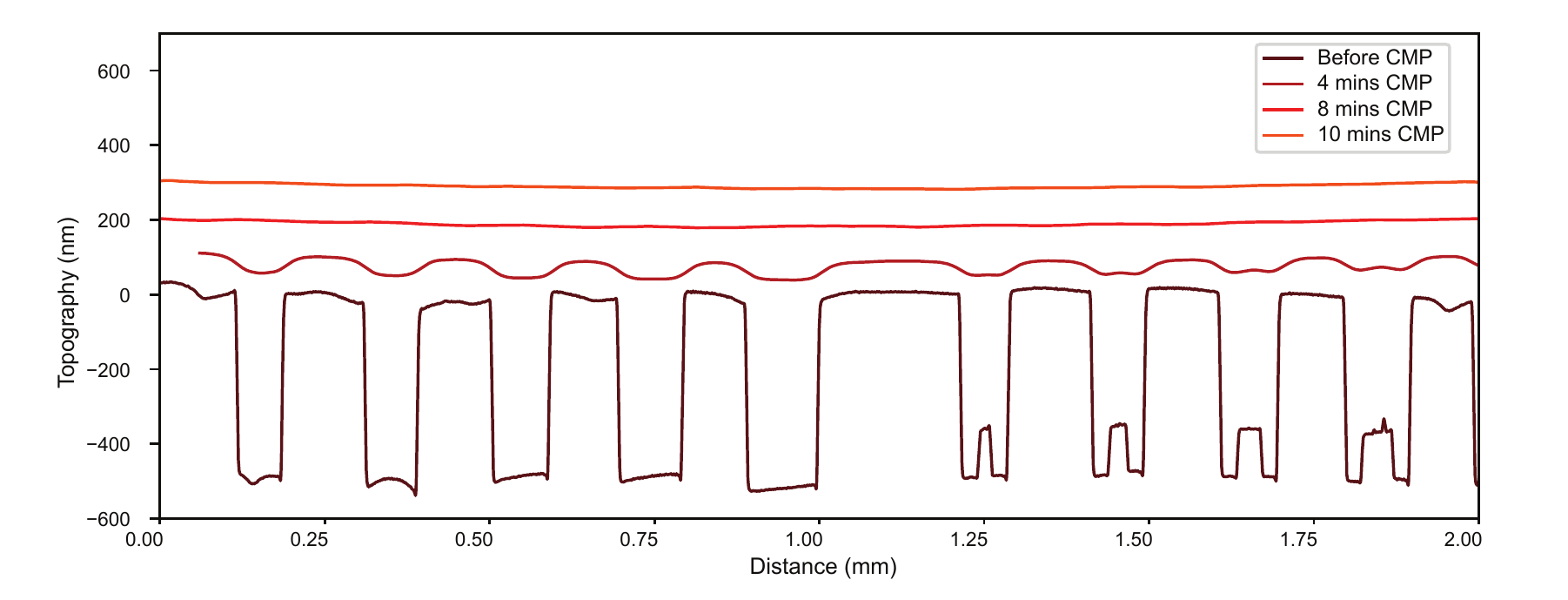}
	\caption{\textbf{Chemical mechanical polishing (CMP) for removing the surface topography}. CMP enables us to reduce the surface topography from $\sim500$ nm to below 10 nm. The figure shows the effect of sequential CMP steps on the topography measured by mechanical profilometry. The final global curve is the wafer bow.}
	\label{fig:CMP}
\end{figure}

\begin{figure}[h]
	\centering
	\includegraphics[width=\textwidth]{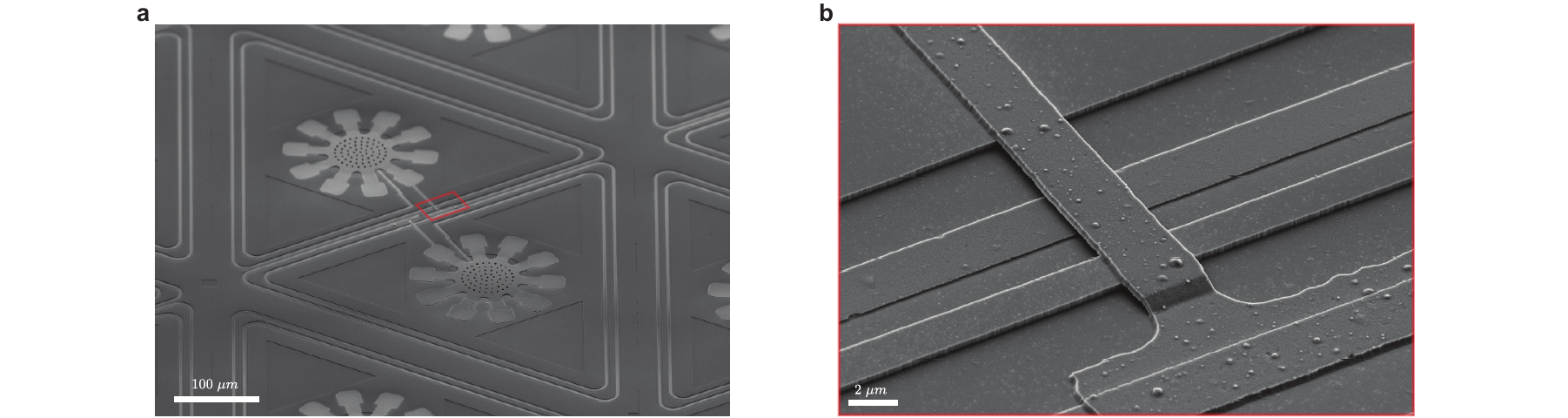}
	\caption{\textbf{Honeycomb lattice}. \textbf{a}, SEM micrograph of the triangular building blocks of the honeycomb lattice. \textbf{b}, The galvanic connection of top and bottom Al layers and the crossovers of the spiral inductor.}
	\label{fig:Cross}
\end{figure}

\section{Experimental setup and measurement techniques}
\subsection{Full experimental setup}
\begin{figure*}[h]
	\includegraphics[scale=1]{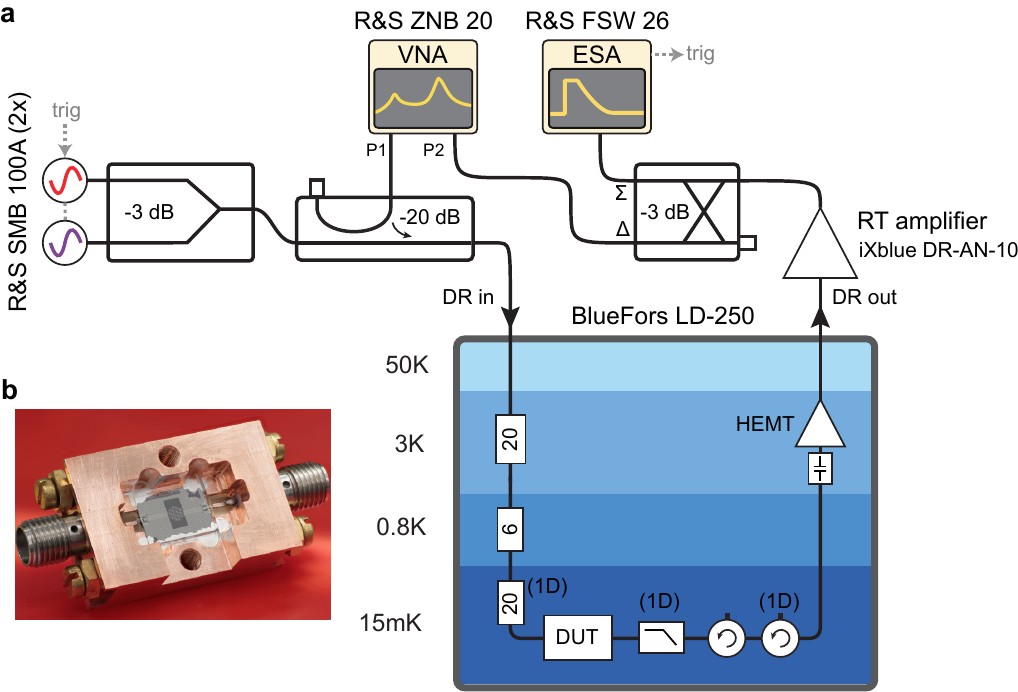}
	\caption{ \textbf{a}, Full experimental setup used for the chips' characterization. Some elements of the input line (inside round brackets) are specific only to the 1D sample. \textbf{b}, Packaged chip inside the copper sample holder.\label{fig:SI_exp_setup}}
\end{figure*}

The full experimental setup (Fig.\ref{fig:SI_exp_setup}a) consists of a room-temperature (RT) and a cryogenic section. At RT, two Rohde \& Schwartz (R\&S) SMB 100A analog microwave sources generate the cooling/probe and excitation pumps, a R\&S ZNB20 Vector Network Analyzer (VNA) and a R\&S FSW 26 Electronic Spectrum Analyzer (ESA) are used for the measurement itself.
The VNA measures the coherent response of the devices, in our case the transmission scattering parameter $S_{21}(\omega)$. The ESA is not used in its usual frequency-domain mode, instead, for our experiment we use it in the so-called 'zero-span'. In this mode the instrument demodulates at a fixed CW frequency and effectively implements a time-domain power measurement over a large dynamic range, which is typical of spectrum analyzers. The ESA is able to trigger the microwave sources to start the time-domain sequence, which is programmed in the sources. All the instruments are locked together and to a SRS FS725 Rubidium Frequency Standard to achieve reliablie frequency stability and accuracy.
The two microwave sources are combined at RT through a 3dB microstrip coupler. The VNA is then directionally-coupled with 20dB of insertion loss and all the signals are finally sent to the Dilution Refrigerator (DR).

A series of cryogenic attenuators are used at the different flanges to thermally anchor the input wiring and more importantly to remove the room temperature noise from the input signal. For the 1D sample a total of 46 dB of nominal attenuation were used. For the 2D samples a total of 26 dB of nominal attenuation were used. In this experiment we are not interested to probe the devices at the single photon level, hence the not-so-high attenuation values. We are more interested in being able to reach the regime where the effective mechanical damping rate greatly exceeds the intrinsic one $\Gamma_{\text{eff}} \gg \Gamma_{\text{m}}$ through optomechanical sideband cooling, which requires high on-chip powers (\added{$\mathcal{O}(-30~\mathrm{dBm})$}).
For the 1D chip an additional K\&F 18GHz lowpass filter and circulator were used.

The output signal from the chips is firstly amplified with a cryogenic High Electron Mobility Transistor (HEMT) amplifier at the 4K flange of the DR. The HEMTs were both from Low-Noise-Factory, model numbers are LNF-LNC4\_8C and LNF-LNC1\_12A for the 1D and 2D sample respectively. The typical gain is 40 dB.
The circulators placed after the chips are crucial in preventing the back-propagating high-amplitude signals of the HEMT amplifier from reaching the chip.
A room temperature amplifier, model number iXblue DR-AN-10 is placed as close to the fridge as possible to further amplify the signal and make it robust against injected noise along the cables until they reach the measurement equipment. This signal is then split with a 180-hybrid coupler and sent back to port 2 of the VNA and the ESA.

The chip holder (Fig.\ref{fig:SI_exp_setup}b) is mounted at the MXC flange of the DR, which has a temperature of $T\approx 15 \; \text{mK}$. The chip holder is constructed with Oxygen-free copper. The CPW signal lines are wire bonded, as well as the ground plane close to them, while the perimeter ground plane has been contacted using conductive silver glue.

\subsection{Ringdown data analysis and cavity shifts}
\begin{figure*}[h]
	\includegraphics[scale=1]{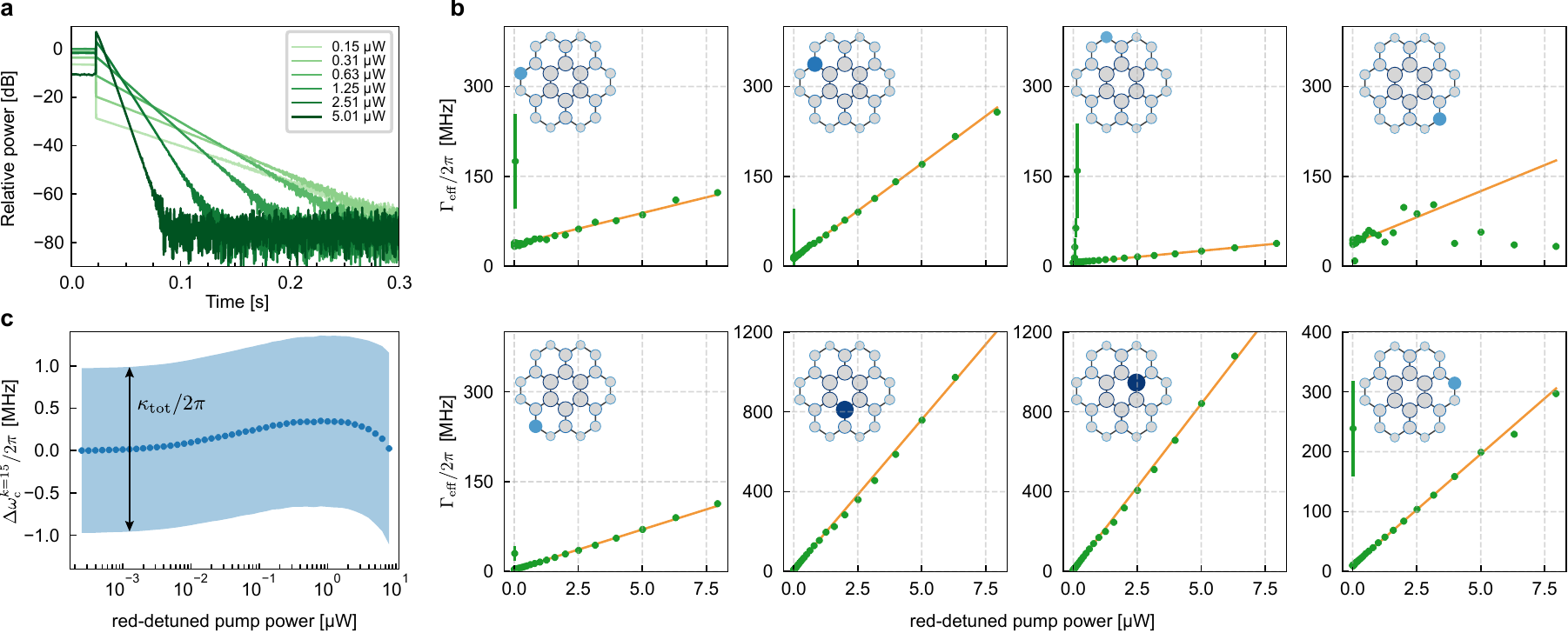}
	\caption{ \textbf{Data analysis.} \textbf{a}, Example raw data set of ringdowns with increasing cooling power. \textbf{b}, $\Gamma_\mathrm{eff}$ vs cooling power data for some sites of the $k=24$ collective mode of the 2D sample. \textbf{c}, Collective mode resonance frequency shift versus cooling power for the average-linewidth mode with $\kappa/2\pi = 2\:\mathrm{MHz}$. The shaded area indicates the mode's linediwth.
		\label{fig:SI_ringdown_fits}}
\end{figure*}

The core of the modeshape measurement is the extraction of the energy participation ratio $\eta_i^k$ of site $i$ to collective microwave mode $k$ by fitting the mechanical oscillator damping rate trend $\Gamma_\mathrm{eff,i}$ at site $i$ with changing power. We extract the effective mechanical damping rates by measuring how fast each mechanical oscillator rings down from a high-phonon occupancy state. \replaced{}{This is the so-called ringdown measurement.}
We fit the ringdown data, an example data set of which can be seen in Fig.~\ref{fig:SI_ringdown_fits}a, with an exponential profile $P(t) = P_0 \exp(-\Gamma_\mathrm{eff}/2\pi) + N$ and extract $\Gamma_\mathrm{eff}$ from such fit.
Given the very large amount of ringdown traces (\# powers)$\times$(\# sites)$\times$(\# collective modes) = ($5'100$ for the 1D and $29'236$ for the 2D sample), a robust automated fitting algorithm has been advised to reliably find good initial parameters for the fits. Another challenge arises because, due to the large variation of $\eta_i^k$ across the different modes and sites, there can be more than 2 orders of magnitude in the exponential decay rate of the measured signal, as can bee seen in Fig.~\ref{fig:SI_ringdown_fits}b. This makes it difficult to have a one-for-all starting condition for the fits. Moreover, the initial decay of a highly-excited mechanical oscillator can exibit non-linear behaviour due to the high-amplitude oscillations and it should not be included in the exponential fit.
Finally, since for the optomechanical damping effect the relative detuning between the cavity and mechanical frequency is relevant, we measured the collective modes' resonance frequency shift with the cooling powers used for the experiment. Figure~\ref{fig:SI_ringdown_fits}c shows an example of such analysis for mode $k=15$ which has a linewidth closest the average among all the modes. We can see that for this mode the cavity shift is negligible compared to the linewidth. This might not be the case for the modes with the lowest linewidths, and it could justify restricting the fitting to a lower-power range. This effect can also be taken into account by plugging in the correct $\tilde\omega_\mathrm{c}^k$ and $\kappa_\mathrm{tot}^k$ into the optomechanical damping rate equation \eqref{eq:Gam_ki}.

\subsection{Modeshape extraction methods comparison}
\begin{figure*}[h]
	\includegraphics[scale=1]{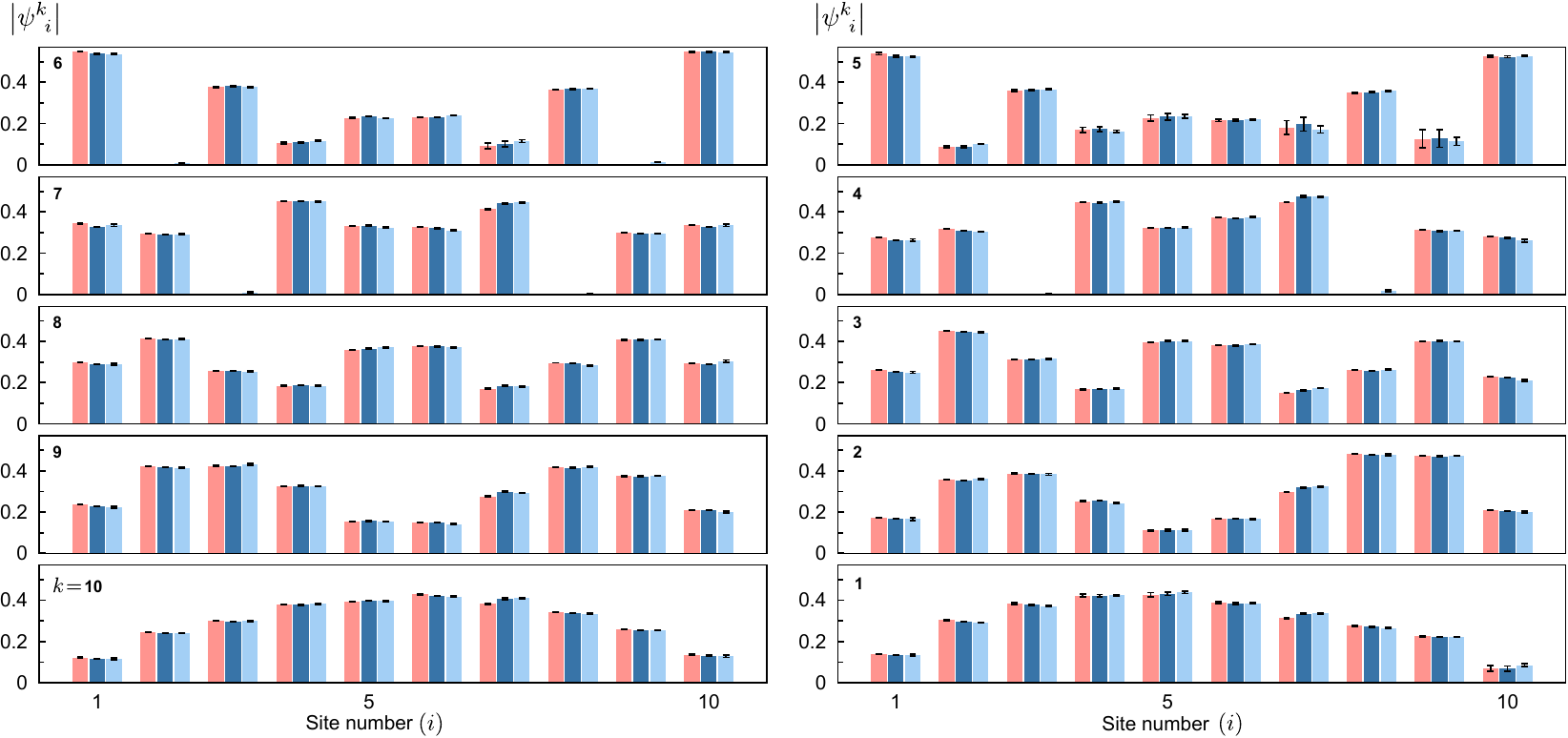}
	\caption{ \textbf{Three different modeshape extraction methods comparison.} The red bars are with the single-step normalization for each collective mode. The dark blue bars are with the iterative normalization method (presented in the main text). The light blue bars are modeshapes corrected based on the orthogonalization property of the unitary matrix. \label{fig:SI_modeshapes_steps}}
\end{figure*}
As explained in section \ref{subsec:eta_extraction}, to extract the modeshapes we have to go through at least one normalization step.
\begin{equation*}
	\widetilde{\eta_i^k} \to \frac{\widetilde{\eta_i^k}}{\sum_i \widetilde{\eta_i^k}} \quad \forall \: k
\end{equation*}
This step is very intuitive as we are summing over all sites $(i)$ in a given collective mode $(k)$: the total input power gets distributed along the sites according to their participation ratio $\eta_i^k$, and all the participation ratios in a given collective mode must sum up to $1$.

Next, we make use of the normalization conditions of the unitary matrix, and perform the iterative normalization process described in section \ref{subsec:eta_extraction}. These are the modeshapes presented in Figs.~3 and 4 of the main text.

As described in Methods, in order to accurately reconstruct the Hamiltonian shown in Fig.~3 of the main text, the modeshapes obtained from the iterative normalization method are further corrected based on the orthogonalization property of the unitary matrix, providing the modeshapes satisfying both the normalization and orthogonalization conditions.
In Fig. \ref{fig:SI_modeshapes_steps} we present the evolution of the modeshapes along these three steps.

\begin{table}[h!]
	\begin{tabular}{c|cccccccccc}
		Collective mode, $i$                                            & 10   & 9   & 8   & 7   & 6   & 5   & 4   & 3   & 2   & 1   \\
		\hline
		$1-F(\psi_i^\mathrm{R},\psi_i^\mathrm{N})$ $[\times 10^{-3}]$   & 0.8  & 0.6 & 0.4 & 1.2 & 0.3 & 0.5 & 1.2 & 0.3 & 0.6 & 0.6 \\
		$1-F(\psi_i^\mathrm{N},\psi_i^\mathrm{N,O})$ $[\times 10^{-3}]$ & 0.06 & 0.3 & 0.5 & 0.4 & 0.7 & 1.3 & 0.7 & 0.4 & 0.2 & 0.5
	\end{tabular}
	\caption{Infidelities between different subsequent normalization methods}
	\label{table:1D_infidelities_comparison}
\end{table}

\added{To verify that our iterative normalization method does not perturb the modeshapes we compute the infidelities $1-F(\psi_i,\psi_i ')$, where $F(\psi_i,\psi_i ') = |\langle \psi_i|\psi_i '\rangle)|^2$, between the same mode with different normalization methods. Table \ref{table:1D_infidelities_comparison} shows the comparison between a one-step (site-wise) normalization as the raw data, $\psi_i^\mathrm{R}$, and the iterative normalization process, $\psi_i^\mathrm{N}$. It also shows the comparison between the iterative normalization method, $\psi_i^\mathrm{N}$, and the modeshapes computed from the orthogonalization after iterative normalization, $\psi_i^\mathrm{N,O}$, (introduced in the SI section 1 C). The small infidelities calculated demonstrates the minimal disorder in the system parameters.}

\begin{table}[h!]
	\begin{tabular}{c|cccccccccc}
		Collective mode                                                & 10  & 9   & 8   & 7   & 6   & 5   & 4   & 3   & 2  & 1  \\
		\hline
		$1-F(\psi_i^\mathrm{N},\psi_i^\mathrm{th})$ $[\times 10^{-3}]$ & 2.1 & 2.2 & 1.5 & 0.3 & 1.4 & 4.6 & 5.4 & 4.2 & 17 & 17
	\end{tabular}
	\caption{Infidelities between the measured modeshapes and the theoretical ones}
	\label{table:1D_infidelities_comparison_exp_th}
\end{table}
\added{For completeness, in table \ref{table:1D_infidelities_comparison_exp_th} we also report the infidelities between the experimental modeshapes (iterative normalization, $\psi_i^\mathrm{N}$) and the theoretical ones, $\psi_i^\mathrm{th}$, presented in Fig. 3 of the main text.}

\subsection{Statistical analysis of orthogonality and measurement error}

\begin{figure*}[h!]
	\includegraphics[scale=1]{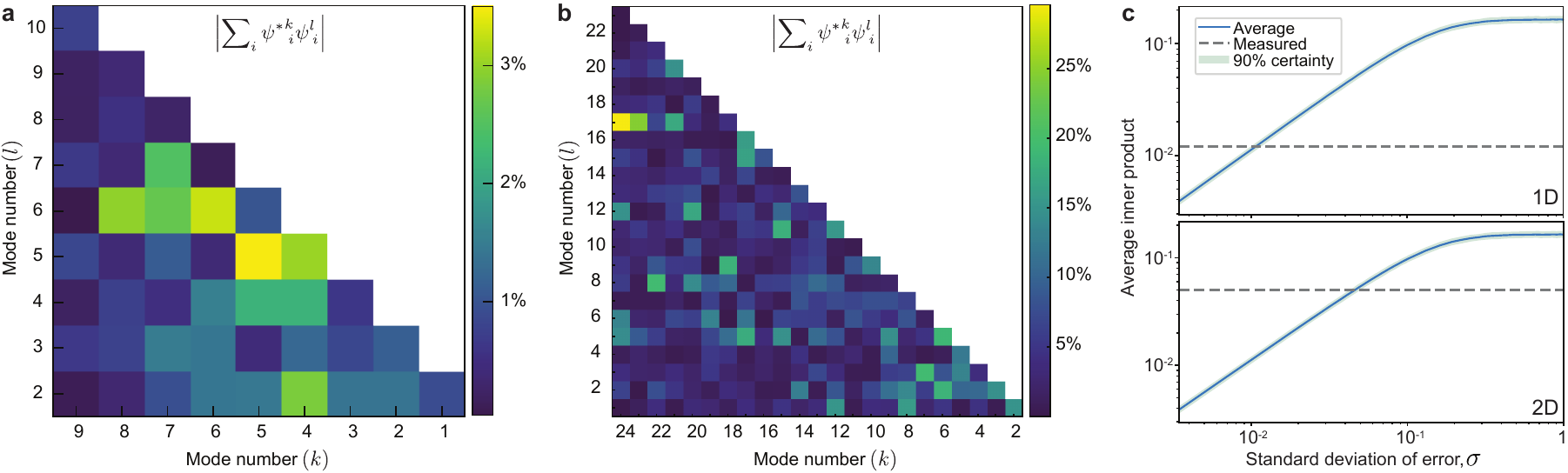}
	\caption{\textbf{Inner products of the measured eigenstates as an indicator for measurement accuracy.} \textbf{a}, \textbf{b}, Scalar products between the measured mode functions after iterative normalization (introduced in the theory section) for 1D and 2D devices correspondingly. The inner products are a quantitative orthogonality verification of the measured modeshapes (eigenvectors) in those systems. \textbf{c} The stochastic analysis of the standard deviation of the equivalent Gaussian noise ($\sigma$) induced in the measurement on the eigenmodes. The expected value of the average inner products in a 10 (1D), and 24 (2D) dimensional space is shown versus $\sigma$. The grey lines indicate the measured average inner product. \label{fig:SI_inner_product}}
\end{figure*}

We discussed the optomechanical modeshape measurement technique in the main text to directly extract the eigenmodes (eigenvectors of the system Hamiltonian). In the theory section of the SI, we explained the iterative normalization method used to extract the modeshapes without the need to measure some hardly-accessible parameters of our system.
To check the accuracy of our measurements and quantitatively verify the orthogonality of the modeshapes as eigenstates of the system, we show the inner product of every pair of modes,$|\Sigma_i\psi_i^{*k}\psi_i^{l}|$ in Fig.~\ref{fig:SI_inner_product}a,b for 10 site 1D chain as well as 24 site honeycomb lattice. Since the measurement does not retrieve the phase in the modeshapes (i.e. positive or negative signs because such a system always results in real eigenvector and eigenvalues), we inferred the phases from the theoretical modeshape estimates. As it is shown in the figure, the 1D dataset manifests high orthogonality with a maximum $\sim 3\%$ inner product. The 2D dataset also shows small values except for a single mode (mode 17) which has a non-zero product with a few other modes. As a figure-of-merit, the average value of all inner products, $\frac{\sum_{k\neq l} \langle\psi^k|\psi^l\rangle} {2N\times (2N-1)} $, is 1.2\% for the 1D chain and 5\% for the 2D lattice.

To estimate the average error induced by the measurement process on the extracted eigenstates, we performed a simple numerical stochastic analysis. In a $2N$-dimensional space, one can consider an ideal orthonormal set of vectors, $\{\mathbf{e}_k\}$, and add a $2N$-dimensional Gaussian random variable with a standard deviation of $\sigma$ to every element in order to create a disturbed set of $\{\mathbf{e'}_k\}$. Figure \ref{fig:SI_inner_product}c shows the stochastic expected value and the 90\% certainty range of the average inner product of the disturbed set, $\frac{\sum_{k\neq l} \mathbf{e'}_k \cdot \mathbf{e'}_l}{2N\times (2N-1)}$ for $2N=10$ (1D) and $2N=24$ (2D). Considering the measured average inner products, we conclude $\sigma\simeq1\%$ for 1D and $\sigma\simeq5\%$ for 2D devices.

\section{Full device characterisations}
\subsection{Samples parameters}
The hybridized modes frequencies $\omega_c$  and line-widths $\kappa$ (Tab. \ref{table:1D_modes} and \ref{table:2D_modes}) were measured at the VNA then fitted with a complex $S_{21}^\text{exp}(\omega) = S_{21}^\text{Lor}(\omega) + x + iy$ expression that also takes into account any non-ideal Fano effect. $s_{21}^\text{Lor}(\omega)$ is a Lorentzian lineshape and $x +iy$ is a complex displacement that even though it doesn't physically interpret the origin of the 'Fanoness', it enables us to model it. For the cavity and mechanical resonance frequencies, the uncertainty in the values is many orders of magnitude smaller than the least significant reported digit.

The mechanical frequencies (Tab. \ref{table:1D_mechanics} and \ref{table:2D_mechanics}) were measured in a similar way, doing and OMIT measurement, and fitting the VNA trace with an
$S_{11}^\text{exp}(\omega)$ reflection Lorentzian experimental profile.
The bare mechanical damping rates $\Gamma_\text{m}$ were extracted by taking the intercept of the $\Gamma_\text{eff}$ fits versus power. For a fixed mechanical site, more than one independent value of $\Gamma_\text{m}$ was extracted by looking at the different hybridized modes. What we are reporting here is the weighted average of such values. The error in these measurements is of the same order of magnitude of the digit in round brackets.

\begin{table}[h!]
	\begin{tabular}{c|cccccccccc}
		Hybdirized mode       & 10    & 9     & 8     & 7     & 6     & 5     & 4     & 3     & 2     & 1     \\
		\hline
		$\omega_c/2\pi$ [GHz] & 8.463 & 8.139 & 7.767 & 7.435 & 7.016 & 6.922 & 6.590 & 6.404 & 6.285 & 6.216 \\
		$\kappa/2\pi$ [MHz]   & 3.904 & 4.556 & 4.176 & 4.668 & 4.964 & 7.09  & 0.696 & 0.384 & 0.239 & 0.080
	\end{tabular}
	\caption{1D hybridized cavity modes parameters}
	\label{table:1D_modes}
\end{table}

\begin{table}[h!]
	\begin{tabular}{c|cccccccccc}
		Site number                  & 1      & 2      & 3      & 4      & 5      & 6      & 7      & 8     & 9      & 10      \\
		\hline
		$\Omega_\text{m}/2\pi$ [MHz] & 2.142  & 2.165  & 2.202  & 2.238  & 2.267  & 2.315  & 2.616  & 2.405 & 2.448  & 2.506   \\
		$\Gamma_\text{m}/2\pi$ [Hz]  & 4.3(3) & 4.2(3) & 12.(1) & 11.(1) & 15.(4) & 12.(2) & 15.(2) & 8.(0) & 16.(6) & 10.6(3)
	\end{tabular}
	\caption{1D mechanical modes parameters}
	\label{table:1D_mechanics}
\end{table}

\begin{table}[h!]
	\begin{tabular}{c|cccccccccccc}
		Hybridized mode       & 24    & 23    & 22    & 21    & 20    & 19    & 18    & 17    & 16    & 15    & 14    & 13    \\
		\hline
		$\omega_c/2\pi$ [GHz] & 8.250 & 7.926 & 7.851 & 7.667 & 7.644 & 7.512 & 7.494 & 7.481 & 7.401 & 7.374 & 7.224 & 7.189 \\
		$\kappa/2\pi$ [MHz]   & 4.79  & 4.601 & 4.485 & 1.09  & 6.468 & 2.165 & 2.504 & 1.027 & 3.31  & 2.002 & 0.349 & 0.362
	\end{tabular}
	
	\vspace*{0.2 cm}
	
	\begin{tabular}{c|cccccccccccc}
		Hybridized mode       & 12    & 11    & 10    & 9     & 8     & 7     & 6     & 5     & 4     & 3     & 2     & 1     \\
		\hline
		$\omega_c/2\pi$ [GHz] & 7.100 & 7.068 & 6.911 & 6.896 & 6.871 & 6.837 & 6.829 & 6.797 & 6.772 & 6.750 & 6.717 & 6.671 \\
		$\kappa/2\pi$ [MHz]   & 0.265 & 0.766 & 4.36  & 1.121 & 0.241 & 0.174 & 2.635 & 1.603 & 1.261 & 0.532 & 1.308 & 0.456
	\end{tabular}
	\caption{2D hybridized cavity modes parameters}
	\label{table:2D_modes}
\end{table}

\begin{table}[h!]
	\begin{tabular}{c|cccccccccccc}
		Site number                  & 1      & 2     & 3     & 4     & 5      & 6      & 7      & 8      & 9      & 10    & 11     & 12     \\
		\hline
		$\Omega_\text{m}/2\pi$ [MHz] & 2.106  & 2.127 & 2.158 & 2.179 & 2.208  & 2.233  & 2.260  & 2.291  & 2.314  & 2.347 & 2.380  & 2.413  \\
		$\Gamma_\text{m}/2\pi$ [Hz]  & 43.(7) & 9.(0) & 6.(9) & 4.(1) & 15.(6) & 12.(6) & 14.(1) & 6.0(9) & 29.(8) & 8.(1) & 20.(5) & 3.2(7)
	\end{tabular}
	
	\vspace*{0.2 cm}
	
	\begin{tabular}{c|cccccccccccc}
		Site number                  & 13    & 14     & 15     & 16    & 17     & 18    & 19     & 20     & 21     & 22     & 23    & 24     \\
		\hline
		$\Omega_\text{m}/2\pi$ [MHz] & 2.435 & 2.469  & 2.501  & 2.539 & 2.571  & 2.611 & 2.648  & 2.672  & 2.708  & 2.749  & 2.796 & 2.836  \\
		$\Gamma_\text{m}/2\pi$ [Hz]  & 5.(5) & 10.(8) & 6.7(7) & 6.(4) & 16.(3) & 9.(8) & 39.(0) & 20.(5) & 18.(6) & 10.(3) & 8.(9) & 18.(1)
	\end{tabular}
	\caption{2D mechanical modes parameters}
	\label{table:2D_mechanics}
\end{table}


\subsection{Measurement of optomechanical coupling rate}
In order to measure $\eta_i^k\: g_{0,i}$, the effective optomechanical coupling rate between collective microwave mode $k$ and mechanical oscillator $i$, we characterize the mechanical sideband induced by a resonant microwave drive.
In contrast to the measurement of the optomechanical damping rate, we set the drive power so that the cooperativity can be about 1 to minimize the measurement backaction on the phonon occupation.
We measure the power spectrum density~(PSD) of the upper sideband signal and integrate the PSD to obtain the total power.
From Eq.~(\ref{eq:langevin_ki_linearized}), the sideband power~\cite{SI_toth2017dissipative} scaled to photon flux is given by
\begin{equation}
	\label{eq:nsb}
	n_\mathrm{sb} = G^k \frac{\kappa_2^k}{\kappa_\mathrm{tot}}\frac{\kappa_\mathrm{tot}^k(\eta_i^k\: g_{0,i})^2 n^k_\mathrm{c}}{\Omega_{\mathrm{m},i}^2 + {\kappa_\mathrm{tot}^k}^2/4} \: n_{\mathrm{m},i},
\end{equation}
where $G^k$ is the gain of the full measurement chain from the device, $\kappa_2^k$ is the external coupling rate to the output line, and $n_{\mathrm{m},i}$ is the phonon occupation of mechanical oscillator $i$.
\begin{figure*}[t]
	\includegraphics[scale=1]{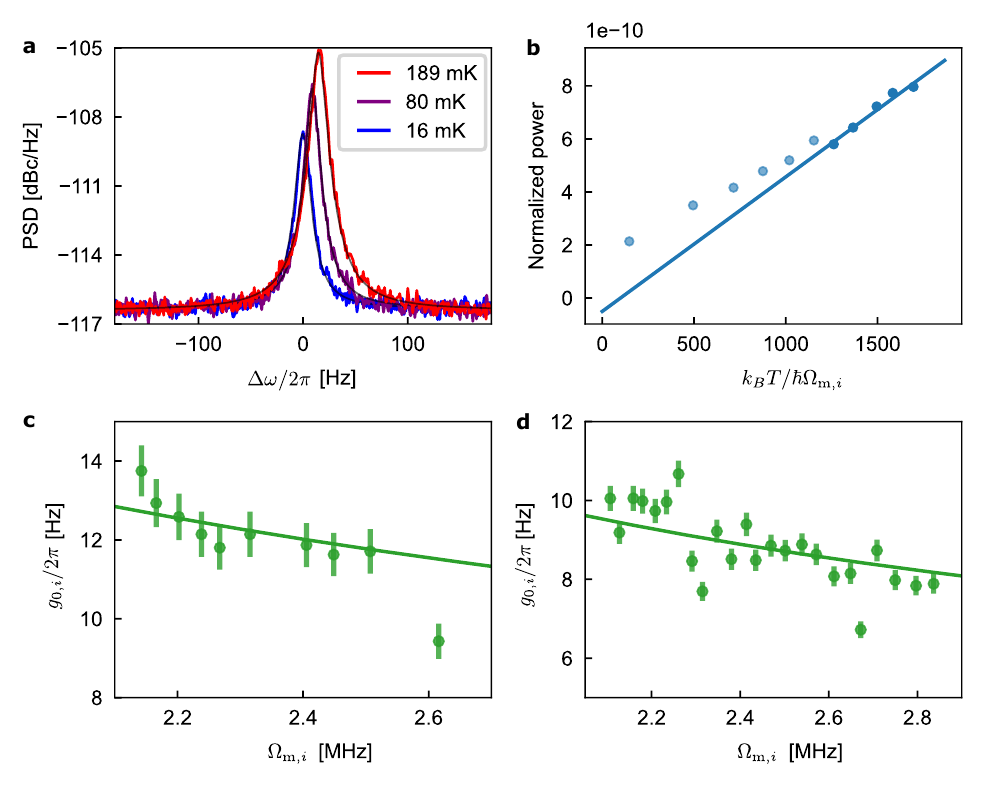}
	\caption{\textbf{Measurement of optomechanical coupling rate.} \textbf{a}, Power spectrum density~(PSD) of the upper mechanical sideband with the different base temperatures of the dilution refrigerator. The black lines are the Voigt function fits to extract the total power of the sideband. \textbf{b}, Normalized total power of the mechanical sideband as a function of the nominal phonon occupation, calculated from the base temperature. The dots are the experimental results and the line is the linear fit to the data in the higher temperature region (dark blue). \textbf{c,d} Single-photon optomechanical coupling rate at site $i$ for the 1D chain and 2D lattice, respectively. The lines are inverse square root fits. \label{fig:SI_g0}}
\end{figure*}
The intracavity photon number with the resonant drive is explicitly described as
\begin{equation}
	\label{eq:nck}
	n^k_\mathrm{c} = \frac{4\kappa_1^k}{{\kappa_\mathrm{tot}^k}^2}\:R^k \: n_\mathrm{d,in},
\end{equation}
where $R^k$ is the transmittance between the device and a microwave source used for the drive, $\kappa_1^k$ is the external coupling rate to the input line, and $n_\mathrm{d,in}$ is the drive power scaled to photon flux at the output of the microwave source.
From the scattering parameters based on the input-output formalism~\cite{SI_clerk2010introduction}, the transmitted drive power scaled to photon flux is give by
\begin{equation}
	\label{eq:ndout}
	n_\mathrm{d,out} = G^k \frac{4\kappa_1^k\kappa_2^k}{{\kappa_\mathrm{tot}^k}^2}R^k \: n_\mathrm{d,in},
\end{equation}
where we assume that the gain $G^k$ and transmittance $R^k$ does not have frequency dependence between the drive and the mechanical sideband.
Using Eqs. (\ref{eq:nsb}), (\ref{eq:nck}), and (\ref{eq:ndout}), we obtain
\begin{equation}
	\frac{n_\mathrm{sb}}{n_\mathrm{d,out}} = \frac{(\eta_i^k\: g_{0,i})^2}{{\Omega_{\mathrm{m},i}}^2 + {\kappa_\mathrm{tot}^k}^2/4} \: n_{\mathrm{m},i}.
\end{equation}
Note that all the parameters that would be challenging to obtain experimentally, except for the effective optomechanical coupling rate $\eta_i^k\: g_{0,i}$, are canceled out in this expression, enabling us to determine the coupling rate.

We calibrate the phonon occupation $n_{\mathrm{m},i}$ by increasing the temperature of the environment for the device so that the mechanical oscillator can be thermalized, i.e.\ $n_{\mathrm{m},i} \approx k_\mathrm{B}T/\hbar\Omega_{\mathrm{m},i}$, where $T$ is the base temperature of the dilution refrigerator.
Figure \ref{fig:SI_g0}a shows the PSD of the mechanical sideband of mechanical mode $i=6$ induced by the resonant drive to collective microwave mode $k=10$ of the 1D chain.
Since the minimum resolution bandwidth of the spectrum analyzer is comparable with the mechanical linewidth, we extract the total power of the mechanical Lorentzian peak by fitting the PSD with a Voigt function with the Gaussian bandwidth corresponding to the resolution bandwidth.
Figure \ref{fig:SI_g0}b shows the sideband power normalized by the transmitted drive power as a function of the nominal phonon occupation calculated by $ k_\mathrm{B}T/\hbar\Omega_{\mathrm{m},i}$.
In the region of the higher temperature, the normalized sideband power follows linearly the nominal phonon occupation, so that we can assume it is thermalized.
By fitting the slope in this region, the effective optomechanical coupling rate can be extracted as
\begin{equation}
	(\eta_i^k\: g_{0,i})^2 = \frac{\partial(n_\mathrm{sb}/n_\mathrm{d,out})}{\partial n_{\mathrm{m},i}}\left( {\Omega_{\mathrm{m},i}}^2 + {\kappa_\mathrm{tot}^k}^2/4 \right),
\end{equation}
where all the remaining parameters are determined from independent measurements.
Using the data shown in Fig.~\ref{fig:SI_g0}b, the optomechanical coupling rate between the collective microwave mode and the mechanical oscillator is found to be $(\eta_{i}^{k} g_{0,i})/2\pi = 2.2$~Hz for $k=10$ and $i=6$. Using the known participation ratio $\eta_{i}^{k}$, the optomechanical coupling rate at site $i$ is found to be $g_{0,i}/2\pi = 12$~Hz for $i=6$.

Since all the participation ratio $\eta_i^k$ are determined as discussed in Sec.~\ref{subsec:eta_extraction}, all the optomechanical coupling rate $g_{0,i}$ are determined as follows.
Form the optomechanical damping rate, the unnormalized participation ratio $\widetilde{\eta_i^k}$ is experimentally obtained by Eq.~(\ref{eq:unnorm_eta}).
Using the relation described in Eq.~(\ref{eq:eta_eta}), the relative optomechanical coupling rate at site $i$ can be determined as
\begin{equation}
	\bar{g}_{0,i} = \frac{g_{0,i}}{\sum_i g_{0,i}} = \frac{\left(\widetilde{\eta_i^k}/\eta_i^k\right)}{\sum_i \left(\widetilde{\eta_i^k}/\eta_i^k\right)}.
\end{equation}
We obtain the relative coupling rate using the modeshapes for $k=10$ of the 1D chain.
From the mechanical sideband measurement, we have already known one of the optomechanical coupling rate, i.e.\ $g_{0,i'}$ for $i' = 6$ in our case. Thus, the optomechanical coupling rate at site $i$ is determined as
\begin{equation}
	g_{0,i} = \frac{\bar{g}_{0,i}}{\bar{g}_{0,i'}} \: g_{0,i'}.
\end{equation}
Figure \ref{fig:SI_g0}c shows all the optomechanical coupling rate between the bare microwave mode and the mechanical oscillator at site $i$ of the 1D chain.
We apply the same measurement and analysis for the 2D lattice. All the optomechanical coupling rates in the 2D lattice are determined as shown in Fig.~\ref{fig:SI_g0}d.

As shown in Figs.~\ref{fig:SI_g0}c and d, the single-photon optomechanical coupling rates are fitted well to the inverse of the square root of the mechanical frequency. This can be interpreted by the fact that the zero-point fluctuation of motion of a mechanical oscillator is proportional to the inverse of the square root of the mechanical frequency, i.e.\ $x_\mathrm{ZPF} = \sqrt{\hbar/(2m_\mathrm{eff}\Omega_\mathrm{m})}$, where $m_\mathrm{eff}$ is the effective mass~\cite{SI_aspelmeyer2014cavity}.

\begin{figure*}[h]
	\includegraphics[scale=1]{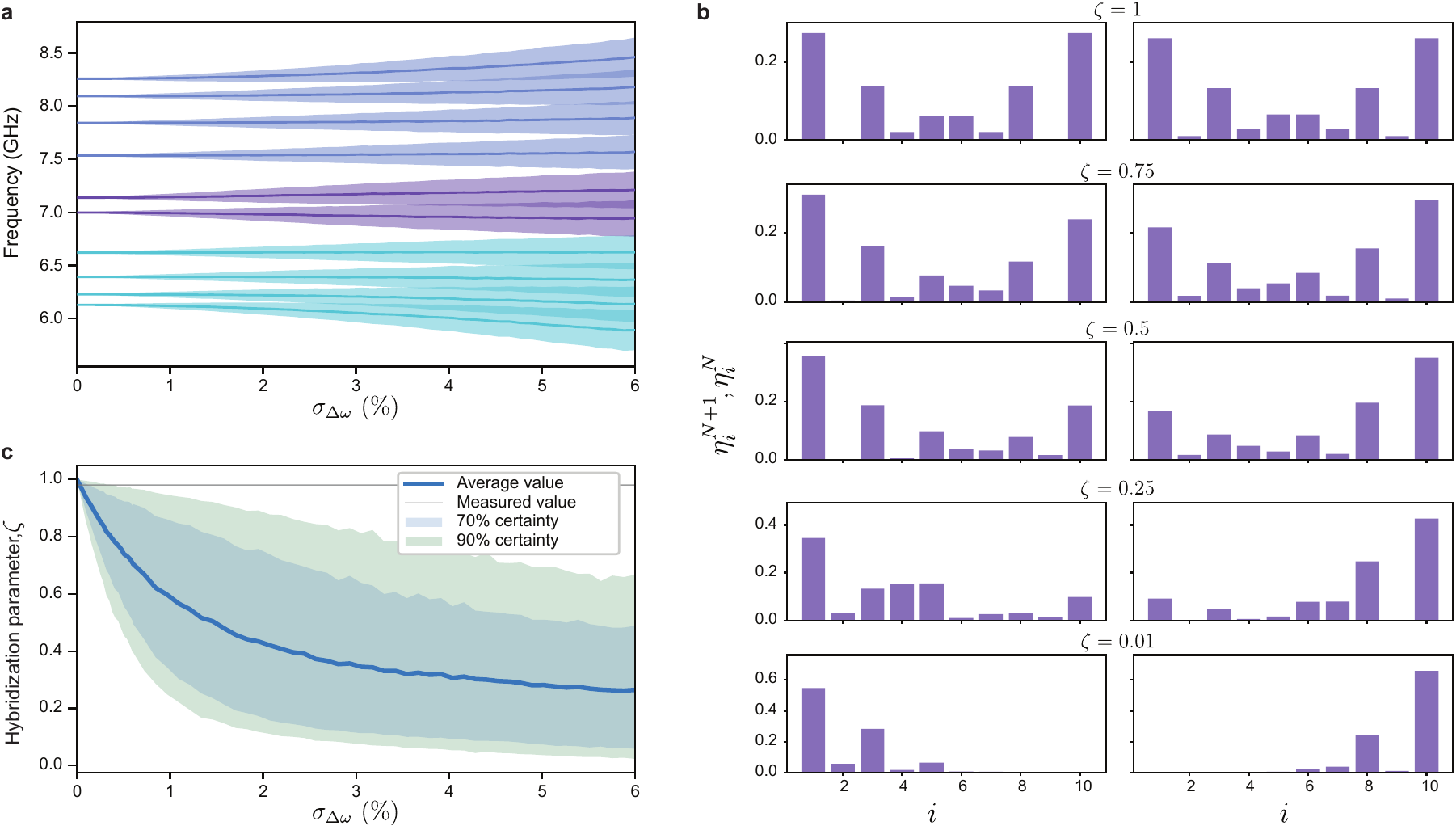}
	\caption{ \textbf{Disorder effect in SSH chains}. \textbf{a}, Effect of cavity frequency disorder on the mode spectrum of a 10 cell SSH chain considering the same parameters as the device discussed in the main text. For every relative frequency disorder standard deviation, $\sigma_{\Delta\omega}$, 4000 random cases were generated. The solid lines show the average mode frequency. The shades show the statistical standard deviation of mode frequencies. \textbf{b}, Example of disordered topological edge states with different hybridization factors, $\zeta$. \textbf{c}, Plot of hybridization factor versus frequency disorder standard deviation. The solid blue line shows the statistical average over 4000 random points for each error percentage. The shades show bonds for 70\% and 90\% statistical certainty. The gray line identifies the measured hybridization factor in the 1D chain discussed in the main text. \label{fig:SI_disorder}}
\end{figure*}

\begin{figure*}[h]
	\includegraphics[scale=1]{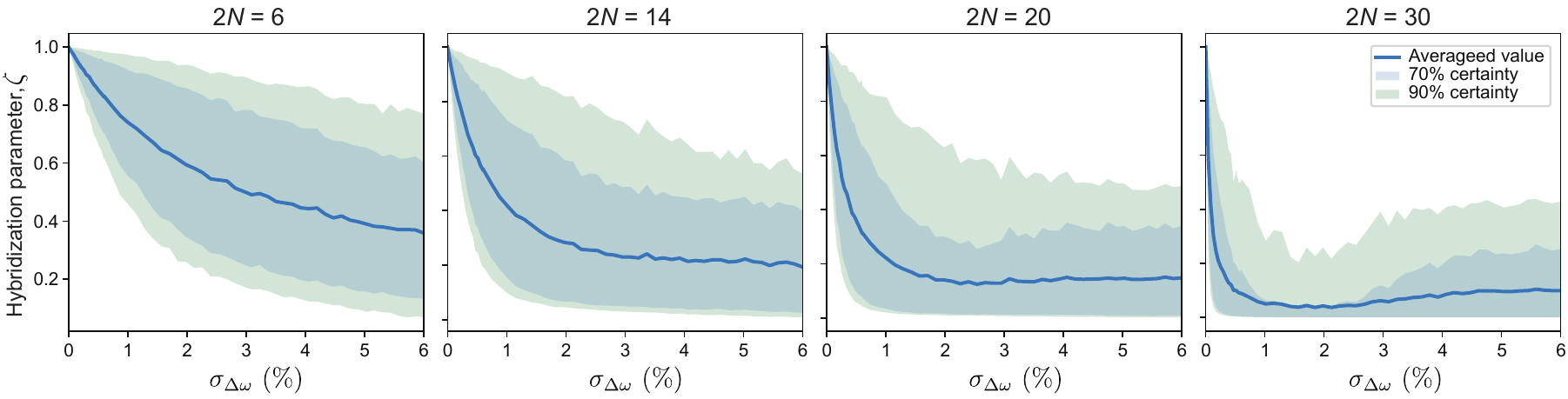}
	\caption{ \textbf{Disorder effect on hybridization for different chain lengths}. The hybridization factor $\zeta$ in relation to the frequency disorder is shown for several chain lengths with same the parameters. In longer chains, the hybridization is more sensitive to disorder. \label{fig:SI_disorder_length}}
\end{figure*}

\section{Numerical analysis}
\subsection{Disorder effect and edge state hybridization in SSH chains}
\label{subsection:disorder}
Here we discuss the effect of disorder in the finite SSH chains. In any realized system, disorder and fluctuations can distort the system's parameters and consequently deviate the response from the ideal model. There are two possible types of disorder in the coupled circuit chains: cavity frequency fluctuation and mutual coupling rate fluctuation. In the case of electromechanical arrays introduced in this work, the coupling rate fluctuation is related to the -lithographically defined- self and mutual inductances disorder. In contrast to capacitive coupling where the coupling rate depends on the local electric field between closely spaces electrodes, here the coupling rate depends on the longer-range magnetic field generated by spirals and is more robust to the geometry distortions that arise from the lithographic accuracy ($\sim 500$ nm). The dominant source of disorder is the capacitor's gap size fluctuation, which is strongly depending on CMP planarization uniformity as well as the etched trench depth variation. The gap size fluctuation was observed up to $1\%$ using mechanical profilometry (which corresponds to 0.5\% cavity frequency fluctuation), \added{hence $g_0$ variations are small.}

To study the effect of disorder on the energy spectrum and mode structure of the SSH chains, here we consider a stochastic error on the cavity frequencies with normal distribution:
\begin{equation}
	\omega_{c, i} = \omega_c \ ( 1 + \mathcal{N}(0,\sigma_{\Delta\omega}))
\end{equation}
Where $\mathcal{N}$ is a Gaussian \replaced{random }{stochastic} variable \added{with zero mean} and \added{standard deviation} $\sigma_{\Delta\omega}$ expressing the relative cavity frequency error. Assuming coupling rates and resonance frequencies same as the 10 cell SSH chain discussed in the main text, we numerically calculate the energy spectrum and modeshapes for various relative disorder standard deviations, $\sigma_{\Delta\omega}$. Figure \ref{fig:SI_disorder}a shows the result for frequency disorder up to $6\%$, averaged over 4000 random distributions for each standard deviation. The uncertainty ($\pm$ standard deviation) in energies increases with the disorder as shown with shaded areas in Fig.~\ref{fig:SI_disorder}a and leads to the regime that two eigenmodes are not distinguishable anymore, as they overlap. As mentioned in the main text, the ideal SSH model in finite chains always results in two topological edge states, both with co-localized modeshapes on two edges regardless of the size of the chain (See Fig.~1h in the main text). In contrast to eigen energies that are robust against disorder, the topological modeshapes hybridization will break down sooner due to disorder and result in two edge states each one only localized on one end of the chain. To quantitatively describe this effect, we defined the edge states hybridization factor, $\zeta$, defined by:

\begin{equation}
	\zeta = \frac{1}{2} \left(\frac{\min\{\eta_1^N,\eta_{2N}^N\}}{\max\{ \eta_1^N,\eta_{2N}^N\}} + \frac{\min\{\eta_1^{N+1},\eta_{2N}^{N+1}\}}{\max\{\eta_1^{N+1},\eta_{2N}^{N+1}\}} \right)
\end{equation}
\added{where again $\eta_i^k$ is the energy participation ratio of site $i$ to the collective microwave mode $k$, hence} $\eta_i^N$ and $\eta_i^{N+1}$ are the energy participation ratios of the topological edge modes in a $2N$ cell SSH chain. For a fully co-localized topological modeshape $\zeta = 1$, while it reduces to 0 for the fully non-hybridized (single side-localized) case (Fig. \ref{fig:SI_disorder}b). We calculated the average and standard deviation of $\zeta$ over the statistical pool for various $\sigma_{\Delta\omega}$. Figure \ref{fig:SI_disorder}c shows that hybridization is dramatically affected by disorder. Based on the modeshape measurement data presented in the main text, the hybridization factor for the 10 cell SSH device is $\zeta^\mathrm{meas} = 0.98$ which corresponds to the the stochastic frequency disorder in the range of $\sigma_{\Delta\omega} = (0.01\%, 0.38\%)$ with 90\% certainty. This indicates small relative disorder in the gap size ($\sigma_{\Delta\mathrm{gap}} = \frac{1}{2} \sigma_{\Delta\omega}$). It's worth mentioning that hybridization is also strongly affected by the length of the SSH chain, where for an example of 20 cell chain with the same parameters as the 10 cell case, the hybridization factor reduces sharply at smaller $\sigma_{\Delta\omega}$ as shown in Fig.~\ref{fig:SI_disorder_length}.

\section{Experimental observation of edge state localization}
\begin{figure*}[h!]
	\includegraphics[scale=1]{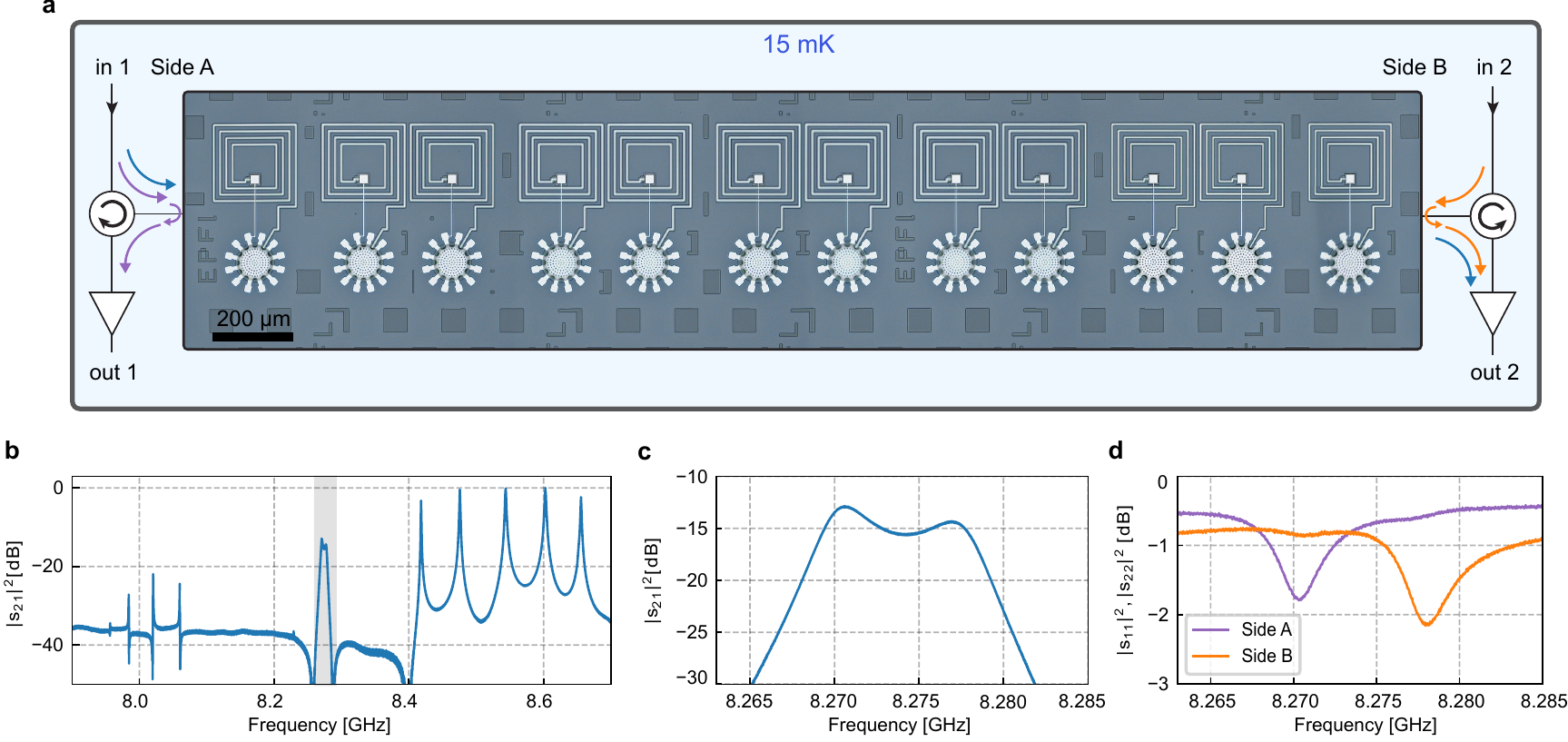}
	\caption{ \textbf{Experimental evidence of edge state localization.} \textbf{a}, First generation 12-site topological SSH array measured with reflection capabilities from both sides. \textbf{b,c}, VNA $S_{21}$ transmission measurement of the device. \textbf{d}, VNA $S_{11}$ measurements from both sides of the device. The horizontal axis is the same as in \textbf{c}.
		\label{fig:SI_localized}}
\end{figure*}

As explained in Section \ref{subsection:disorder}, in order to observe well-hybridized edge states instead of localized ones, the disorder in the cavity frequencies needs to be low enough. The physical quantity to compare the disorder to are the coupling rates $J,J'$ between the array sites. We say that the disorder is low enough when it's negligible compared to the coupling rates. The device shown in the main text belongs to a second generation design. The first generation design, visible in Fig. \ref{fig:SI_localized}a featured a different inductor design and the coupling rates were lower: $J'/2\pi = 127~\mathrm{MHz}$, $J/2\pi = 260~\mathrm{MHz}$ by fitting the hybridized modes' frequencies with an SSH model. The superconducting optomechanical array was also inductively coupled to input-output waveguides, albeit with a different, microstrip, geometry. In Fig. \ref{fig:SI_localized} we show a 12 site topological realization of this generation of devices. Thanks to the use of two circulators on the two input/output ports of the chip this device could be measured either in transmission or in reflection from both sides. Firstly, we see in the transmission spectrum (\ref{fig:SI_localized}b-c) that the edge modes have abnormally low transmission, even though their current density at the edges should be very big, resulting in a great external coupling rate. This is because the edge states are actually localized at one of the edges each. This can be seen in the reflection measurements in Fig. \ref{fig:SI_localized}d where when we look at the reflection from one side we see only one of the topological peaks, and the other one is only visible at the opposite side of the array. The horizontal axis in subpanels c and d are the same.

\pagebreak

\section*{Supplementary References}
\bigskip

\end{document}